 \documentclass[iop,revtex4]{emulateapj}
\usepackage{natbib}
\usepackage{amsmath}
\usepackage{graphicx}
\usepackage{color}
\usepackage{ulem,color}
\definecolor{mycolor}{rgb}{0.99, 0., 0.5}

\begin{document}
\title{Radio Counterparts of Compact Binary Mergers detectable in
  Gravitational Waves:
A Simulation for an Optimized Survey}
\author{K. Hotokezaka$^1$, S. Nissanke$^2$, G. Hallinan$^3$,  T. J. W. Lazio$^4$, 
E. Nakar$^5$, and T. Piran$^1$}

\affil{
$^1$Racah Institute of Physics, Hebrew University, Jerusalem, 91904, Israel\\
$^2$Institute of Mathematics, Astrophysics and Particle Physics, Radboud University, Heyendaalseweg 135, 6525 AJ Nijmegen, The Netherlands\\
$^3$Cahill Center for Astronomy, MC 249-17, California Institute of Technology, Pasadena, CA 91125, USA\\
$^4$Jet Propulsion Laboratory, California Institute of Technology, Pasadena, CA 91109, USA\\
$^5$Raymond and Beverly Sackler School of Physics \& Astronomy, Tel Aviv University, Tel Aviv 69978, Israel\\
}

\begin{abstract}
Mergers of binary neutron stars and black hole-neutron star binaries
produce gravitational-wave~(GW) emission and outflows with significant kinetic energies. These outflows result in
radio emissions through synchrotron radiation.
We explore the detectability of these synchrotron generated radio signals
by follow-up observations of GW merger events lacking a detection of
electromagnetic counterparts in other wavelengths.
We model radio light curves arising from (i) sub-relativistic merger ejecta and (ii)
ultra-relativistic jets. The former produces radio remnants on timescales of a few years
and the latter produces $\gamma$-ray bursts in the direction of the jet and orphan-radio afterglows 
extending over wider angles on timescales of weeks.
Based on the derived light curves, we suggest an optimized survey at $1.4$\,GHz 
with five epochs separated by a logarithmic time interval.
We estimate the detectability of 
the radio counterparts of simulated GW-merger events to be detected by advanced LIGO and Virgo 
by current and future radio facilities. The detectable distances for these GW merger events 
could be as high as 1~Gpc.  
20--60\% of the long-lasting radio remnants will be detectable in the case of the moderate kinetic energy 
of $3\cdot~10^{50}$~erg and a circum-merger density of
$0.1{\rm~cm^{-3}}$  or larger, while $5$--$20\%$ of the orphan radio afterglows with kinetic energy of $10^{48}$~erg will be detectable.
The detection likelihood increases if one focuses on the well-localizable GW events.
We discuss the background noise due to radio fluxes of host galaxies
and false positives arising from extragalactic radio transients and
variable Active~Galactic~Nuclei and we show that the quiet radio transient sky is
of great advantage when searching for the radio counterparts. 
\end{abstract}


\section{Introduction}\label{sec:intro}


Joint electromagnetic~(EM) and gravitational-wave~(GW) observations of
compact object mergers could allow for unprecedented measurements of
astrophysical processes in strongly-curved 
spacetimes. Such observations are possible thanks to
a suite of advanced GW detectors and multi-wavelength wide-field
surveys that recently came online last year. A hundred
years after Einstein's initial prediction of GWs, the first observations of GWs
from binary black hole mergers were measured simultaneously by the two LIGO
detectors on September 14th~2015~\citep{PRL_GW:2016} and December 26th~2016. These
first detections mark a new era of transient-GW
astronomy (see~\citealt{EmfollowupGW150914:2016} for an overview and references
therein). Aside from binary black holes, the other most
numerous sources predicted for kilohertz GW detectors, such as LIGO \citep{LIGOPR:2009}
and Virgo \citep{Virgo:2015}, are the mergers of double neutron
star (DNS) and black hole-neutron star (BH-NS) systems
\citep{LIGOprospects:2013}. Through their dimensionless wavestrain
$h(t)$, GWs encode key information about the
progenitors' physical and geometric properties that probe the sources'
bulk dynamic properties. These
complicated mergers should also produce EM signatures from energetic
matter outflows at different timescales. GW-EM detections will increase our confidence in
detections of the GW signal and are essential for the identification of the hosts and measuring their distances~\citep{KochanekPiran93}. Additionally,
EM counterparts of mergers inform us about the environment, and the matters' thermodynamic and composition
properties.  We require both EM and GW measurements to fully
understand neutron star binary mergers. In this paper, we focus on how to
observe both GW and late radio emissions from neutron star binary mergers. 

Based on population synthesis models calibrated by
observed Galactic DNS systems, advanced versions of GW detectors operating at their design sensitivity,
such as LIGO, Virgo and KAGRA, expect to
detect from 0.4 to 400 (with a mean value of 40) DNS mergers per year out to several hundred
Mpc \citep{abadieetal:2010, LIGOprospects:2013, Dominik:2015}. Such an
expected rate is consistent with the observed short-$\gamma$ ray burst
(sGRB) rate assuming a half-opening beaming angle of $\approx 10^{\circ}$
\citep{GuettaPiran05, GuettaPiran06,nakar2006ApJ,Coward+12,wanderman2015MNRAS} and solar system
abundance patterns of $r$-process elements \citep{Piran+14,bauswein2014ApJ,shen2015ApJ,vandevoort2015MNRAS,wehmeyer2015MNRAS,hotokezaka2015NatPh,vagioni2016MNRAS}.
Based entirely on population synthesis models as we have yet to observe a BH-NS system, we expect to detect 0.2 to 
200 BH-NS systems per year with detectable distances $\sim 1$ Gpc. 

For EM follow-up observations of compact object binary mergers, a fundamental challenge
is the poor sky resolution of a single GW interferometer. Localizing a
transient source on the sky depends primarily on triangulating the
GW signal's arrival times using networks of three or more
GW detectors (e.g., \citealt{Fairhurst:2010}). We must anticipate GW errors that span from 0.5 to
100s square degrees sometimes with multimodal islands (e.g.,
\citealt{Wen:2010,Nissanke:2011, Klimenko:2011,
  Schutz:2011,Veitch:2012,Nissanke:2013,Rodriguez:2014,
  Grover:2014}). The wide range depends on the sources'
signal-to-noise ratio, their sky position, the number of detectors in the network and whether they are being used
coherently. By exploiting the detectors' quadrupolar antenna
functions, two or more GW detectors can also localize events from several
hundreds to a thousand square degrees within arcs of the
sky \citep{Kasliwal:2014,Singer:2014}. Therefore EM observatories must
be prepared for triggers
from GW detectors that span these huge swaths of the sky of several
hundred degrees released within minutes to thirty minutes of the
merger being detected in GWs (e.g., \citealt{Cannon:2012,Singer:2014}).

At present several EM counterparts of DNS and BH-NS
mergers have been proposed at different wavelengths and emission
timescales. These include sGRBs and their afterglows \citep{Eichler:1989,Paczynski:1991,Narayan:1992}, optical-near IR
counterparts called macronova or kilonova (e.g.,
\citealt{LP98,Kulkarni:2005,Metzger:2010,roberts2011ApJ,Metzger:2012,Barnes:2013,Tanaka:2013,
Tanvir:2013, Berger:2013,Yang+15,jin2016}) and
long-lasting radio merger remnants \citep{NakarPiran11,Piran+13,HotokezakaPiran15}. Additional radio
emission and, in particular, a coherent prompt
radio pulse from a magnetically-driven, relativistic plasma outflow prior to the DNS
merger have also been
proposed
\citep{Hansen:2001,Pshirkov:2010,paenzuela2013PRL,totani2013PASJ}. These
prompt radio counterparts are both harder to detect and less certain than the above three
candidates to DNS mergers, so we do not discuss these in this
paper; see \cite{Chu:2015} for a detailed study.



The radio counterparts of neutron star merger events promise to 
uniquely probe their energetics and environment. 
The interaction of the merger ejecta with the circum-merger medium 
results in the radio remnant with timescales of a few months to
years \citep{NakarPiran11}. The radio luminosity can be brighter by a
couple of orders of magnitude than that
of a typical radio supernovae. Although such radio remnants have not been observed yet
after sGRBs~(e.g., \citealt{metzger2014MNRAS,horesh2016ApJ}), they may be detectable by
follow-up observations for GW merger events since they will take place at much
closer distances than sGRBs. In addition, the sGRB jet
produces the radio afterglow with timescales of a few weeks,
which can be seen provided that the viewing angle
is not too large. In fact, radio afterglows have been already detected 
for four sGRBs~(e.g., \citealt{fong2015ApJ}) and they will be detectable counterparts
of GW merger events~\citep{feng2014}.
Recent observations suggest that the radio transient 
sky is far more quiet compared to the optical one, so we expect far fewer false
positives by a factor of a hundred or more depending on the sky location than in the optical-IR (e.g., \citealt{frail2012ApJ,mooley2013ApJ,Mooley16}). 

This paper focuses on strategies for detecting
and identifying, following a detection of a GW event, two different radio post-merger counterparts:
i) the longer duration radio remnants that may last from months to years and
ii) the faster ultra-relativistic radio afterglows components that last weeks to months 
\citep{NakarPiran11,Piran+13,HotokezakaPiran15,margalit2015MNRAS}. Importantly, we do
not consider here the ideal case in which  a sGRB or an optical counterpart to a GW merger
have been detected. Note that a sGRB and its bright afterglow are expected only for
on-axis events, which are most likely a small fraction of all events. 
In this case, the high-energy signal will immediately
provide a much better localization enabling very accurate radio 
identification of the merger's position.  

To define an optimal strategy for detecting radio counterparts, we run a simulation
that considers both GW and radio detectability of neutron star mergers with a
slew of GW and radio telescopes.  Our simulation comprises six steps and the structure of the paper follows these:
\begin{enumerate} 
\item[i)] We construct underlying astrophysically-motivated distributions of DNS
and BH-NS (with a 5 M$_{\odot}$ BH) mergers
(Sec.~\ref{sec:catalog}).
\item[ii)] We now 
simulate GW wavestrains $h (t) $ for each binary in our underlying
catalog and ask whether the wavestrain
is detectable with different GW networks (Secs.~\ref{sec:GWdetector}
and~\ref{sec:GWPE}).
\item[iii)] For a random subset of
GW detectable sources we estimate the entire set of source parameters
including the sky position, distance and inclination angle to each
binary (Secs.~\ref{sec:GWdetector} and~\ref{sec:GWPE}).
\item[iv)] We perform an
approximate mapping from the binary progenitor to the dynamical
mass ejecta and ultra-relativistic outflows using a range of numerical-relativity and 
smooth particle hydrodynamic (SPH) simulations (Sec.~\ref{sec:radiomap}). 
\item[v)] We compute different radio afterglow and remnant signatures for
each GW detectable binary using our set of ejecta models (Sec.~\ref{sec:radiolightcurves}). 
\item [vi)] We check
whether the different radio signatures are detectable by a
slew of radio telescopes at different frequencies (Sec.~\ref{sec:detectability}). 
\end{enumerate}

Following these steps we outline the
challenges to ensure a secure identification of a GW-radio counterpart amongst
other astrophysical transients and suggest an optimal strategy to overcome them  (Sec.~\ref{sec:identification}). 
We  summarize the results, compare them with previous work  (Sec.~\ref{sec:Discussion}) and conclude with a discussion of its implications (Sec.~\ref{sec:Conclusion}).


\section{GW detectability of mergers}
\label{sec:GWdetectability}

We begin describing  how we construct neutron star binary merger
catalogs and the methods used to derive sky location,
distance and inclination angle measures for populations of DNS or BH-NS
binary mergers detectable in GWs. We first outline the schema  of our
method. Based on Sections 2 -- 5 of \cite{Nissanke:2013}
(henceforth denoted NKG13), we then describe technical aspects of
simulating the anticipated sky positions and distance measurements. 
 
\subsection{Catalogs of neutron star binary populations}
\label{sec:catalog}

We construct two underlying astrophysical neutron star binary catalogs with either $4 \times 10^4$ DNS or $3 \times
10^4$ NS -- $5\,M_{\odot}$ BH populations. In this section, we do not
take into account DNS and BH-NS merger rates or GW detection
volumes (see Sec.~\ref{sec:GWdetector}). However, each binary in our two
catalogs is described by the nine parameters that are encoded in the
theoretical predictions of the GW wavestrain (or GW waveform) for
non-spinning compact binary systems (\citealt{Blanchet:2014LRR} and
references therein). As discussed in Sec.~\ref{sec:GWPE},
these parameters are critical to the GW detectability as the GW
waveform depends on them. They include the two compact object masses $m_1$ and $m_2$, a sky position
$\bf{n} \equiv (\theta,\phi)$, a luminosity distance $D_L$, an inclination angle $\iota$,
a polarization angle $\psi$ and a time and phase of GW merger. The
colatitude $\theta$ and longitude $\phi$ are related to the
declination $\delta$ and right ascension $\alpha$, by $\theta = \pi/2
- \delta$ and $\phi=\alpha-$GAST
respectively, where GAST is Greenwich Apparent Sidereal Time. Apart
from the component masses, the luminosity distance and the time of merger, we assume random
distributions in the other parameters: $p \, (\cos \iota )
\propto \mathrm{const}$ with $\cos \iota \in [-1,1]$, and a
random sky position such that $p \, (\cos \theta) \propto
\mathrm{const}$  with $\cos \theta \in [-1,1]$,  and $p \, (\phi) \propto
\mathrm{const}$ with $\phi \in [0, 2 \pi]$. We set the time of
merger to be a constant value.

We specify neutron star and black hole component masses of $ 1.4 \,M_{\odot}$ and $5
\,M_{\odot}$ respectively. As indicated by SPH and numerical relativity
simulations, we choose these particular values in masses such that the mergers of the two objects
will form some dynamic ejecta and the neutron star will not be
swallowed entirely by the gravitational potential of a too massive
BH (see e.g., \citealt{Foucart:2012,kyutoku2015PRD,kawaguchi2015PRD}). The ejecta create matter outflows and
therefore are potentially responsible for radio
afterglows and long-lasting radio remnants \citep{NakarPiran11,Piran+13,HotokezakaPiran15}. For simplicity, we
assume in our simulations that the neutron stars and black holes are non-spinning; in
the physical Universe we expect BHs in
BH-NS systems to have a considerable spin, so there may be more
dynamical ejecta produced for a neutron star orbiting in a prograde
orbit around a black hole with a maximal
spin. Furthermore, modulo selection effects and the small sample number, observations of Galactic DNS systems imply a narrow mass
distribution for DNS systems. 
We assume that the DNS and BH-NS systems chosen in our catalogs
will merge within the Hubble time.

For systems with distances $< 200$ Mpc, we assume that the
spatial distribution of neutron star binaries traces host galaxy light. As
described in NKG13, we use and correct for B-band luminosity
using the  ``Census of the Local
Universe" (CLU) with information
compiled from different galaxy catalogs (e.g., HyperLEDA, NED, EDD;
see \citealt{KasliwalPhD:2011} and \citealt{gehrels2016ApJ} for details). For those systems located with distances $> 200$
Mpc, we assume that the neutron-star binary merger distribution follows a constant comoving volume density
in a $\Lambda$CDM Universe \citep{komatsu09}.

\subsection{Gravitational-wave detector networks}
\label{sec:GWdetector}

We consider different GW networks comprising the advanced versions of
LIGO, Virgo, KAGRA and LIGO India; see \cite{Aasi:2013} for a
observing scenario roadmap
of the LIGO and Virgo detectors over the next decade. Each detector will operate in staggered science modes with increasing
sensitivity until they reach their target design sensitivity. For instance, advanced LIGO began its first observation run
in September 2015, advanced Virgo could begin its first observation run
as early as September 2016 and KAGRA could begin as early as 2019. We
make several assumptions in this work; for instance, we assume Gaussian, stationary, and zero-mean noise that is independent and uncorrelated between detectors. We
also take the anticipated noise sensitivity curve for a
single advanced LIGO detector, given in \cite{Aasi:2013}, for
broadband tuning, to be representative of all our detectors, imposing
a low-frequency cut off of 10 Hz; see the recent \cite{Berry:2015} for parameter
inference methods using non-Gaussian, non stationary noise. In reality, LIGO, Virgo
and KAGRA will have different noise sensitivities in
different frequency bands because of variations in each instrument's
design, and systematic calibration effects must be taken into effect.

In the rest of the paper, we use the following notation to
describe different GW networks with $N$ detectors: \emph{GW Net3} or
GW network 3 comprises LIGO Hanford, LIGO Livingston and Virgo, 
and \emph{Net5} or network 5 consists of LIGO Hanford, LIGO
Livingston, Virgo, LIGO India and KAGRA. See the recent works of
\cite{Singer:2014,Berry:2015,Kasliwal:2014} for sky location works
using either only two LIGO detectors or the three LIGO-Virgo detectors
at staggered (and not full) design sensitivities.

To implement a detectability criterion, we assume that each merger in
the two catalogs is detectable in GWs if its
GW expected network SNR\,$>$\,8.5 (see Sec.~3.3 in NKG13 for different triggering
criteria and GW networks).

\subsection{Extracting the binaries' sky location, luminosity distance and
  inclination angle from the GW signal}
\label{sec:GWPE} 

Based on optimal matched filtering
(\citealt{finn1992PRD,cutler1994PRD}), we extract the luminosity distance
$D_L$, the inclination angle $\cos \iota$, and sky position ${\bf n}$ for
each DNS or BH-NS binary merger using knowledge of the expected GW
signal.

Regarding the predicted GW waveform, we use only the early inspiral
(pre-merger) portion of the
waveform, which for low-mass systems provides most of the signal for
advanced detectors \citep{Flanagan:1998}. The inspiral waveform is modeled accurately using post-Newtonian (PN) expansions in
general relativity and is based on an expansion in $\sim v^2/c^2$, where $v$ is
the characteristic orbital speed for gravitationally-bound systems \citep{Blanchet:2014LRR}. Specifically, we use the
non-spinning restricted 3.5PN waveform in the frequency domain for the
two GW polarizations $h_+$ and $h_{\times}$; see
Eqn.~(1) in NKG13 and Eqns.~(7)-(14) in \cite{Nissanke:2010}. The overall
Newtonian amplitude of the GW waveform encodes the source's orientation, sky location,
luminosity distance and redshifted chirp mass. The GW phase depends
on the redshifted chirp mass, redshifted reduced mass, the phase and time of
merger. The detector antenna functions depend on
${\bf n}$ and the binary's polarization angle. The time of flight from source at direction
${\bf n}$ to detector at location ${\bf r}$ involves the scalar
product ${\bf n}\cdot{\bf r}$, and differences in time of flight between
detectors in the network dominate how well we can localize the event
on the sky.

To infer the sky position $(\cos\theta, \phi)$, luminosity distance
$D_L$, and $\cos \iota$, we explicitly
map out the posterior PDF for all source parameters (including chirp
mass, orientation, etc.) after simulating a data stream at a detector
using Markov Chain Monte Carlo (MCMC methods). The Metropolis--Hastings MCMC algorithm used is
based on a generic version of CosmoMC, described in \cite{lewis02},
and is detailed in Sec.~3.3 of \cite{Nissanke:2010}. Other parameter
estimation methods used frequently in the LIGO-Virgo analysis pipelines are summarised in
\cite{Veitch:2015} and \cite{PEGW150614:2016}

We take prior distributions in all source parameters to be flat over the region of sample space where
the binary is detectable at an expected network signal-to-noise ratio (SNR) = 3.5. The
expected network SNR is defined as the root-sum-square of the expected
individual detector SNRs. For each MCMC simulation, we derive distance
and inclination angle measures, and solid angle areas over $(\cos
\theta, \phi)$ for 68$\%$, 95$\%$, and 99$\%$ confidence regions.

\section{Radio signatures of compact binary mergers}
\label{sec:radiosignal}

Radio emission from post-merger events are produced via synchrotron radiation
of accelerated electrons in shocks formed between expanding outflows and 
circum-merger material~\citep{NakarPiran11,Piran+13,HotokezakaPiran15}. 
 We briefly review synchrotron radiation from sub-relativistic and ultra-relativistic outflows.
We then provide our models of radio emission arising from these outflows.

\subsection{Synchrotron radiation of expanding outflows}
\label{subsec:synchrotron}

We turn now to estimate  the properties of radio signals arising from
outflows expanding into homogeneous circum-merger material.
We consider first long-lasting radio remnants arising from sub-relativistic merger ejecta
 and then orphan GRB afterglows arising from ultra-relativistic jets.

{\it Long-lasting radio remnants}: 
The outflow expands with an initial velocity until 
the kinetic energy of the swept-up material is comparable to the ejecta's own kinetic energy. 
For ejecta thrown out at mildly and sub-relativistic speeds, 
the deceleration timescale is given by\,~(e.g., \citealt{NakarPiran11})
\footnote{One may find that the value of $t_{\rm dec}$ estimated here is slightly different from
the one in \cite{NakarPiran11}. This is because we round off the number whilst \cite{NakarPiran11}
round the value down.}:
\begin{eqnarray}
t_{\rm dec} \approx 80~{\rm day}~E_{50}^{\frac{1}{3}}n^{-\frac{1}{3}}\beta_{0}^{-\frac{5}{3}},\label{t1}
\end{eqnarray}
where $E$ is the kinetic energy of the ejecta,
$\beta_{0}$ is the ejecta's initial velocity in units of
the speed of light, and $n$ is the circum-merger density.
Here and elsewhere the notation $Q_{x}$ indicates the value of the variable $Q/10^{x}$ in cgs units.
The peak times of the light curves arising from the ejecta are scaled with Eqn.~(\ref{t1}).

Electrons are accelerated in shocks between the ejected outflow and the circum-merger material
and emit synchrotron radiation. Here we assume a power-law electron distribution with an index $p$. 
The characteristic frequency of the synchrotron radiation is:
\begin{eqnarray}
\nu_{m} \approx 1~{\rm GHz}~n^{1/2}\epsilon_{B,-1}^{1/2}\epsilon_{e,-1}^{2}\beta^{5},\label{num}
\end{eqnarray}
where $\beta$ is the ejecta's velocity, $\epsilon_{e}$ and $\epsilon_{B}$ are
the conversion efficiencies from the shock's internal energy into the energy of the accelerated electrons and magnetic
field respectively. The synchrotron spectra have a maxima at $\nu_{m}$ as long as $\nu_{m}$ is above the
synchrotron self-absorption frequency~(see below).
After $t_{\rm dec}$, $\nu_{m}$ decreases with time and the flux density at a given frequency above 
$\nu_{m}$ declines with time. 
The peak flux density above $\nu_{m}$ is estimated as: 
\begin{eqnarray}
F_{{\rm peak},~\nu>\nu_{m}} \approx  8~{\rm mJy}~E_{50} \, n^{\frac{p+1}{4}} \, \epsilon_{B,-1}^{\frac{p+1}{4}}
\, \epsilon_{e,-1}^{p-1} \, \beta_{0}^{\frac{5p-7}{2}} \nonumber \\
~~~~~~~~~~~~~~~\times \left(\frac{D_{L}}{200~{\rm Mpc}}\right)^{-2}
\left(\frac{\nu}{1.4~{\rm GHz}}\right)^{-\frac{p-1}{2}},\label{f1}
\end{eqnarray} 
where $\nu$ is the observation frequency, and
$D_{L}$ is the luminosity distance to the source.
In this work we set the values of the parameters: $\epsilon_{e}=\epsilon_{B}=0.1$ and $p=2.5$.
This choice is motivated by observations of late radio afterglows in long
GRBs and typical radio supernovae \citep{chevalier1998ApJ,frail2000ApJ,frail2005ApJ}.

The above estimates of the peak timescale and flux density are valid only when
synchrotron self-absorption is unimportant. In the case of either
sufficiently high circum-merger densities or low observation frequencies,
such absorption can be important. 
The synchrotron self-absorption frequency $\nu_{a}$ at $t_{\rm dec}$ is estimated as:
\begin{eqnarray}
\nu_{a{\rm ,dec}} \approx 1.3~{\rm GHz}~E_{50}^{\frac{2}{3(p+4)}}
n^{\frac{3p+14}{6(p+4)}}
\epsilon_{B,-1}^{\frac{2+p}{2(p+4)}} \nonumber \\
~~~~~~~\times \epsilon_{e,-1}^{\frac{2(p-1)}{p+4}}
\beta_{0}^{\frac{15p-10}{3(p+4)}}.
\end{eqnarray} 
Below $\nu_{a,{\rm dec}}$,
the peak flux density and peak timescale are estimated as:
\begin{eqnarray}
F_{{\rm peak}, \, \nu<\nu_{a}}\approx 0.6~{\rm mJy}~E_{50}^{\frac{4}{5}}n^{\frac{1}{5}}\epsilon_{B,-1}^{\frac{1}{5}}
\epsilon_{e,-1}^{\frac{3}{5}} \nonumber \\
~~~~~~~~\times \left(\frac{D_{L}}{200~{\rm Mpc}}\right)^{-2}
\left(\frac{\nu}{150~{\rm MHz}}\right)^{\frac{6}{5}},\label{f2} 
\end{eqnarray}
and
\begin{eqnarray}
t_{\nu < \nu_{a} \, (t_{\rm dec})}\approx 570~{\rm day}~E_{50}^{\frac{5}{11}}n^{\frac{7}{22}}\epsilon_{B,-1}^{\frac{9}{22}}\label{t2}
\epsilon_{e,-1}^{\frac{6}{11}}\\ \nonumber
~~~\times \left(\frac{\nu}{150~{\rm MHz}}\right)^{-\frac{13}{11}}. 
\end{eqnarray}
\vspace{1.0cm}

\begin{table}[t]
\caption[]{The energetic properties of 
  DNS and BH-NS merger ejecta from different sets of numerical simulations (the
  notation {\bf GR} refers to a fully general relativistic simulation,
{\bf CF} to a conformally-flat simulation, {\bf Newton} to a non-relativistic
simulation, {\bf Mesh} to a grid-based hydrodynamics and {\bf SPH} to smooth-particle hydrodynamics).
\\ }
\label{tab:simulation}
\scalebox{0.9}{\begin{tabular}{lcccc}
\hline \hline
Type & Range of $\beta_{\rm ave}$  & Range of $E~[10^{50}{\rm erg}]$ & Scheme & Ref. \\\hline
DNS & $0.15,~0.3$ & $0.1,~10$ & GR,~Mesh & [1]\\
DNS & $0.15,~0.4$ & $0.5,~10$ & CF,~SPH & [2]\\
DNS & $0.1,~0.15$ & $2,~10$ & Newton,~SPH & [3]\\ \hline
BH-NS  & $0.2,~0.3$ & $10^{-3},~60$& GR,~Mesh & [4]\\
BH-NS & $0.2,~0.25$& $10,~40$ & CF,~SPH & [5]\\
BH-NS & $0.15,~0.2$& $6,~20$ & Newton,~SPH & [3]\\
\hline \hline\\
\end{tabular}}
{\scriptsize \\
References;\\
$[1]$ \cite{hotokezaka2013PRDa,sekiguchi2015PRD,radice2016},
$[2]$ \cite{bauswein2013ApJa},
$[3]$ \cite{rosswog2013RSPTA,Piran+13},
$[4]$ \cite{foucart2013PRD,kyutoku2015PRD,kawaguchi2015PRD},
$[5]$ \cite{just2015MNRAS}.
}
\end{table}

{\it Orphan GRB afterglows}: 
Compact binary mergers produce ultra-relativistic jets that result in sGRBs~\citep{Eichler:1989,Nakar2007}. 
Jets produce not only the prompt gamma-ray emission but also afterglows
at longer wavelength as a result of the interaction 
with the circum-merger material~(e.g.,~\citealt{sari1998ApJ}).
For relativistic jets, the emission is highly
beamed towards the jet axis, which has an important consequence
for the detectability. Observers only on or close to the jet axis can measure its bright emission.
On the contrary, observers far away from the jet axis can measure the faint
emission only after the jet's sufficient deceleration and its
subsequent interaction with its environment. 
Therefore the observed light curves depend strongly on the observers' viewing angle.
Roughly speaking the peak timescale of the relativistic radio afterglow for an
off-axis observer occurs when $\Gamma \sim \theta_{\rm obs}^{-1}$,
where $\Gamma$ is the jet's Lorentz factor and $\theta_{\rm obs}$ is the
observer's viewing angle. 

It is worth noting that the spectral shapes of GRB afterglows 
are different from those of the long-lasting radio remnants.
For GRB afterglows, as the characteristic frequency decreases with
time because of the relation $\nu_{m}\propto \Gamma^{4}$ during the jet's deceleration,
observers detect synchrotron radiation from higher ~(e.g., X ray) to
lower multi-wavelength frequencies.
When $\nu_{m}$ decreases to the radio frequencies,
the Lorentz factor, i.e, the beaming factor, is sufficiently low so that
off-axis observes can detect the late-time radio signals from the jet.
For observers on or close to the jet axis, observing at higher radio
frequencies (e.g., $5$~GHz) is preferable for avoiding flux losses due to synchrotron self-absorption.
On the contrary, for the long-lasting radio remnants and off-axis orphan afterglows,
observing at lower radio frequencies is preferable because the characteristic frequency $\nu_{m}$ 
is typically lower than $1$~GHz for sub-relativistic ejecta and off-axis afterglows.
Therefore, we focus mainly on frequencies of $150$~MHz and $1.4$~GHz
in the rest of this work.

\begin{table*}[t]
\begin{center}
\caption[]{
 The mean energetics chosen for our different ejecta models in the case
 of DNS, BH-NS, and sGRB-jet driven events. For the jet models, we assume
a viewing angle of $45^{\circ}$. The 4th--6th columns show the
radio peak luminosities at $1.4$~GHz with the circum-merger densities of $1,~0.1,$ and $0.01~{\rm cm^{-3}}$.\\
}
{\small
 \begin{tabular}{lccccccc} 
\hline 
\hline
Model & $E_{K}$~[erg] & $\langle \beta_{0}\rangle$~[c] & $L_{1.4\rm{GHz}}^{n=1}$~[{\footnotesize${\rm erg~s^{-1}Hz^{-1}}$}]
 & $L_{1.4{\rm GHz}}^{n=0.1}$
& $L_{1.4{\rm GHz}}^{n=0.01}$\\ \hline
DNS$_{h}$ & $10^{51}$ & $0.3$ & $4\cdot 10^{29}$ & $8\cdot 10^{28}$ & $10^{28}$\\
DNS$_{m}$ & $3\cdot 10^{50}$ & $0.25$& $8\cdot 10^{28}$ & $10^{28}$ & $2\cdot 10^{27}$\\
DNS$_{l}$ &$10^{50}$&  $0.2$ & $10^{28}$ & $2\cdot10^{27}$ & $3\cdot 10^{26}$\\
BH-NS$_{h}$ &$5\cdot10^{51}$ & $0.3$&  $2\cdot 10^{30}$ &$5\cdot 10^{29}$ & $7\cdot 10^{28}$\\
BH-NS$_{m}$ & $2\cdot 10^{51}$ & $0.25$& $5\cdot 10^{29}$& $8\cdot 10^{28}$ & $10^{28}$ \\
BH-NS$_{l}$& $5\cdot 10^{50}$ & $0.2$& $7\cdot10^{28}$& $9\cdot 10^{27}$ & $10^{27}$\\
{\sl strong-jet} & $10^{49}$ & $\sim 1$ & $3\cdot 10^{28}$& $10^{28}$ & $2\cdot 10^{27}$  \\
{\sl canonical-jet} & $10^{48}$ &$\sim 1$ & $4\cdot 10^{27}$& $10^{27}$& $2\cdot 10^{26}$ \\
\hline \hline\\
\label{tab:models}
\end{tabular}
}
\end{center}
\end{table*}

\subsection{An approximate mapping from progenitor to outflow models}
\label{sec:radiomap}

Various components of merger outflows may be produced with different velocities and
kinetic energies.
We consider here the expected range in energetics for the two post-merger radio counterparts that give rise to the most
  luminous radio remnants and are arguably the most robust~(see \citealt{HotokezakaPiran15})
  : 
  i) the long-lasting isotropic radio remnant that varies from a year to ten
  year timescale, and ii) the afterglow of an
  ultra-relativistic GRB jet that varies on a week to month timescale. 
Below we detail how we model these two outflows using the results of
numerical relativity and SPH simulations of the neutron star mergers themselves.

{\it Dynamical Mass Ejection:}

{\it DNS mergers}: We expect mergers to produce tidal tails which result in
gravitationally-unbound {\sl dynamical mass ejecta} that undergoes
shock heating processes. For a binary with given component masses and
different neutron stars' equation of
state, we estimate the range of ejectas' kinetic energy and velocity using
results that span a diverse set of numerical relativity and SPH
simulations~\citep{rosswog2013RSPTA,hotokezaka2013PRDa,bauswein2013ApJa,sekiguchi2015PRD,sekiguchi2016,radice2016}. 
Table~\ref{tab:simulation} lists the range of the kinetic energy and
average velocity for DNS mergers from the literature.
Within these uncertainties, the majority of the
simulations exhibit ejecta kinetic energies in the range of $10^{50}\lesssim E \lesssim 10^{51}$~ergs
and average velocities of $0.2\lesssim \beta \lesssim 0.3$. 
To incorporate model uncertainties in $E$ and $\beta$, 
we henceforth define three models named DNS$_{h}$, DNS$_{m}$, and
DNS$_{l}$, where the indices $_h$, $_m$ and $_l$ stand for   
typical high, median and low values of the energy and
velocity parameter space that our representative set of numerical simulations span (see Table~\ref{tab:simulation}).  
Table~\ref{tab:models} shows the kinetic energy and average velocity
for each model.

{\it  BH-NS mergers:}
  In a BH-NS merger we expect dynamical ejecta through the
  tidal disruption of the neutron star by the black hole's gravitational potential. The amount of ejecta depends on the asymmetry of
  the BH-NS system, namely the individual masses, the black hole's spin and
  the neutron star equation of state~(e.g., \citealt{rosswog2005ApJ,foucart2013PRD,
rosswog2013RSPTA,Piran+13,just2015MNRAS,kyutoku2015PRD,kawaguchi2015PRD}). 
  A high spin parameter and a
  large neutron star radius 
 should result in a larger amount of ejecta.
  It is worth emphasizing that BH-NS mergers could eject larger amounts of mass than DNS
  mergers. For instance, recent numerical relativity simulations show that for extreme tidal disruption cases, masses of $\sim 0.05M_{\odot}$ can
  be ejected with velocities in the range of $\beta \sim
  0.2$~--~$0.3$~\citep{foucart2013PRD,kyutoku2015PRD}. Here we focus
  only on the cases where the tidal disruption is sufficiently strong 
and define three different models to encompass the range in energetics: BH-NS$_h$, BH-NS$_{m}$, and BH-NS$_l$ (see the 
kinetic energy and average velocity of each model in Table~\ref{tab:models}).
Note that although we consider only NS-$5M_{\odot}$ BH mergers for simulating GW detections,
the kinetic energies and velocities used here somewhat cover the ejecta of 
massive BH-NS mergers, e.g., a BH mass of $10M_{\odot}$.

\begin{figure*}[t]
\begin{center}
\includegraphics[bb=0 -4 355 201,width=85mm]{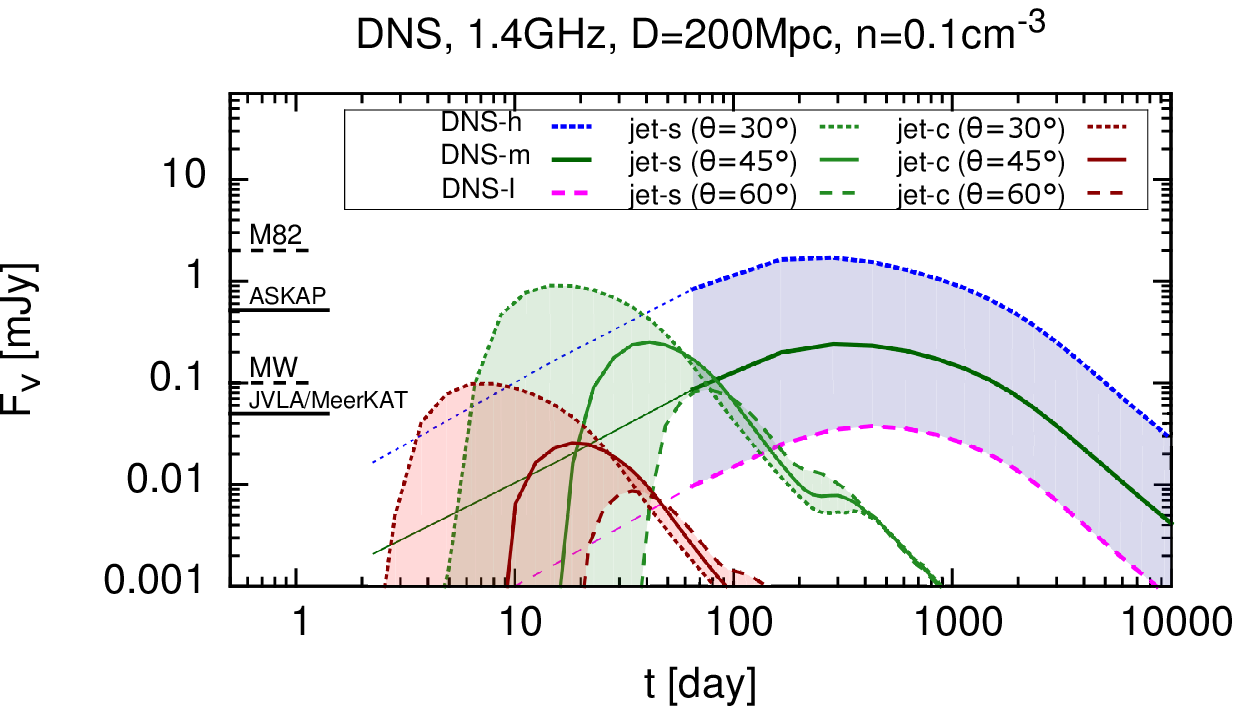}
\includegraphics[bb=0 -4 355 201,width=85mm]{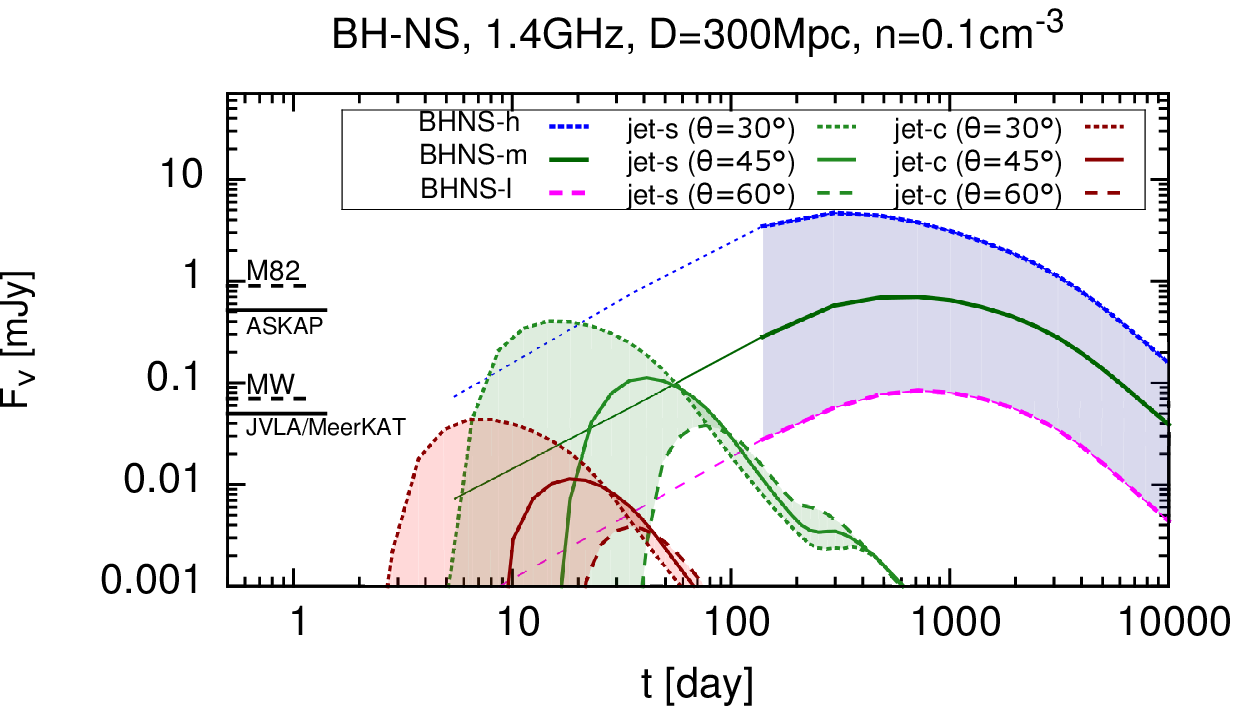}\\
\includegraphics[bb=0 -4 355 201,width=85mm]{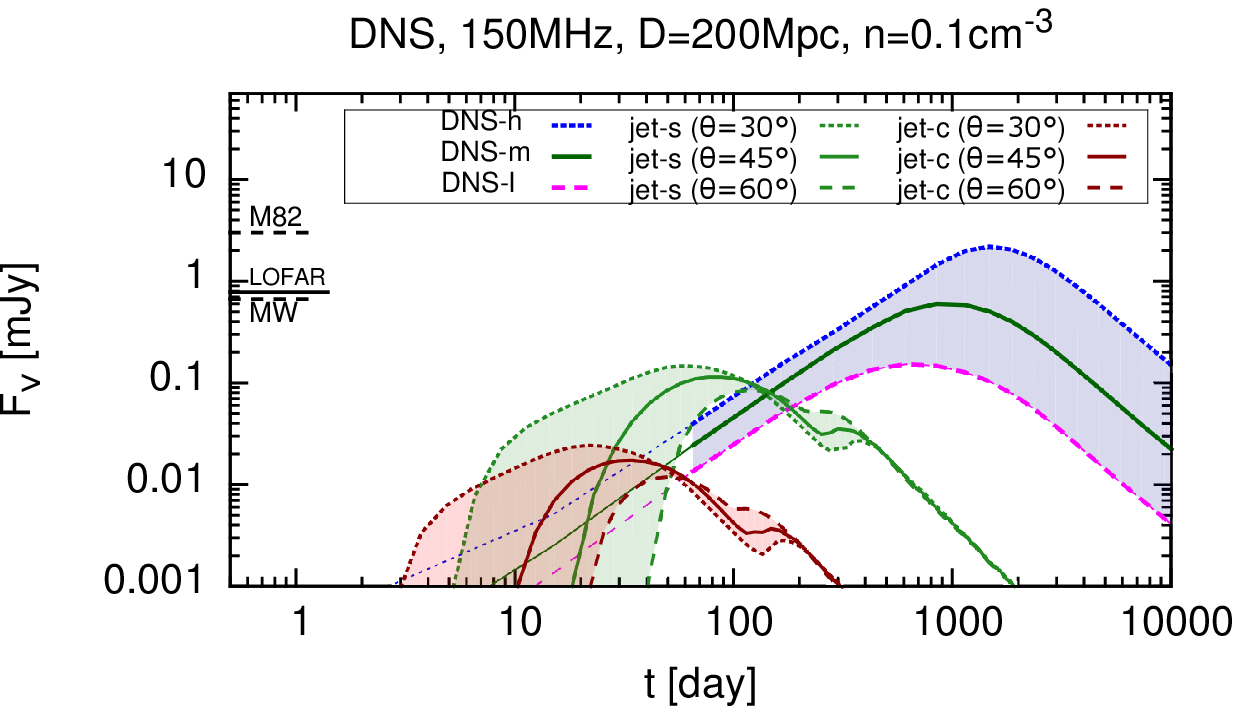}
\includegraphics[bb=0 -4 355 201,width=85mm]{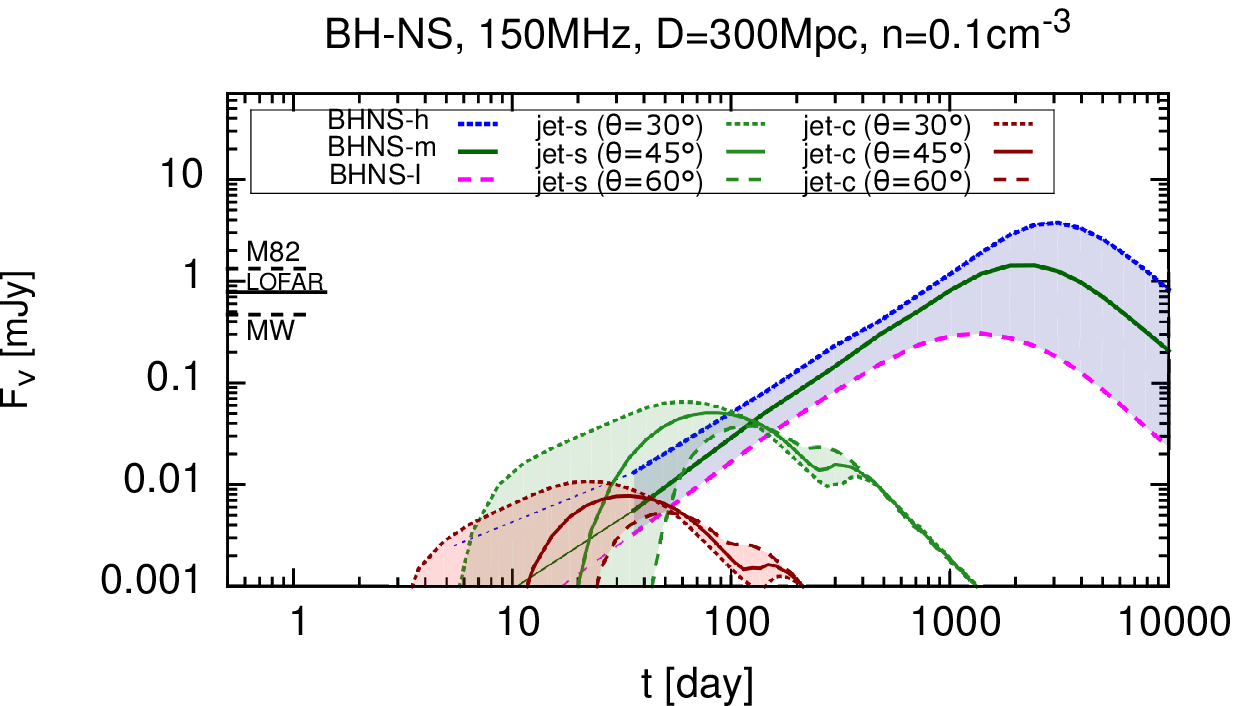}
\end{center}
\caption{Expected radio light curves at $1.4~{\rm GHz}$~(upper panels) and
$150~{\rm MHz}$~(lower panels) of a DNS merger 
at 200~Mpc~(left panels) and a BH-NS merger at 300~Mpc~(right panels).
The circum-merger density is set to be $0.1~{\rm cm^{-3}}$.
The blue, green, and magenta curve in the blue shaded region correspond to DNS$_{h}$,
DNS$_{m}$, and DNS$_{l}$ model, respectively.
Also shown are the orphan radio afterglows assuming a {\sl canonical-jet}~(red shaded region: {\sl jet-c})
and a {\sl strong-jet}~(green shaded region: {\sl jet-s}) with viewing angles of
$30^{\circ}$~(dotted), $45^{\circ}$~(solid), and $60^{\circ}$~(dashed). 
The horizontal solid bars represent the detection limits~($7$-$\sigma$ noise rms with integration of
one hour) of the different radio facilities.
The radio flux densities of the galaxies, M82 and the Milky Way, are shown
as the horizontal dashed bars assuming a distance of 200~Mpc for DNS and
of 300~Mpc for BH-NS systems. For the Milky Way, the peak flux density in the
edge-on case for an angular resolution of $7^{\prime \prime}$ is shown~(see Sec.~\ref{sec:host}). 
}
\label{fig:lightcurve1}
\end{figure*}

Numerical simulations show that the ejecta velocity is not a single value 
but follows a power law towards the high-velocity limit $\beta \gtrsim 0.5$.
Here we assume a velocity distribution
function which can broadly describe the results of numerical simulations as~(see Appendix):
\begin{eqnarray}
\frac{dM}{d\beta} = \frac{M_{0}\left(\frac{\beta}{\beta_{a}}\right)^{-\alpha}}{1+\exp((\beta-\beta_{c})/\sigma_{c})},\label{eq:pheno}
\end{eqnarray}
where we choose the parameters $\alpha=-1$ for $\beta<\beta_{a}$, $\alpha=2.5$ 
for $\beta\geq \beta_{a}$, $\beta_{c}=2\beta_{a}$, and
$\sigma_{c}=0.035$, as motivated by \cite{hotokezaka2013PRDa}. 
$M_{0}$ and $\beta_{a}$ are the parameters
which control the values of the kinetic energy and average velocity.
The exact form of the high-velocity cutoff 
is unclear because it is difficult to resolve the dynamics of
such a small amount of ejecta. 
Recently, from the results of a GR-SPH simulation
by~\cite{bauswein2013ApJa}, \cite{metzger2015MNRAS} found free neutron components of the ejecta mass extending
to high velocities $\beta\gtrsim 0.8$ in
DNS ejecta ~(see also
\citealt{kyutoku2014MNRAS} for an analytic argument). 
If this component physically exists, radio luminosities of DNS ejecta
should be brighter than our results given here at earlier times~\citep{HotokezakaPiran15}.

As the ejecta's velocity distribution is non uniform, we estimate the emission from each shell of matter
and combine the results~\citep{Piran+13}.
For a given kinetic energy distribution in velocity space,
we divide the outflow into spherical shells.  
The circum-merger material with a mass of $M(R)$ swept up at a radius $R$ can be associated
with each shell such that this mass slows down the shells:
\begin{eqnarray}
M(R)(c\beta \Gamma)^{2} = E(\geq \beta \Gamma).
\label{eq:MR}
\end{eqnarray}
Once we implicitly solve Eqn.~(\ref{eq:MR}), we are able to determine
the observable light curves.
We then combine the contributions of the different shells to obtain
the total light curve.
Note that, in this work, we do not take non-spherical geometry of the ejecta into account.
The asphericity does not affect the radio fluxes significantly 
but it will delay the peak timescales~\citep{margalit2015MNRAS}.

\begin{figure*}[t]
\begin{center}
\includegraphics[bb=0 -4 355 201,width=85mm]{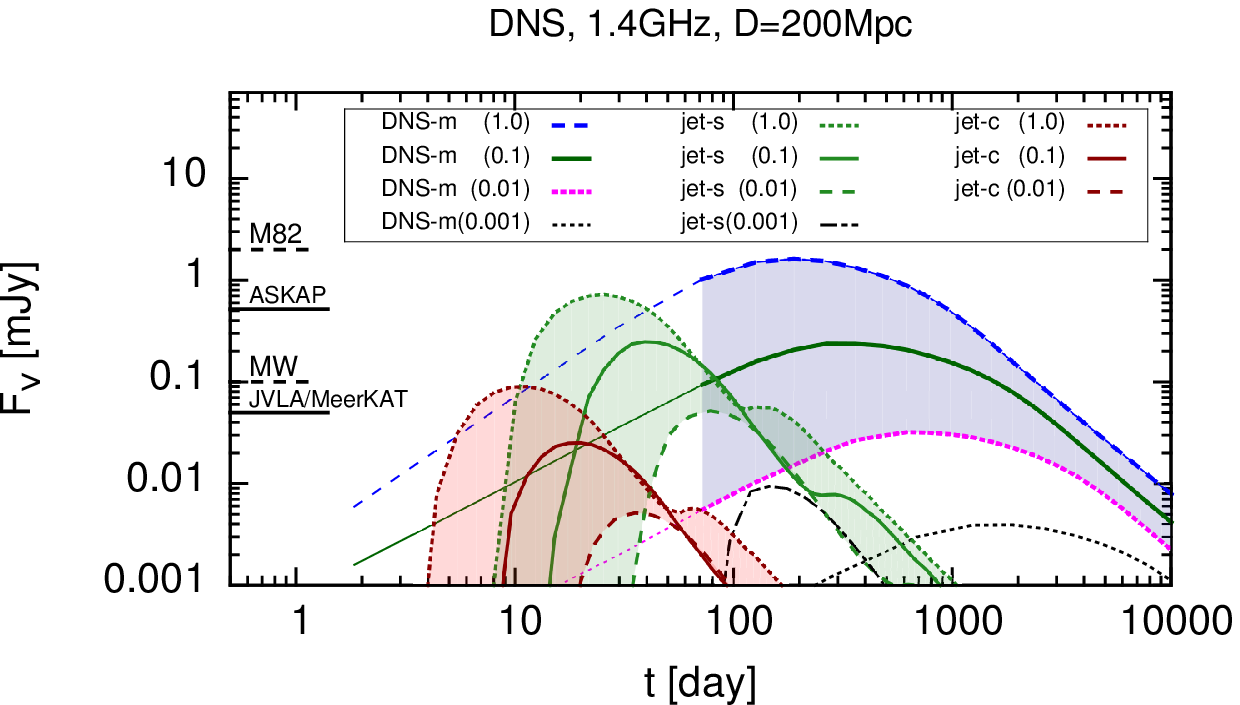}
\includegraphics[bb=0 -4 355 201,width=85mm]{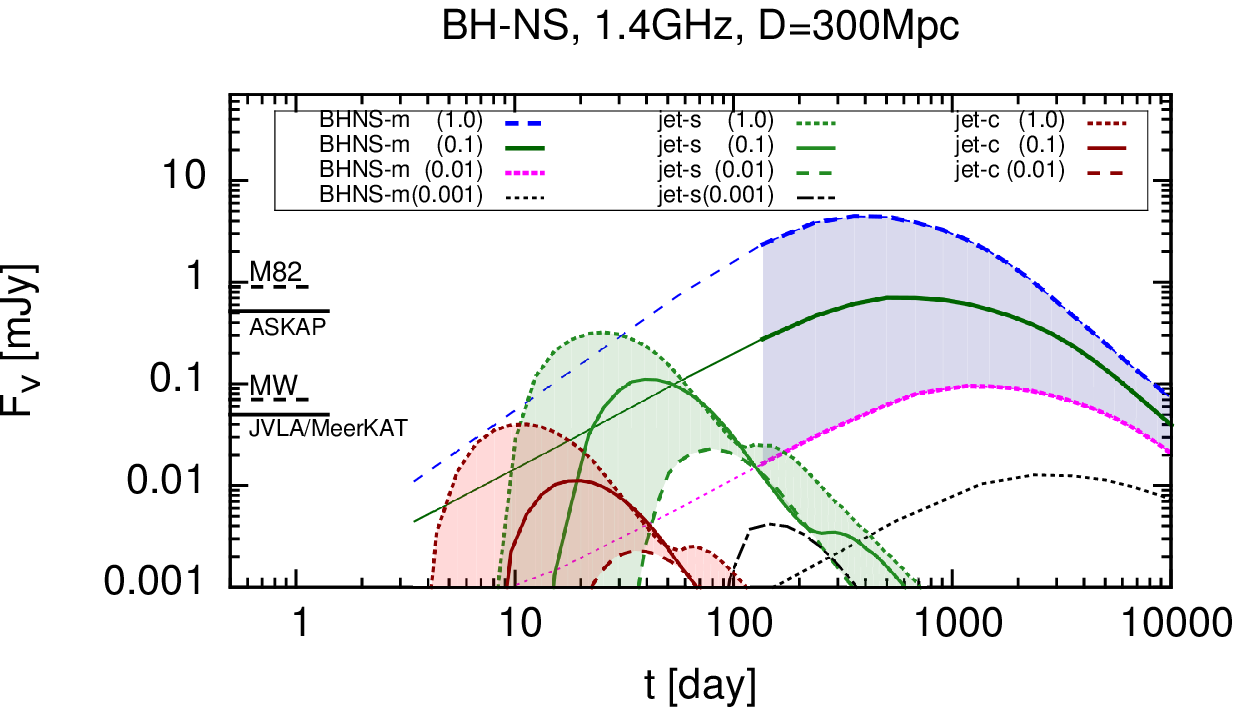}\\
\includegraphics[bb=0 -4 355 201,width=85mm]{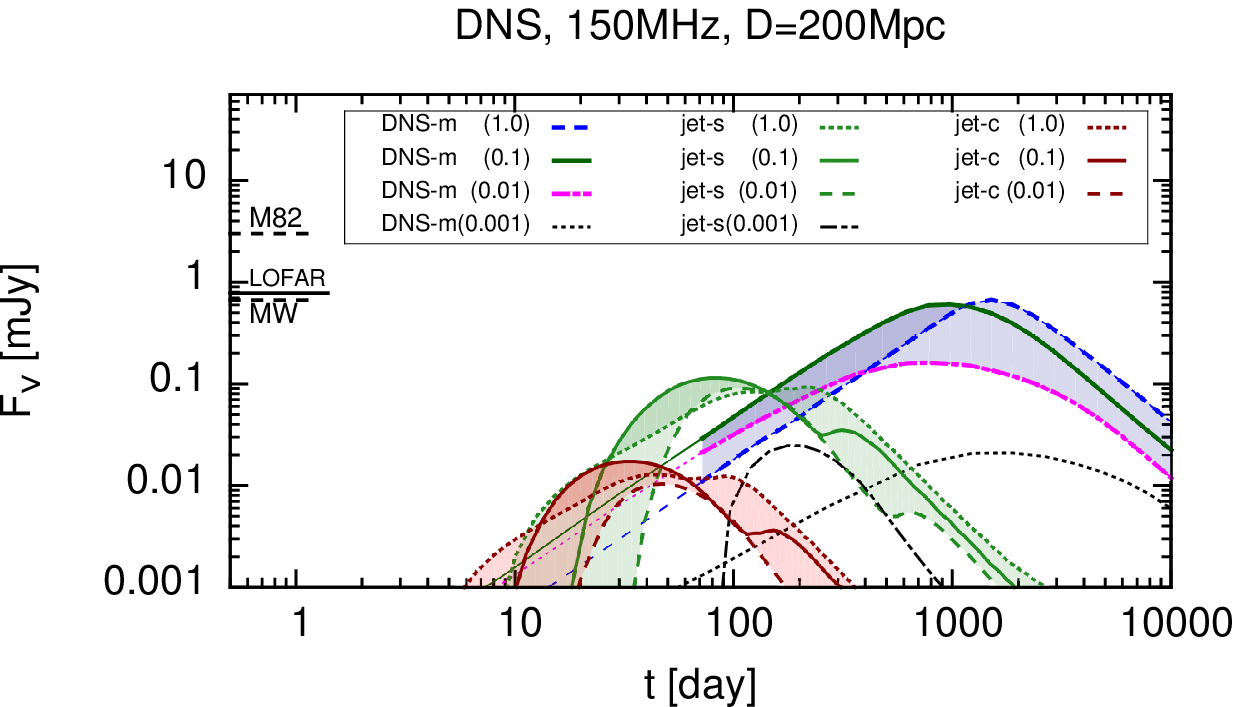}
\includegraphics[bb=0 -4 355 201,width=85mm]{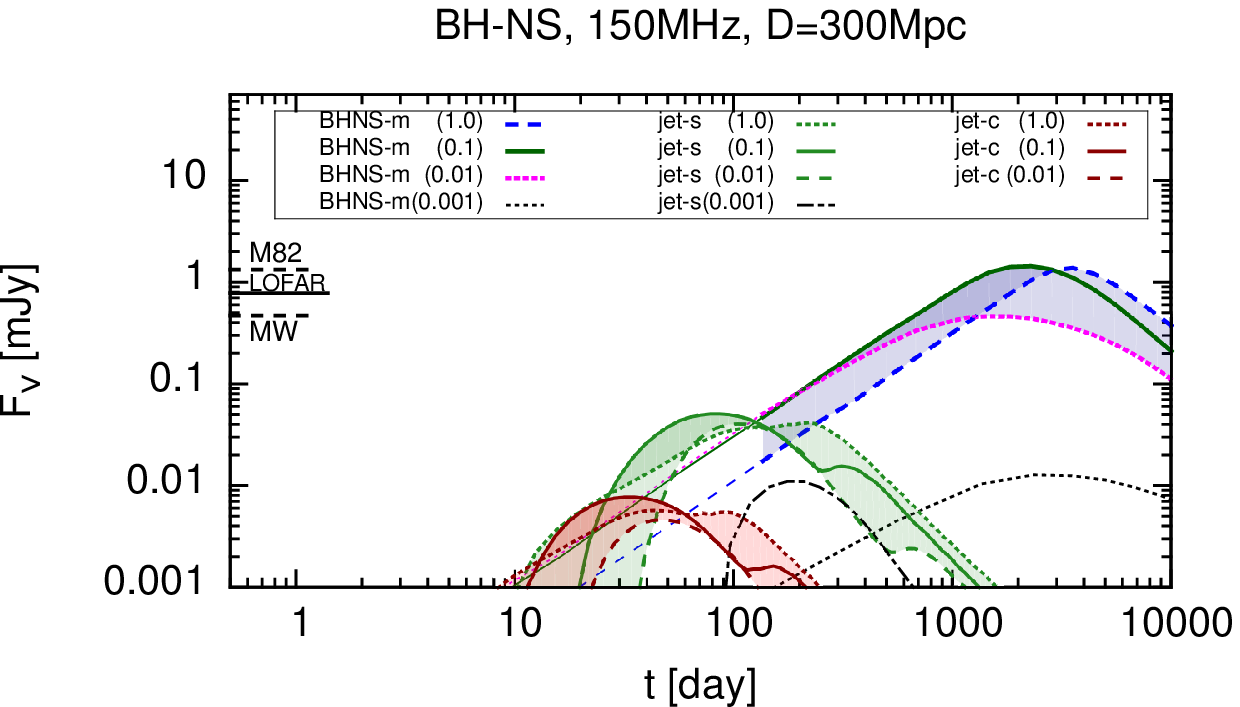}
\end{center}
\caption{
Same as Fig.~\ref{fig:lightcurve1} but shown above are the dependence of the light curves
on the circum-merger densities. In the parentheses, circum-merger densities in units of ${\rm cm^{-3}}$
are shown. Here we show the light curves for DNS$_{m}$~(left), BH-NS$_{m}$~(right), 
{\sl canonical-jet} and {\sl strong-jet} afterglows with a viewing angle of $45^{\circ}$. 
}
\label{fig:lightcurve2}
\end{figure*}

{\it Ultra-relativistic jet:}
An ultra-relativistic jet travels with the initial
Lorentz factor and the initial jet half-opening angle
in the external medium until the total energy of
the swept up material becomes comparable to the
jet's initial kinetic energy. After this stage, the 
jet slows down but remains relativistic and maintains the initial jet's opening angle. The radiation from
the jet is thus still collimated. 
Once the jet's Lorentz factor becomes roughly
$\theta_{j}^{-1}$, the jet starts the sideway
expansion and approaches to a fully spherical expansion. 
During this stage, the jet's radiation begins to be 
decollimated and detectable to off-axis observers
(see \citealt{granot2002ApJb,nakar2002ApJ,vaneerten2010ApJ,decolle2012ApJ} for details
of off-axis afterglow light curves).

We follow the jet dynamics using a semi-analytic model
proposed by~\cite{granot2012MNRAS}, which can 
approximately reproduce the jet evolution resulting from a numerical
simulation by~\cite{decolle2012ApJ}. Once we compute the jet dynamics, we derive
the afterglow synchrotron radiation at each observer's time~\citep{sari1998ApJ,granot1999ApJb}.
We choose the jet parameters, the initial jet half-opening
angle and the jet's kinetic energy based on the observations 
of sGRBs. The initial jet's half-opening angle is
measured from the chromatic break in the 
afterglow light curves. 
While there are significant uncertainties in
estimates of $\theta_{j}$ from observations,
we set the initial jet half-opening
angle to be $10^{\circ}$~\citep{fong2014ApJ}.

As with the long-lasting radio remnant, we choose {\sl two} different jet models: the
{\bf canonical-jet} model as a kinetic energy of  $10^{48}~{\rm erg}$
and the {\bf strong-jet} model has a corresponding value of $10^{49}~{\rm
  erg}$ (see Table~\ref{tab:models}). We choose the values for the
kinetic energies because the isotropic equivalent $\gamma$-ray energy of sGRBs
is in the range from  $10^{49}$ to $10^{51}~{\rm erg}$~\citep{Nakar2007}.
Assuming that the kinetic energy of the jet is comparable to
the $\gamma$-ray energy and taking into account a jet beaming 
angle of $10^{\circ}$, the jets' kinetic energies are $10^{47}\sim 10^{50}~{\rm erg}$.
We choose $10^{48}$~erg as a canonical value since there are more events 
in the lower energy range according to the luminosity function of sGRBs~\citep{wanderman2015MNRAS}.

\subsection{Radio light curves}
\label{sec:radiolightcurves}

In this section, we explicitly show the expected light curves for our
radio counterpart models assuming different circum-merger densities
$n=1.0$, $0.1~{\rm cm^{-3}}$, $0.01~{\rm cm^{-3}}$ and $0.001~{\rm cm^{-3}}$ .

Figure~\ref{fig:lightcurve1} shows the radio light curves of
the DNS models~(left panels) and BH-NS models~(right panels)
at $1.4~{\rm GHz}$~(upper panels) and $150~{\rm MHz}$~(lower panels).
Also shown are the light curves of {\sl strong-jet} and
{\sl canonical-jet} sGRB models with three different viewing angles of 
$30^{\circ}$,~$45^{\circ}$, and $60^{\circ}$. 
We set the luminosity distances to be
$200$~Mpc and $300$~Mpc for DNS and BH-NS respectively (e.g.,
NKG13 and \citealt{Aasi:2013}).
In addition, as we discuss in Sec.~4, we show the $7$-$\sigma$
root-mean-square (rms) noise level of the radio facilities considered 
with integration of one hour and the
flux densities at $1.4$~GHz of fiducial galaxies: the Milky Way and M82, 
assuming an observer at a distance of $200$~Mpc for DNS
mergers and $300$~Mpc for BH-NS mergers. Here we show the peak flux density
of the edge-on Milky Way for ASKAP~(see Sec.~\ref{sec:host}).

The radio peak flux density of each model is in the range of 
$\sim 0.01~{\rm mJy}$ to a few ${\rm mJy}$.  However, the long-lasting radio
remnants and orphan afterglows have different timescales. 
The orphan afterglows peak at early times,
between a week and a month, depending on the
viewing angle, on the jets' kinetic energy, and
on the circum-merger density. The long-lasting radio remnants
peak at late times~(a few hundred days).
Roughly speaking, for generic observers $\theta_{v}\sim 45^{\circ}$,
the {\sl strong-jet} and
{\sl canonical-jet} afterglows are as bright as DNS$_{m}$ and DNS$_{l}$
at $1.4~{\rm GHz}$ respectively. 
At $150~{\rm GHz}$, the peak flux densities  
from the long-lasting radio remnants are higher than at $1.4$~GHz and their timescales
are longer. On the contrary, the orphan afterglows at $150$~MHz
are significantly fainter because of the synchrotron self-absorption.

Figure~\ref{fig:lightcurve2} shows the dependence of the radio flux density on the circum-merger density. Here we show only DNS$_{m}$,
BH-NS$_m$, and {\sl canonical jet} with a viewing
angle of $45^{\circ}$. At $1.4~{\rm GHz}$~(above the self-absorption frequency in these cases), 
the flux densities are sensitive to the circum-merger density as $F\propto n^{(p+1)/4}$. On the other hand,
at $150$~MHz, these depend only
weakly on the density as long as the self-absorption frequency is higher than $150$~MHz.
At densities below $0.1~{\rm cm^{-3}}$,
the peak fluxes of the long-lasting radio remnants significantly 
decrease with densities since synchrotron self-absorption is less important.

The peak times and flux densities of the long-lasting 
radio remnants are faster and larger by a factor of $\sim 2$
than those in \cite{Piran+13}.
One of the reasons is that we take the faster ejecta velocities
based on numerical relativity simulations~(see Eqn.~\ref{f1} for a strong dependence of
the peak flux density on the velocity), whilst \cite{Piran+13} use a Newtonian
SPH simulation. Furthermore, 
when calculating the light curves, we incorporate
the relativistic effects, the Doppler effect and relativistic beaming,
which also result in slightly faster and brighter light curves
even for mildly relativistic velocities.

In order to optimize the detectability of orphan radio afterglows and
the long-lasting radio remnants of
GW mergers, based on our predicted light-curves (see
Figures~\ref{fig:lightcurve1} and \ref{fig:lightcurve2}), we
provide suggested epochs of follow-up observations that are roughly separated by logarithmic time intervals:
\begin{enumerate}
\item[0).] Reference imaging of the GW
error area or GW area occupied by galaxies within a day after the GW detection.
\item[i).] Observing the peak timescale of the canonical orphan afterglows
and the rise in the light curve of bright orphan afterglows at $\lesssim 10$~days.
\item[ii).] Observing the decline in canonical orphan
  afterglows and the peak in strong afterglow signatures at $\sim 30$~days.
\item[iii).] Observing the fading of strong orphan afterglows and
the rise in the light curves of long-lasting radio remnants at $\sim 100$~days.
\item[iv).] Observing the peak in the long-lasting radio remnants at $\sim 300$~days.
\item[v).] Observing the decline in long-lasting radio remnants at $\gtrsim 1000$~days.
\end{enumerate}
The last epoch will naturally only be required if the long-lasting radio remnant
candidates are detected in former epochs.
By ``radio detection'', we require, in what follows, a $7$--$\sigma$
or greater detection during 
at least one epoch, which corresponds to at least a $40\%$ change in
flux above a $5$--$\sigma$ noise limit that is required to claim detections.
With this detection criterion, most radio variables will be rejected as false positives 
(see Sec. 5.2 for details) and this depends on the nuclear versus
non-nuclear location of the source. We note that Gaussian thermal fluctuations
in the noise are an issue dependent on the number of synthesized
beams.

Note that the above recommendation of observations with
five survey epochs is for surveys at $1.4$\,GHz. 
At 150\,MHz, the epochs i), ii), and iii) are not relevant
because, due to the strong synchrotron self-absorption, both
the orphan afterglows and radio remnants
are too faint to be detected at these epochs and the peak time
of the radio remnants is later.

\subsection{Discussion on circum-merger densities}
\label{sec:circummergerdensities}

A central concern about the brightness and timescales in the
light curves of radio counterparts is 
the ambiguity in circum-merger densities, which
can spread over many orders of magnitude.
Here we try to address the question how likely it is that a merger takes place in a relatively high
circum-merger density of $\gtrsim 0.1~{\rm cm^{-3}}$.
In what follows, we consider this problem using our knowledge of the
Galactic DNS population and sGRB afterglow observations.

{\it Double neutron star population in the Galaxy:}
The interstellar medium~(ISM) is known to have highly inhomogeneous structures.
The Galactic disk of the Milky Way
is filled with three types of gas~\citep{draine2011book}: 
(i) warm neutral medium~($f_{V}\sim 0.4$,~$n\sim 0.6~{\rm cm^{-3}}$), 
(ii) warm ionized medium~($f_{V}\sim 0.1$,~$n\sim 0.3$--$10^{4}~{\rm cm^{-3}}$),
and (iii) hot ionized medium~($f_{V}\sim 0.5$,~$n\sim 0.004~{\rm cm^{-3}}$), 
where $f_{V}$ is a volume filling factor.
Assuming that the Milky Way is typical of 
galaxies hosting merger events\footnote{The volume filling factor of each phase depends on the supernova rate and  
the mean ISM density. 
\cite{li2015ApJ} show that the volume 
filling factor of hot ionized medium decreases
with the star formation rate density, suggesting that the chance that a merger takes place
in a larger ISM density is higher for galaxies with higher star formation rate densities.}
, we estimate that half the volume of the Galactic disk is filled by the ISM with densities of  
$\gtrsim 0.3~{\rm cm^{-3}}$.

\begin{table*}[t]
\caption{Mapping Speed \label{tab:Lspeed}}
\begin{center}
\begin{tabular}{lcccc}
\hline \hline
 Parameter & LOFAR & JVLA & ASKAP &  MeerKAT\\ \hline
Frequency (GHz)      & 0.150            & 1.4            & 1.4  & 1.4\\
SEFD (Jy)            & 31            &  13           &  87 & 7.7  \\
FoV (deg${}^2$)      & 11.35            & 0.25           & 30& 0.86  \\
bandwidth (MHz)      & 90             &  600            &  270 & 690 \\
Survey Speed (deg$^{2}$/hr)  &  8.2~(240)            & 14            &  20   & 140\\
Angular resolution (arcsec)  &  10   &  4.3                &  7    & 5.25\\   \hline \hline
\end{tabular}
\tablecomments{We have assumed a standard correlator quantization loss
of~0.9. 
Note that the values of the JVLA are based on real results while
the other values are based on current predictions of performance.
The bandwidths are the values after taking the loss due to RFI 
into account.
Here the survey speeds at $1$-$\sigma$ noise  rms of $0.1~{\rm mJy}$ are shown
but the one of $0.7$~mJy is also shown in the parentheses for the LOFAR since
flux densities at $150~{\rm MHz}$ of optically thin sources are brighter than those 
at $1.4~{\rm GHz}$ by a factor of $\sim 7$.
The JVLA B configuration using the natural weighting and HBA-Inner Dual 
configuration of the LOFAR using $40$ stations are chosen.}
\end{center}
\end{table*}

The probability that a merger takes place 
in the Galactic disk~(assuming a half thickness of $\sim 250$~pc) can be estimated
based on the spatial distribution of the known Galactic DNS systems.
It is worth noting that most of these systems that will coalesce 
within a Hubble time are located in the Galactic disk
even though their characteristic ages are  $\gtrsim 100$~Myr~(e.g., \citealt{lorimer2008LRR}). 
In particular, PSR~J0737-3039A/B, PSR~B1913+16, and PSR~B1906+0746, which are the known DNS 
systems with the shortest merger times, are located within $300~{\rm pc}$ above the Galactic plane. 
The measured proper motions are $10$~km/s and $75$~km/s for PSR~J0737-3039A/B and 
PSR~B1913+16 respectively (e.g., \citealt{Weisberg:2010,Paz2016}), 
which indicate that these systems have vertical oscillations in the Galactic disk.
Thus we expect that they will be in the Galactic disk when
they will coalesce.
Whilst there are selection biases of pulsar surveys in the Galactic latitude, 
at least these systems, which contribute predominantly to merger rate estimates
calibrated by these known DNS systems, will merge in the Galactic disk.

Also, whilst incorporating the binary's velocity relative to the
Galaxy,
the fraction of DNS mergers taking place in the Galactic disk 
can be estimated based on binary population synthesis studies 
\citep{voss2003MNRAS,belczynski2006ApJ,kiel2010MNRAS}.
\cite{kiel2010MNRAS} found that the scale height of DNS mergers
is $\sim 500$~pc, which implies that about $40\%$ of the mergers
are in the Galactic disk. The expected circum-merger density is 
$n \gtrsim 0.3~{\rm cm^{-3}}$ for $\sim 20\%$ of the mergers
 and  $n\gtrsim 0.1~{\rm cm^{-3}}$ for roughly half of them.

{\it sGRB afterglows:} sGRB afterglow observations allow for
circum-merger density constraints (e.g., \citealt{fong2015ApJ}).
However, good constraints are necessary from detections of sGRB
afterglows because there exists a degeneracy between the density
and $\epsilon_{B}$. Indeed, even for a well observed
sGRB 130603B, the range of the estimated densities are $4.9\cdot 10^{-3}$~--~$30~{\rm cm^{-3}}$~\citep{fong2014ApJ}.
\cite{fong2015ApJ} show that circum-burst densities span several orders
of magnitude, with  the median densities being $\approx3$--$15\times 10^{-3}~{\rm cm^{-3}}$ under the
assumption of $\epsilon_{e}=0.1$ and $\epsilon_{B}=0.01$ or $0.1$.
The fraction of sGRBs with densities of $\gtrsim 0.1~{\rm cm^{-3}}$
are $15\%$ for $\epsilon_{B}=0.1$ and $40\%$ for $\epsilon_{B}=0.01$.
For a subset of bursts for which the observed X-ray frequencies are
larger than the cooling frequencies, the circum-burst densities are
relatively well constrained, with large median densities of 
$\approx 0.04$ -- $1~{\rm cm^{-3}}$.
Within uncertainties in $\epsilon_{B}$, $\epsilon_{e}$ and the cooling frequencies, 
we consider the densities derived from sGRB afterglows are broadly consistent with the ones estimated from
the Galactic DNS distribution.


\section{Radio detectability of GW merger events}
\label{sec:detectability}


We now consider the likelihood of finding a radio counterpart to a GW
event.  Our approach is the following.  We assume that the GW event to
have been detectable by a given network of GW detectors, and we
consider different network configurations.  We assume that the GW
network will have localized the GW event to some area on the sky,
which could be as small as a few tens of square degrees but which
could plausibly be several hundred square degrees in the immediate
future.  We assume that the total amount of observing time available
for a radio telescope to carry out a search for a counterpart to an
individual GW event is 30 hr.  This ``survey allocation" is adopted
based on a combination of how much time current radio telescopes tend
to allocate to similar efforts and our estimate of the importance of
finding radio counterparts.  As will become evident below, if the ``survey allocation" is longer, it will be more likely to find radio
counterparts, if the ``survey allocation" is shorter, it will be more
difficult.

\subsection{The sensitivity of the radio facilities}
\label{sec:radiofacilities}

From the
radiometer equation, the minimum detectable flux density for a radio source is:
\begin{equation}
S_{\mathrm{min}} 
 = \frac{m \, \mathrm{SEFD}}{\eta_c\sqrt{2\Delta\nu\Delta t}},
\label{eqn:rms}
\end{equation}
where SEFD is the system equivalent flux density or the flux density
that a source would have in order to be equivalent to the
noise power in the system, $\eta_c$ is an efficiency factor accounting
for losses during correlation, $\Delta\nu$ is the processed bandwidth
for the observations, and $\Delta t$ is the total integration time for
a given field.  It is assumed that the noise in the image is Gaussian
so that a source can be detected if it is a factor of~$m$ stronger
than the rms noise level (i.e., $m$-$\sigma$).  Formally,
Eqn.~(\ref{eqn:rms}) specifies the thermal noise limit; in
practice, a variety of factors, such as dynamic range limitations due
to calibration errors and low-level radio frequency
interference may all contribute to a larger image noise
level. As long as the instrument is well understood, the resulting PSF of incomplete uv-coverage can be completely deconvolved away.

Also relevant for searches is the telescope's field-of-view (FoV). Traditionally, this value has been taken to be the
half-power pattern of the individual element in the array, which
itself has a circular aperture, 
\begin{equation}
\Omega
\approx \frac{\pi}{4}\left(1.2\frac{\lambda}{D}\right)^2,
\label{eqn:fov}
\end{equation}
for a dish antenna of diameter~$D$ observing at a
wavelength~$\lambda$.  Emerging technologies in which the electric
field is sampled at the focal plane of the antenna, with ``phased array feeds'' (PAFs), offer the potential of much larger
FoVs.  Eqn.~(\ref{eqn:fov}) is the FoV for a single
pointing direction.  Surveying requires careful attention to the
placement of multiple FoVs to ensure that the sensitivity across the
entire region to be surveyed is approximately constant.  For instance,
for a dish antenna-based array, surveys are typically designed so that
the spacing between adjacent pointing centers is approximately half
the nominal beamwidth. The spacing between adjacent pointings is $\sqrt{2}$ or $\sqrt{3}$ for a survey with uniform sensitivity. 
The survey speed for a given sensitivity is:
\begin{equation}
\dot{\Omega} \approx \frac{\Omega}{2\Delta t}.
\label{eqn:surveyspeed}
\end{equation}

Table~\ref{tab:Lspeed} summarizes relevant values for a number of
telescopes expected to be operational in the latter half of this
decade, when a number of ground-based GW detectors are
also coming online. Here we consider Jansky VLA~\citep{JVLA},
ASKAP~\citep{ASKAP}, and MeerKAT~\citep{MeerKAT} at the GHz band.
We also consider the detectability at $150~{\rm MHz}$ with
the LOFAR~\citep{LOFAR}. Note that, apart from these telescopes,
there are other relevant telescopes including:
WSRT/Apertif~\citep{Apertif}, GMRT~\citep{intema2016}, and MWA~\citep{tingay2013}.
Note that the values of the JVLA are based on current data while
the other values are based on current predictions of performance.

\subsection{Detectability for networks of $3$--$5$ GW detectors}
\label{sec:GWradiodet}

In what follows, we investigate the detectability of the
radio counterparts of GW merger events taking into account 
the distances, inclinations and sky localization errors of GW
detections (Sec.~\ref{sec:GWPE}),
the radio light curves (Sec.~\ref{sec:radiolightcurves}), and the survey speeds of current and
future radio facilities (Sec.~\ref{sec:radiofacilities}). As described
in Sec.~\ref{sec:GWPE}, we simulate GW parameter errors and compute
a diversity of radio light curves for each GW-detectable merger.
Given the large uncertainty in estimates of circum-merger
densities from afterglow modelings as discussed in Sec~\ref{sec:circummergerdensities},
we choose relatively high circum-merger densities of $1$, $0.1$, and $0.01~{\rm cm^{-3}}$
based on the Galactic DNS population. In what follows, we assume the microphysics parameters
$\epsilon_e=\epsilon_b=0.1$ and $p=2.5$.

\begin{table*}[ht]
\caption[]{
Radio-GW detection likelihood ~($\%$)  for each radio telescope and GW Net 3
(shown in parentheses are GW Net 5).
Here the detection requires at least a $7$-$\sigma$ detection
during at least one observation epoch~($10$~day, $30$~day, $100$~day, $300$~day, or $1000$~day
after GW detections). 
A total observation time of $30$~hr is assumed in each observation epoch.
In the last column, the comparison of the radio counterpart with the 
contamination of the hosts is shown. 
B: brighter than $10^{29}~{\rm erg/s/Hz}$, 
F: fainter than $5\cdot 10^{27}~{\rm erg/s/Hz}$
M: between B and F.
Note that Northern and Southern hemisphere considerations of the GW sky
localizations are not taken into account, and hence the relative detectability fractions
should be reduced by approximately a factor of two.\\
} 
\begin{center}
\scalebox{1.0}{
\begin{tabular}{lccccccc} \hline \hline
Model & $n~({\rm cm^{-3}})$ & JVLA~($1.4~{\rm GHz}$) & JVLA~($3~{\rm GHz}$) & ASKAP & MeerKAT & LOFAR & host \\ \hline
DNS$_{h}$    & $1.0$    
& $100~(100)$ & $100~(100)$ & $100~(100)$   
& $100~(100)$ & $51~(52)$ & B\\
DNS$_{m}$    & $1.0$    
& $79~(88)$   & $72~(78)$ & $87~(93)$       
& $99~(99)$   & $37~(39)$ & M\\
DNS$_{l}$    & $1.0$    
& $21~(32)$   & $13~(20)$ & $24~(21)$      
& $64~(71)$   & $19~(19)$ & M \\
BH-NS$_h$      & $1.0$  
& $100~(100)$ & $100~(100)$ & $100~(100)$   
& $100~(100)$ & $30~(30)$ & B\\
BH-NS$_m$      & $1.0$  
& $98~(96)$   & $94~(93)$ & $98~(97)$       
& $100~(100)$ & $20~(21)$ & B\\
BH-NS$_l$      & $1.0$  
& $41~(43)$   & $34~(34)$ & $45~(38)$       
& $74~(82)$   & $17~(11)$ & M\\
strong-jet     & $1.0$  
& $49~(65)$   & $58~(68)$ & $53~(55)$       
& $86~(86)$   & $8~(3)$ & M\\
canonical-jet  & $1.0$  
& $11~(13)$   & $10~(14)$ & $8~(6)$         
& $27~(31)$   & $0~(0)$ & F \\ \hline
DNS$_{h}$    & $0.1$    
& $86~(93)$   & $73~(79)$ & $91~(95)$       
& $100~(99)$   & $78~(86)$ & B\\
DNS$_{m}$    & $0.1$    
& $21~(31)$   & $13~(19)$ & $21~(21)$       
& $62~(67)$   & $44~(46)$ & M\\
DNS$_{l}$    & $0.1$    
& $6~(4)$     & $3~(3)$ & $3~(2)$          
& $12~(15)$    & $10~(8)$ & F\\
BH-NS$_h$      & $0.1$  
& $98~(97)$   & $93~(93)$ & $99~(98)$       
& $100~(100)$ & $55~(54)$ & B\\
BH-NS$_m$      & $0.1$ 
& $44~(44)$   & $35~(36)$  & $47~(41)$       
& $77~(83)$   & $42~(43)$ & M\\
BH-NS$_l$      & $0.1$  
& $4~(6)$     & $2~(4)$ & $3~(2)$          
& $21~(27)$   & $19~(18)$ & M\\
strong-jet     & $0.1$  
& $36~(41)$   & $35~(39)$ & $37~(34)$       
& $55~(62)$   & $9~(6)$ &  M\\
canonical-jet  & $0.1$  
& $8~(8)$     & $8~(7)$ & $7~(4)$          
& $20~(19)$   & $2~(1)$ &  F\\ \hline
DNS$_{h}$    & $0.01$   
& $20~(26)$   & $13~(16)$ & $21~(15)$      
& $60~(59)$   & $61~(64)$ & M\\
DNS$_{m}$    & $0.01$   
& $4~(4)$     & $2~(3)$ & $3~(2)$          
& $12~(11)$    & $13~(11)$ & F\\
DNS$_{l}$    & $0.01$   
& $0~(1)$     & $0~(0)$ & $0~(0)$          
& $2~(3)$     & $2~(1)$ & F\\
BH-NS$_h$      & $0.01$ 
& $41~(43)$   & $34~(34)$ & $45~(38)$      
& $74~(82)$   & $67~(70)$ & M\\
BH-NS$_m$      & $0.01$ 
& $7~(8)$     & $3~(4)$ & $4~(2)$          
& $23~(28)$   & $28~(29)$ & M\\
BH-NS$_l$      & $0.01$ 
& $1~(1)$     & $1~(1)$ & $1~(1)$          
& $1~(2)$     & $1~(2)$ & F\\
strong-jet     & $0.01$ 
& $15~(19)$   & $12~(18)$ & $15~(17)$       
& $29~(34)$   & $10~(6)$ & F\\
canonical-jet  & $0.01$ 
& $3~(4)$     & $3~(4)$ & $1~(1)$    
& $10~(8)$     & $1~(1)$ & F\\
\hline \hline\\
\label{tab:detectability}
\end{tabular}}
\end{center}

\end{table*}

As discussed in Sec.~\ref{sec:radiolightcurves}, we assume in our
simulations that each
GW-detectable merger is observable in five observation epochs approximately spaced 
by logarithmic time intervals and with a $30$~hr total observation
time in each epoch. Critically, we have not taken into account
Northern and Southern hemisphere considerations of the GW sky
localizations, and hence the following relative detectability fractions
should be reduced by approximately a factor of two.
Table~\ref{tab:detectability} lists
the derived radio-GW detection likelihood
of each model for each radio facility.
For DNS mergers with $n=1~{\rm cm^{-3}}$, the majority of GW events for DNS$_{h}$ and
DNS$_{m}$ will be detectable in the GHz band. 
Unsurprisingly as shown in Eqn.~\ref{f1} and
Table~\ref{tab:models}, we find that the radio detection likelihood for GW mergers decreases as the density decreases, e.g., for DNS$_{m}$,
$20$--$60\%$ for $0.1~{\rm cm^{-3}}$ and $3$--$10\%$ for $0.01~{\rm cm^{-3}}$. 
For BH-NS cases, similar results exist because we expect an increase
in the intrinsic higher radio luminosity (Sec.~\ref{sec:radiomap}).
For orphan afterglows with $n\gtrsim 0.1~{\rm cm^{-3}}$, 
$30$--$90\%$ of the events will be detectable for those with {\it strong-jets} and
$5$--$30\%$ for those with {\it canonical-jets}.

\begin{figure*}
\begin{center}
\includegraphics[bb=50 50 410 302,width=80mm]{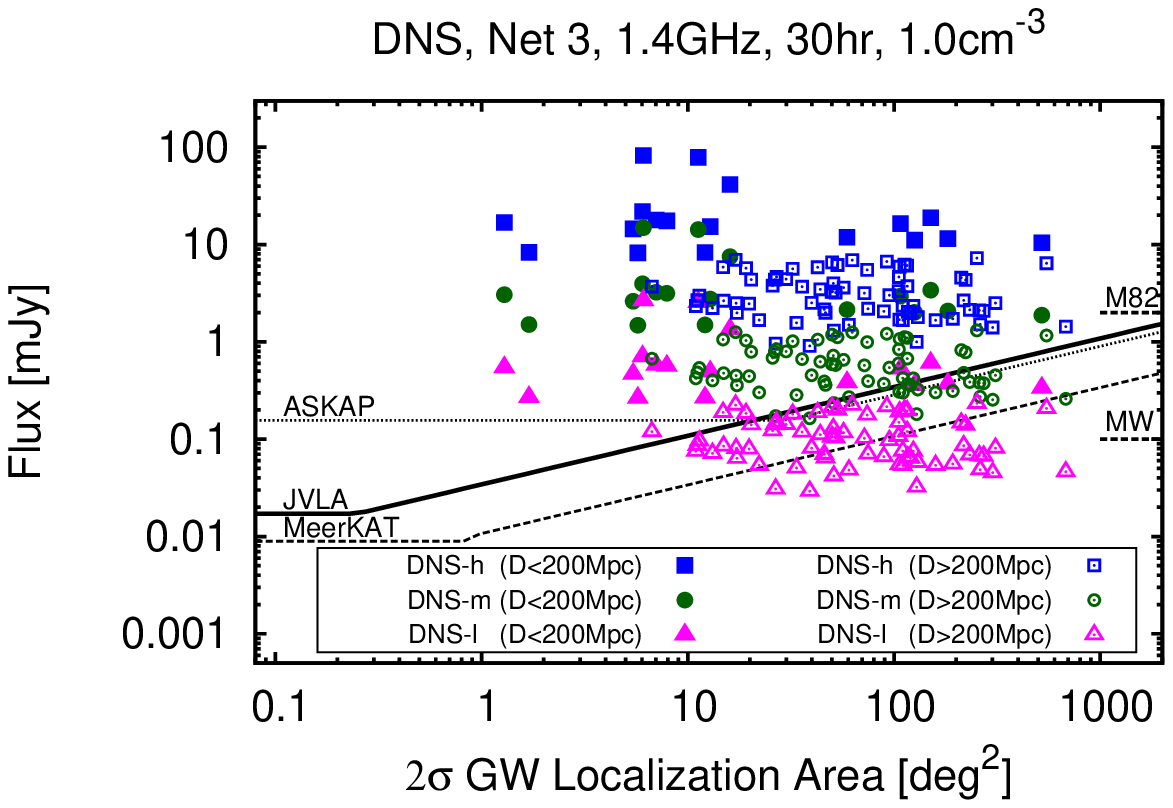}
\includegraphics[bb=50 50 410 302,width=80mm]{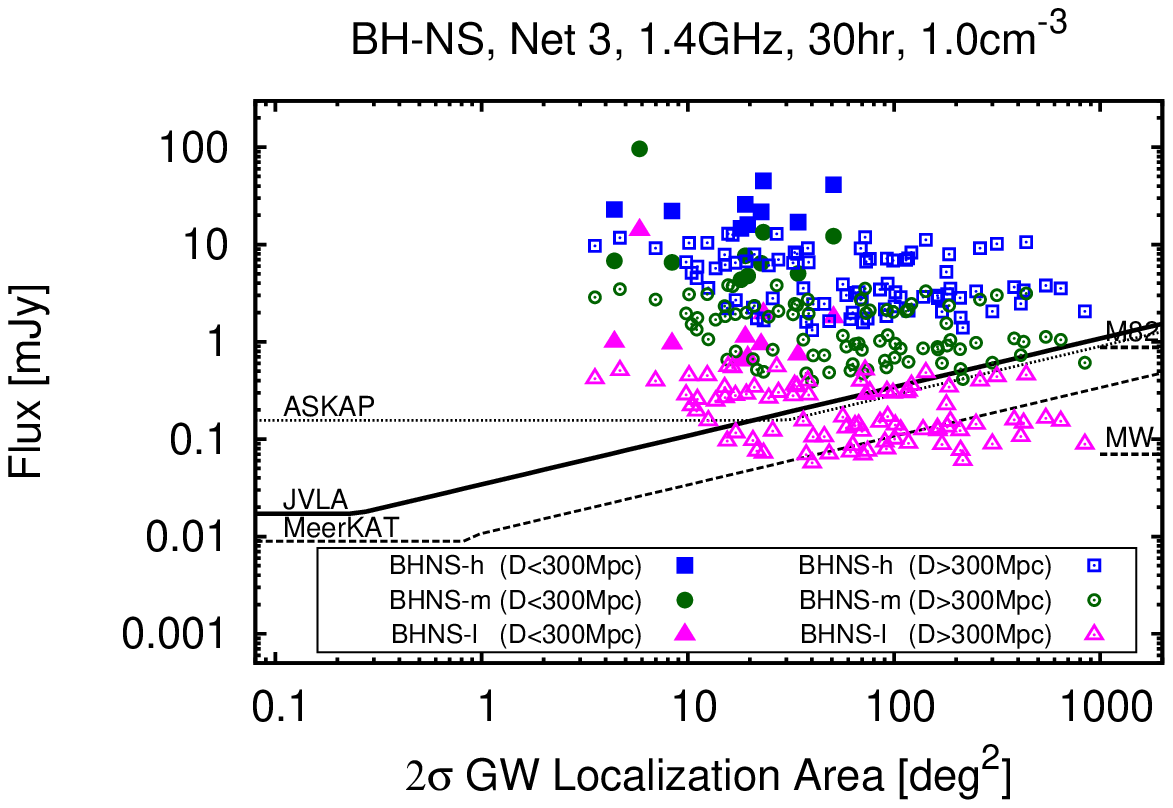}\\
\includegraphics[bb=50 50 410 302,width=80mm]{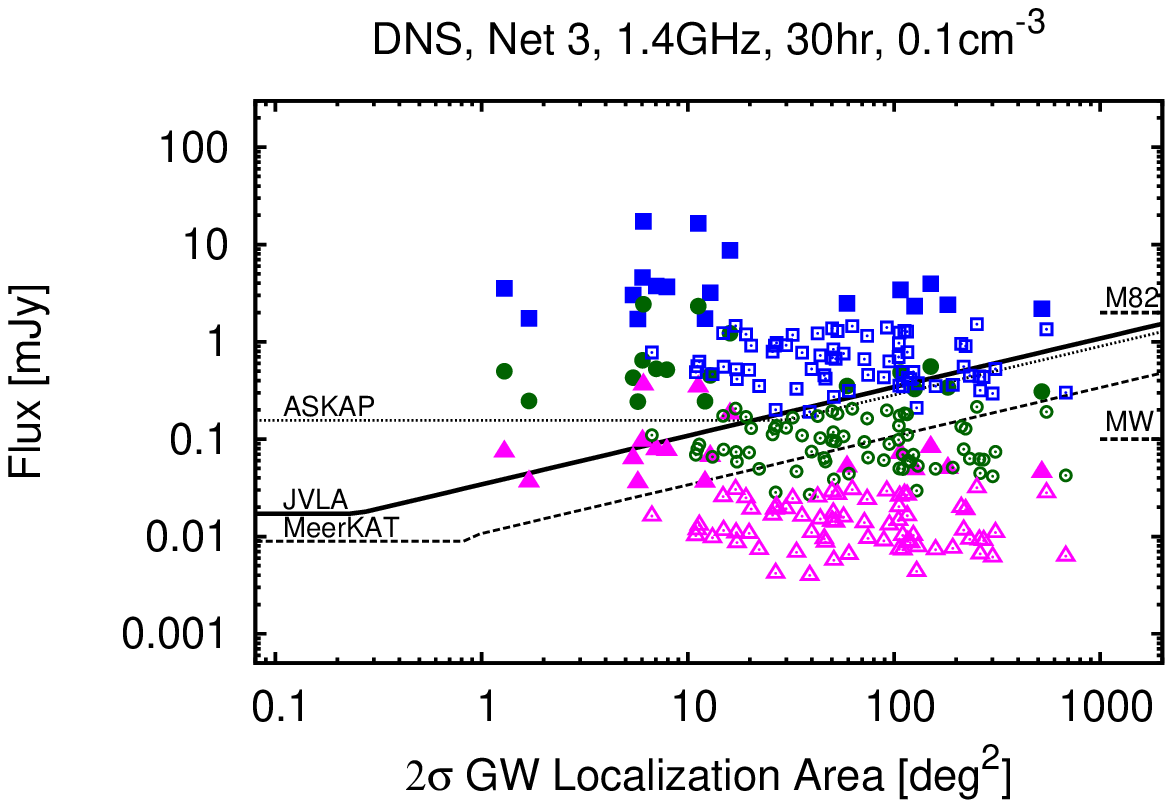}
\includegraphics[bb=50 50 410 302,width=80mm]{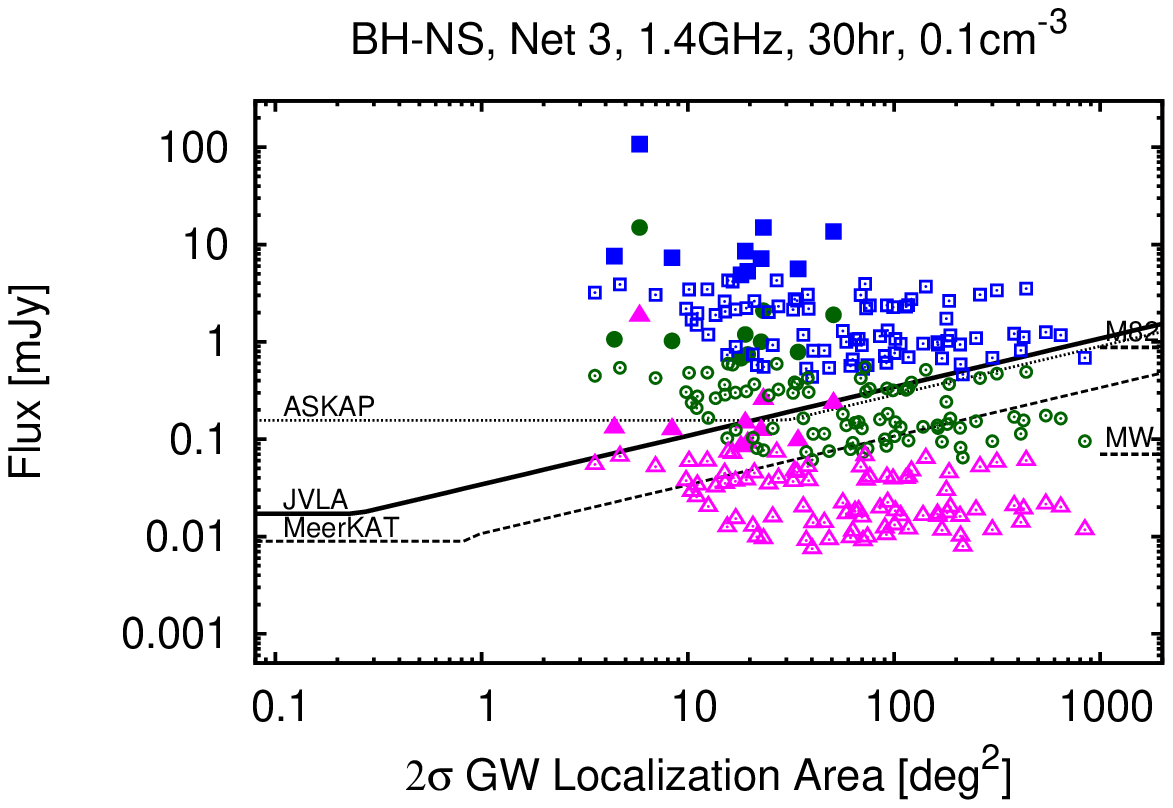}\\
\includegraphics[bb=50 50 410 302,width=80mm]{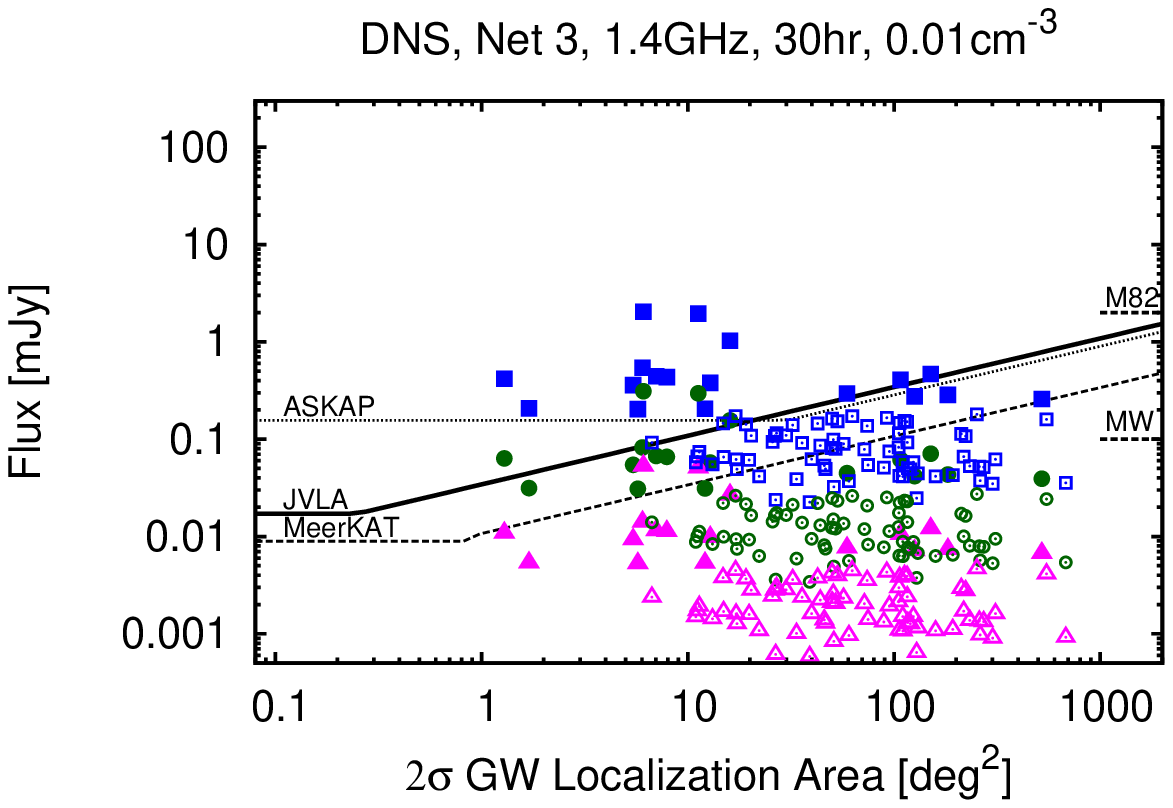}
\includegraphics[bb=50 50 410 302,width=80mm]{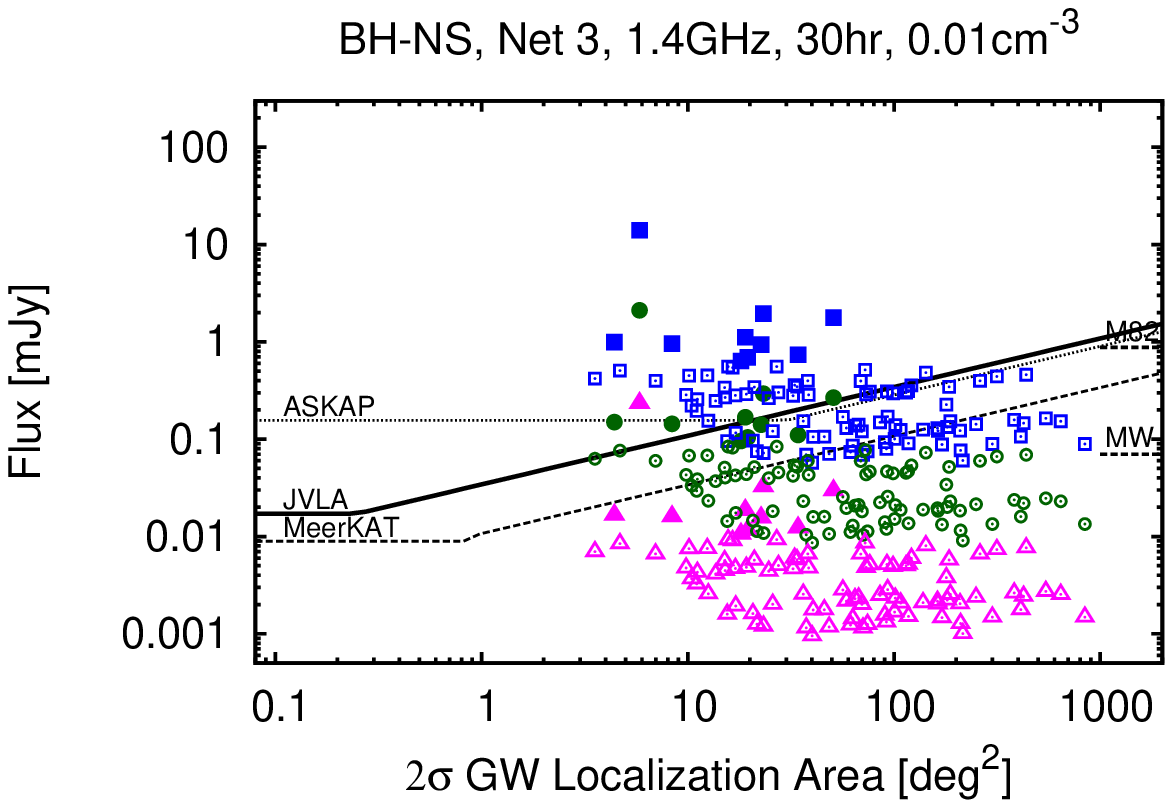}
\end{center}
\caption{ The peak flux densities of long-lasting radio remnants
as a function of the $2$-$\sigma$ GW localization areas using GW Net 3 for DNS mergers~(left panels) and
BH-NS mergers~(right panels). We set the circum-merger densities to be $1.0~{\rm cm^{-3}}$~(upper panels), $0.1~{\rm cm^{-3}}$~(middle panels),
and $0.01~{\rm cm^{-3}}$~(lower panels).
The blue filled squares, green filled circles,
and red filled triangles show the high, medium, low ejecta models within
a distance of $200$~Mpc, respectively. The open ones show those events
that occur greater than $200$ Mpc.
The lines show the $7$-$\sigma$ noise levels of the radio facilities
assuming that the total observation time of each epoch is $30$~hr with
a survey speed given in Sec~\ref{sec:radiofacilities}. 
As examples, the radio flux densities at $1.4$~GHz of the galaxies, 
M82 and the Milky Way, are shown
as the horizontal dashed bars assuming a distance of $200$~Mpc in the
case of DNS and
of $300$~Mpc for BH-NS mergers.
For the Milky Way, the peak flux density in the
edge-on case for an angular resolution of $7^{\prime \prime}$ is shown~(see Sec.~\ref{sec:host}).
Here Northern and Southern hemisphere considerations of the GW sky
localizations are not taken into account, and hence the relative detectability fractions
should be reduced by approximately a factor of two.
}
\label{fig:3det14}
\end{figure*}

\begin{figure*}
\begin{center}
\includegraphics[bb=50 50 410 302,width=80mm]{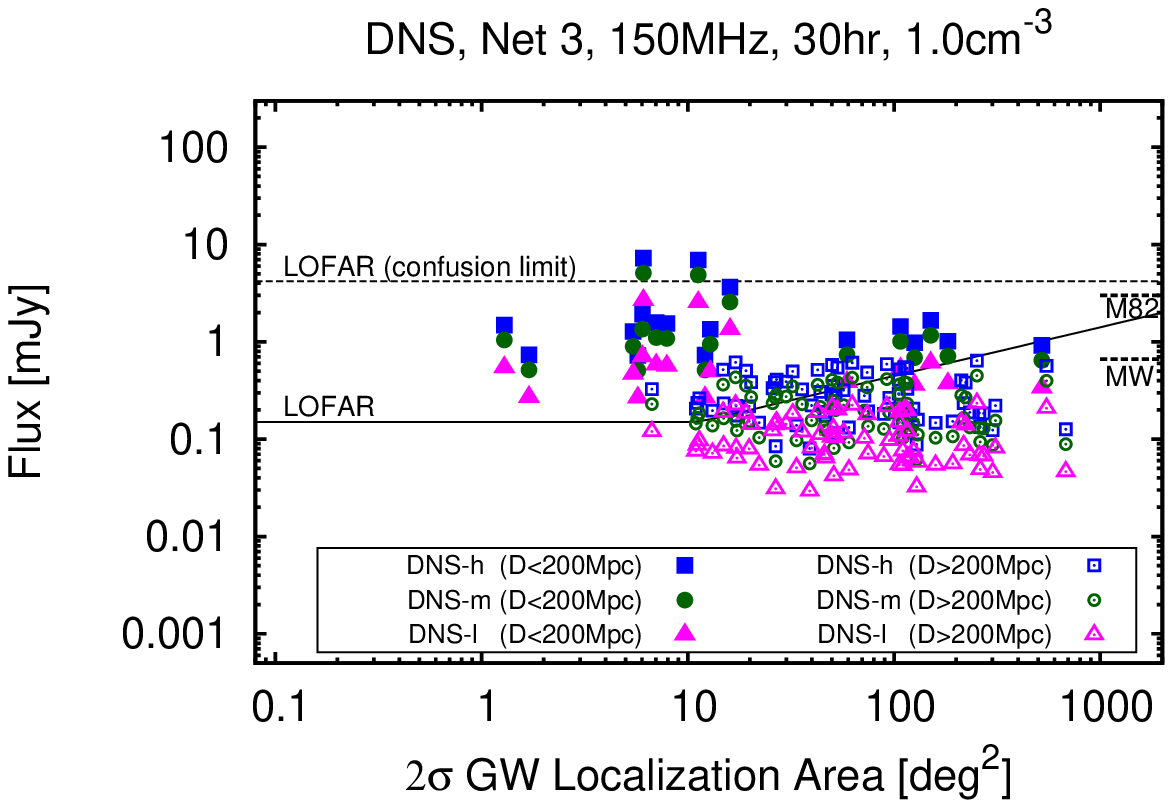}
\includegraphics[bb=50 50 410 302,width=80mm]{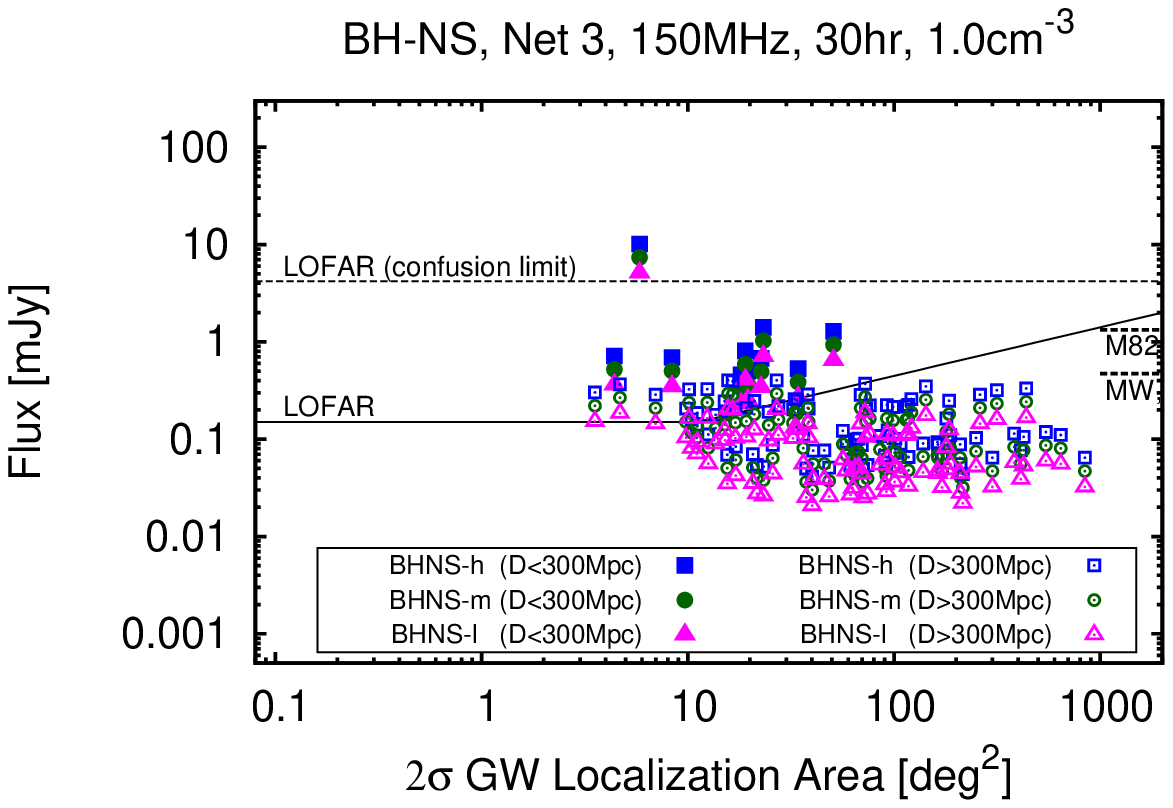}\\
\includegraphics[bb=50 50 410 302,width=80mm]{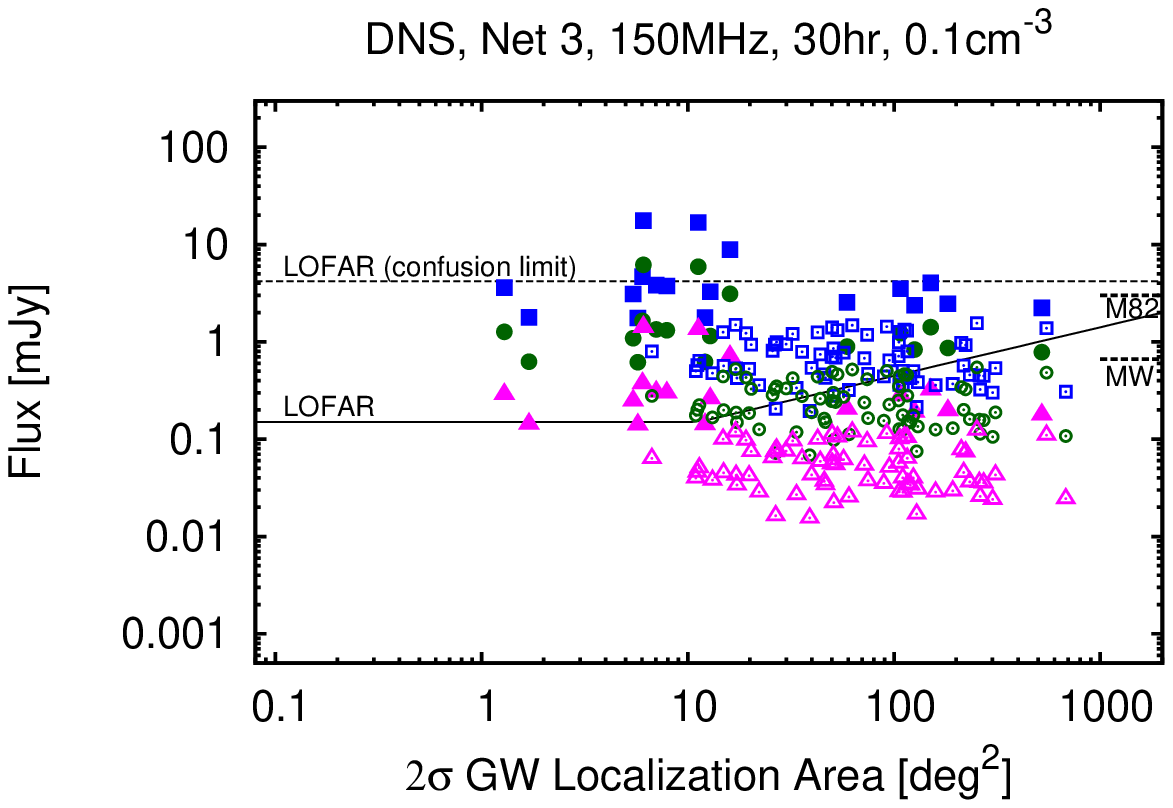}
\includegraphics[bb=50 50 410 302,width=80mm]{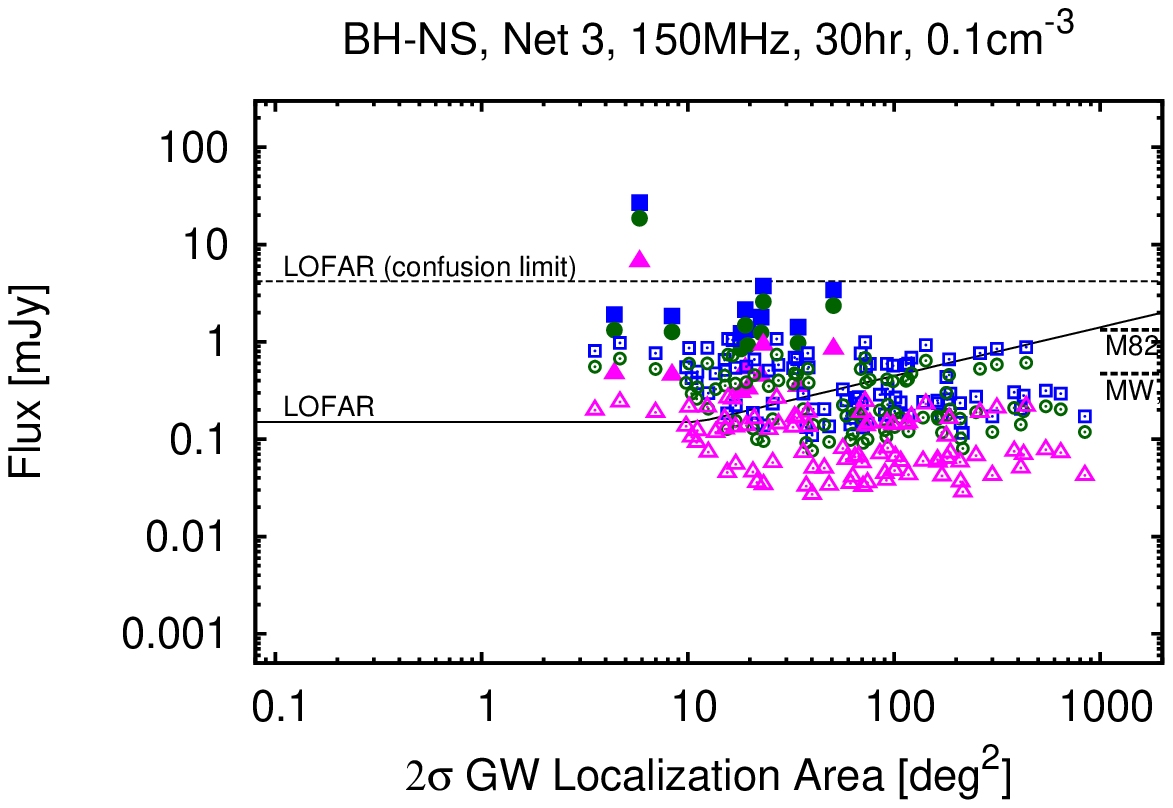}\\
\includegraphics[bb=50 50 410 302,width=80mm]{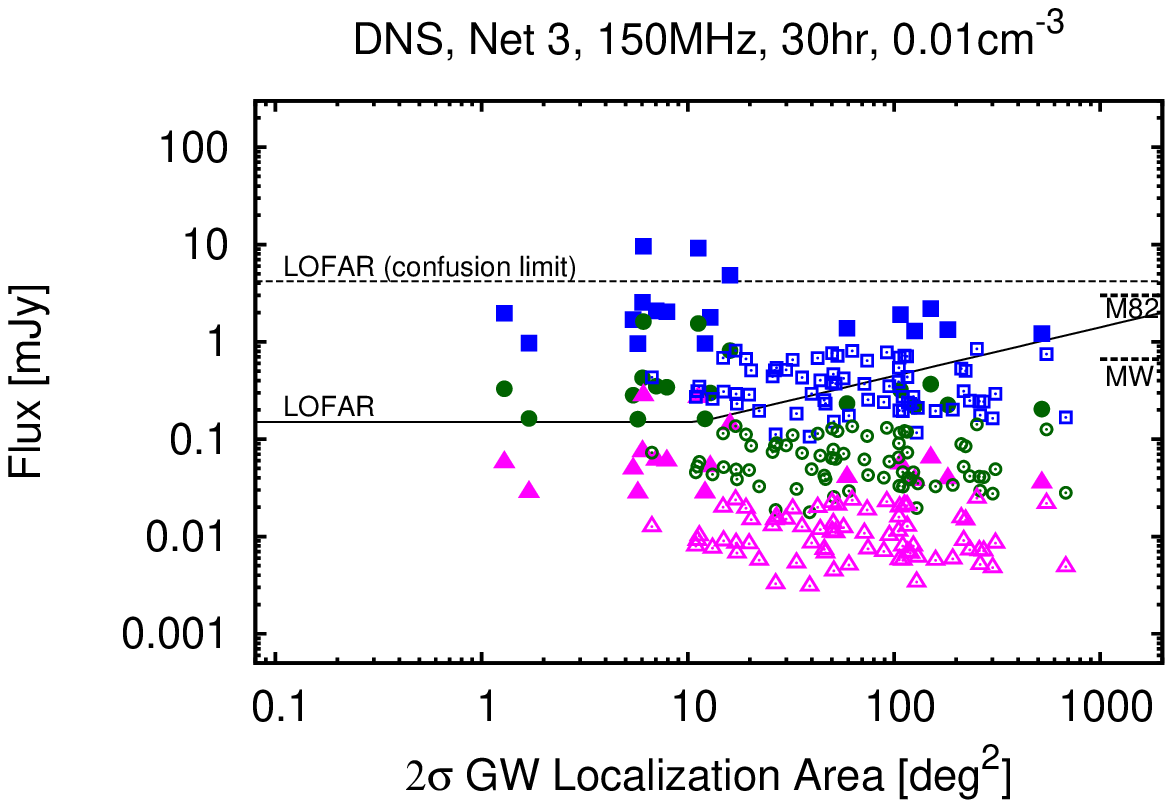}
\includegraphics[bb=50 50 410 302,width=80mm]{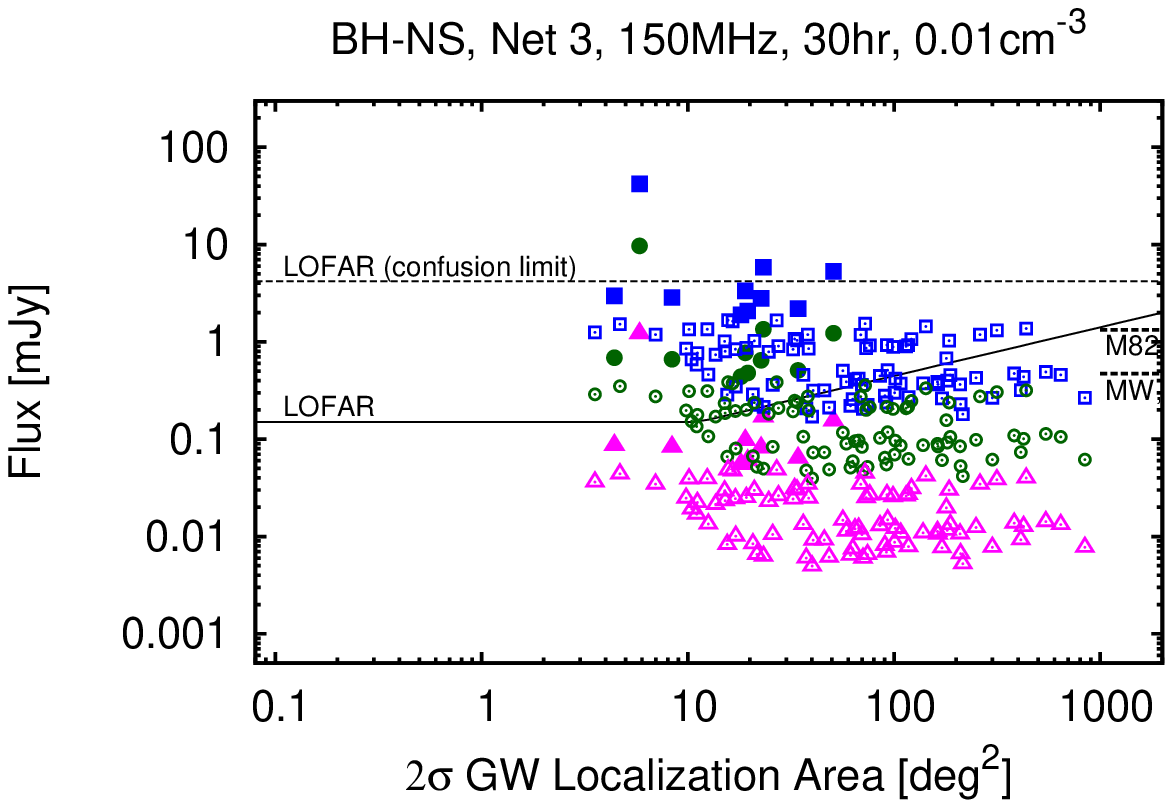}
\end{center}
\caption{The same as Fig.~\ref{fig:3det14} but for $150~{\rm MHz}$.
The expected $7$-$\sigma$ detection limit using the
confusion limit of the LOFAR with an angular resolution of
$10^{\prime \prime}$ is also shown.
}
\label{fig:3det150}
\end{figure*}

We now discuss further details in how to optimize our search in
detecting radio-GW mergers. Figures~\ref{fig:3det14} and \ref{fig:3det150} show the maximum flux density at $1.4~{\rm GHz}$
and $150~{\rm MHz}$ for each event among the five epochs as a function
of the $2$-$\sigma$ GW localization areas using GW Net 3.
The filled points represent the nearby events at distances of $<200$~Mpc. 
The diagonal lines show the 
$7$-$\sigma$ noise rms of the radio facilities corresponding to the detection threshold.
As expected, we find that the detectability of the well-localized GW events, which occur
more often than not at nearby distances, is
much higher than the poorly-localized ones.
The detection likelihood roughly behaves as $\propto \Delta \Omega_{\rm GW}^{-1.5}$,
where $\Delta \Omega_{\rm GW}$ is the GW solid angle measure on the sky.
Furthermore, the integration time of each FoV is longer for such events. 
For instance, for DNS$_{m}$  with $n=0.1~{\rm cm^{-3}}$, the JVLA
detects more than $60\%$ of events localized within a sky area of $20~{\rm deg^{2}}$.
On the contrary, the detection likelihood is less than $10\%$ for the
poorly localized events with $\gtrsim 100~{\rm deg^{2}}$.

\begin{figure*}
\begin{center}
\includegraphics[bb=50 50 410 302,width=80mm]{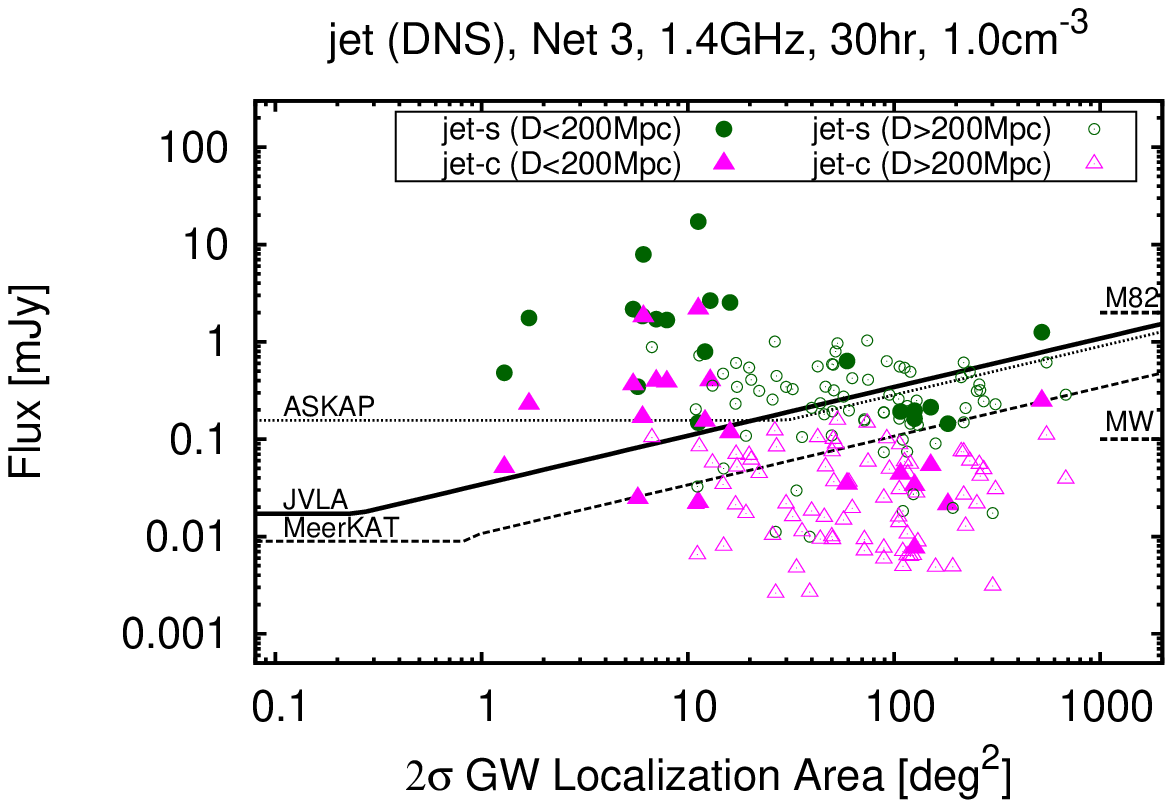}
\includegraphics[bb=50 50 410 302,width=80mm]{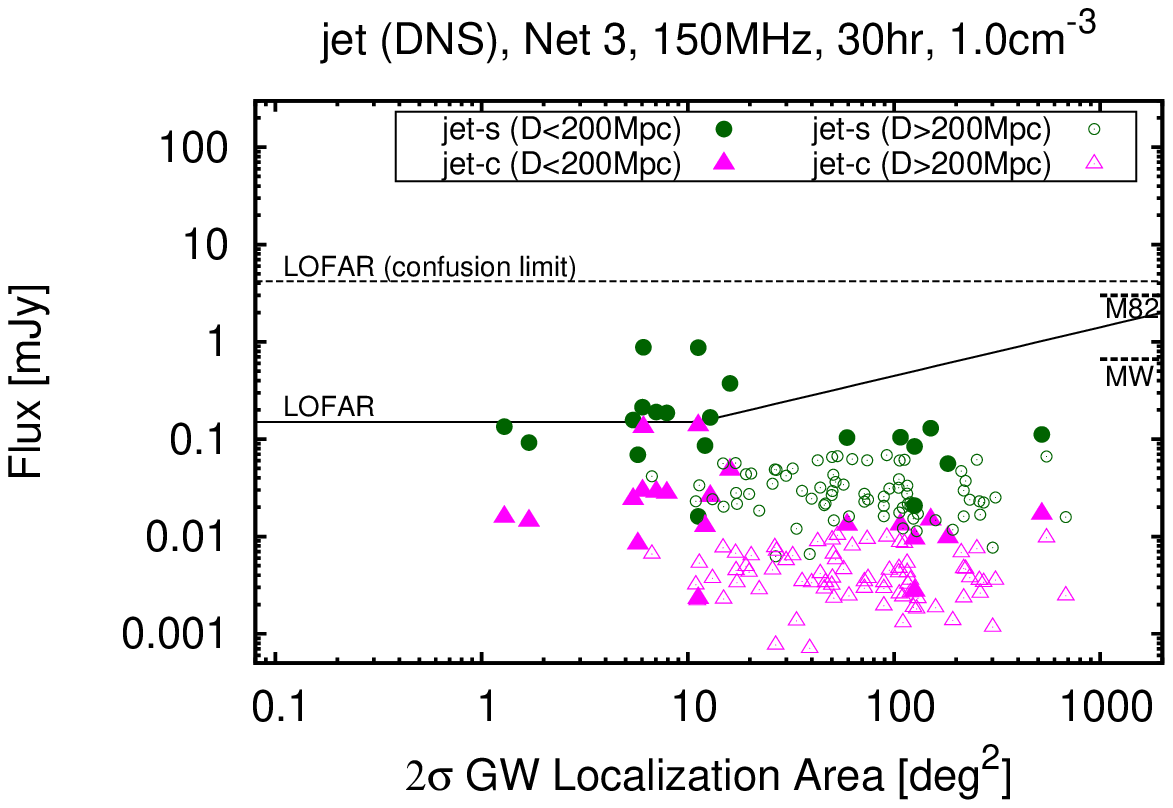}\\
\includegraphics[bb=50 50 410 302,width=80mm]{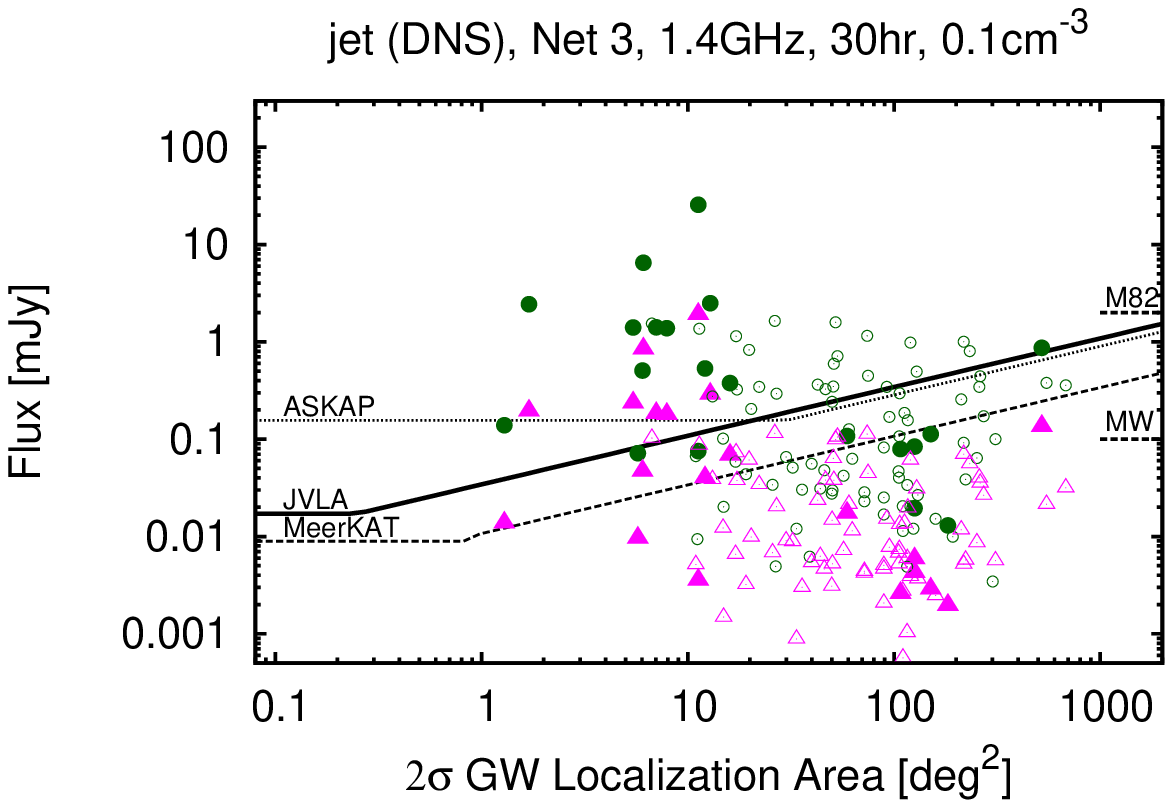}
\includegraphics[bb=50 50 410 302,width=80mm]{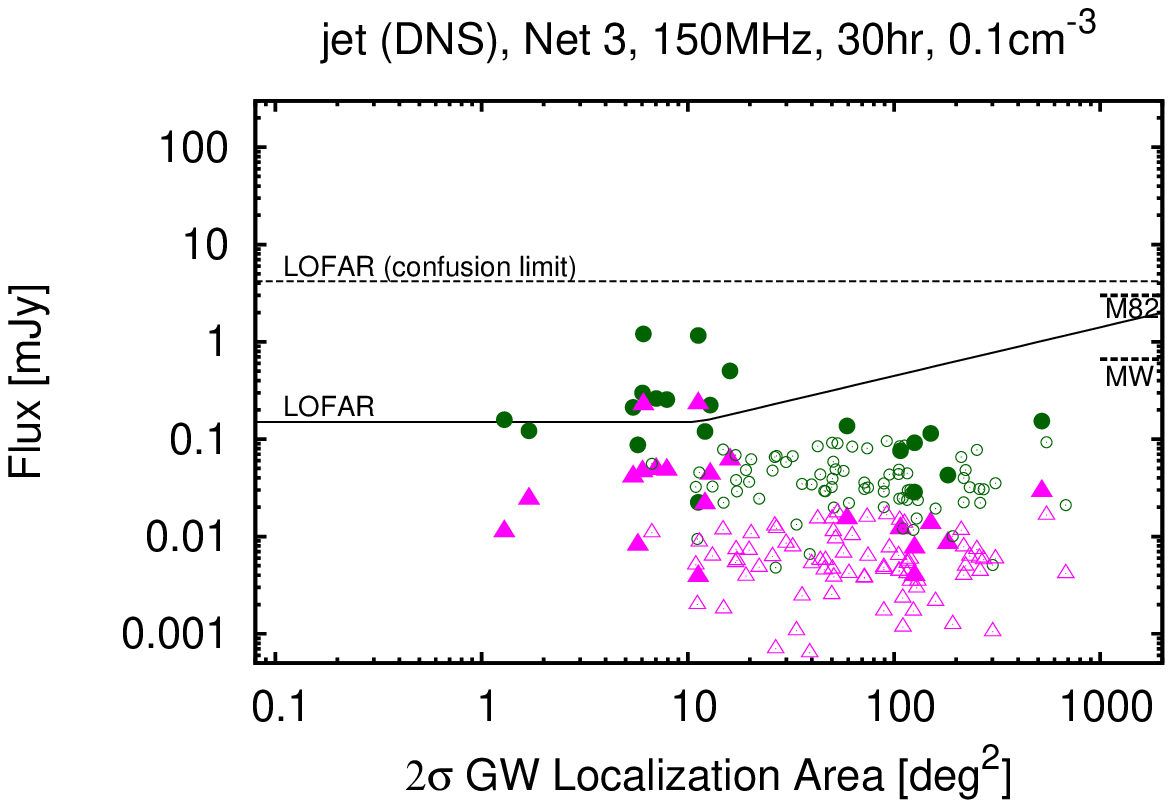}\\
\includegraphics[bb=50 50 410 302,width=80mm]{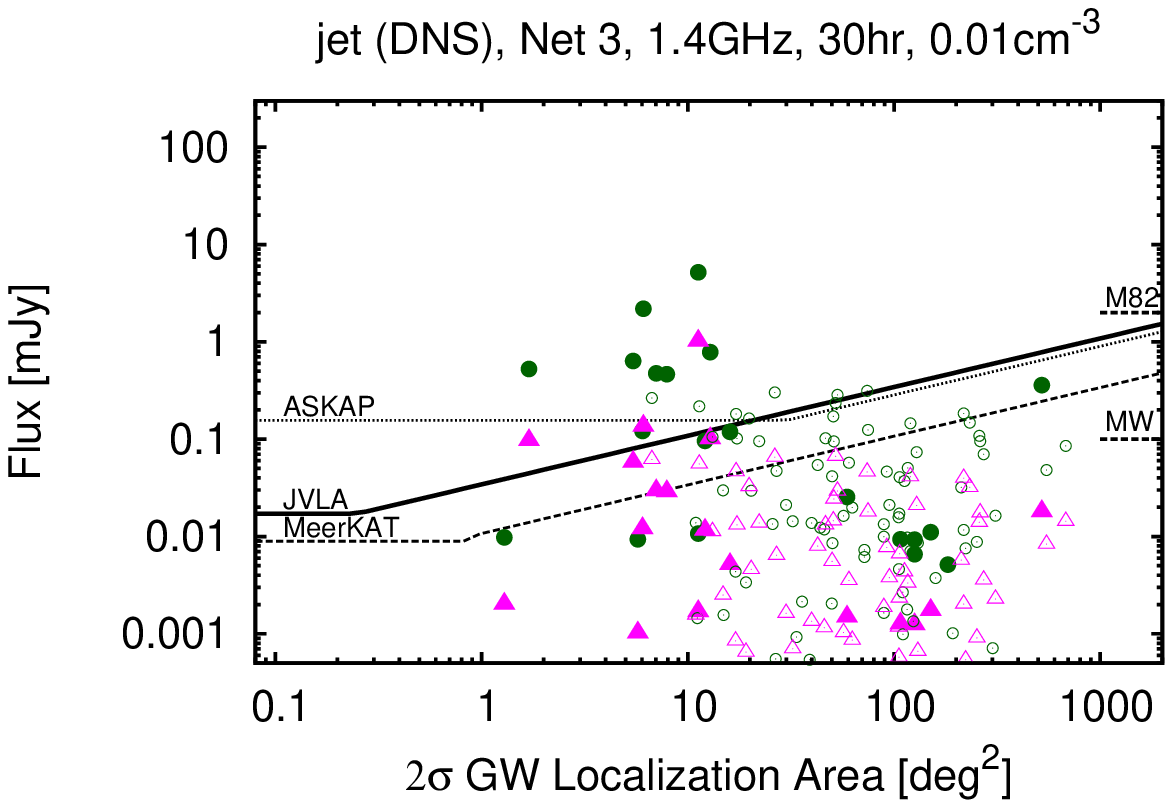}
\includegraphics[bb=50 50 410 302,width=80mm]{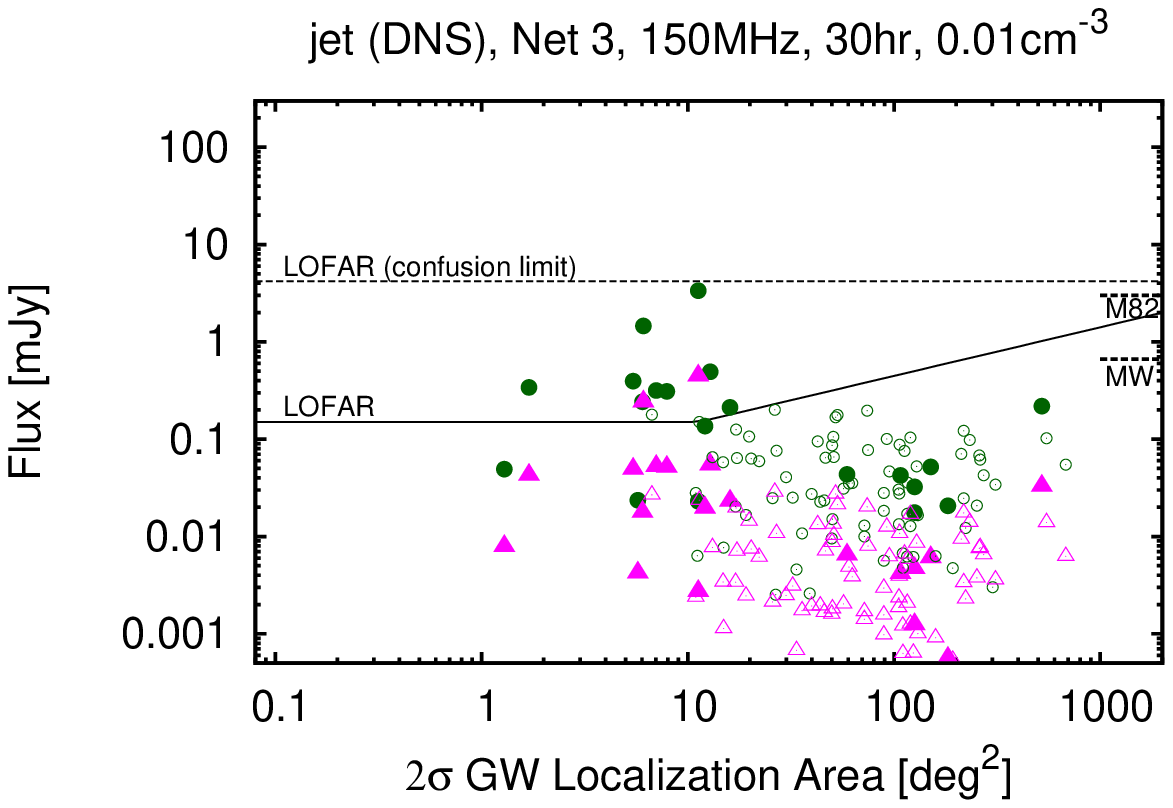}
\end{center}
\caption{The same as Fig.~\ref{fig:3det14} but for the orphan afterglows.
Left and right panels show the result of $1.4$~GHz and of $150$~MHz respectively.
}
\label{fig:3detjet}
\end{figure*}

\begin{figure*}
\begin{center}
\includegraphics[bb=50 50 410 302,width=80mm]{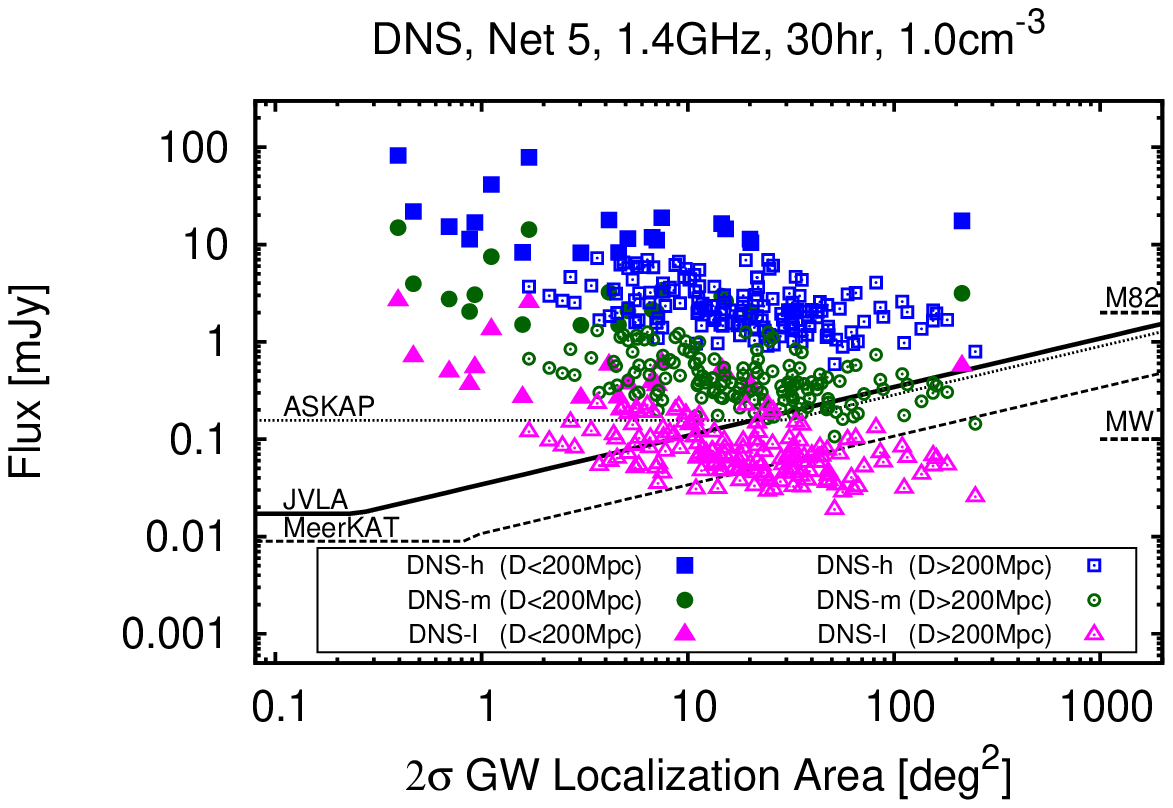}
\includegraphics[bb=50 50 410 302,width=80mm]{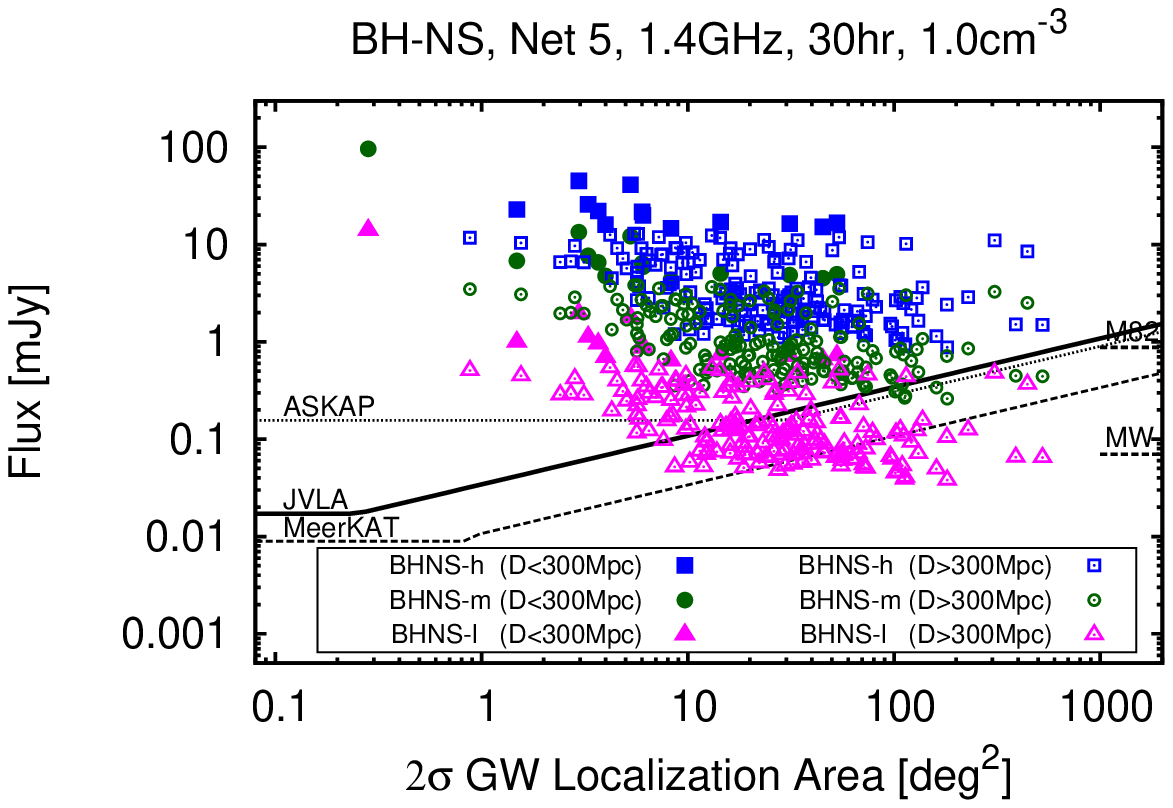}\\
\includegraphics[bb=50 50 410 302,width=80mm]{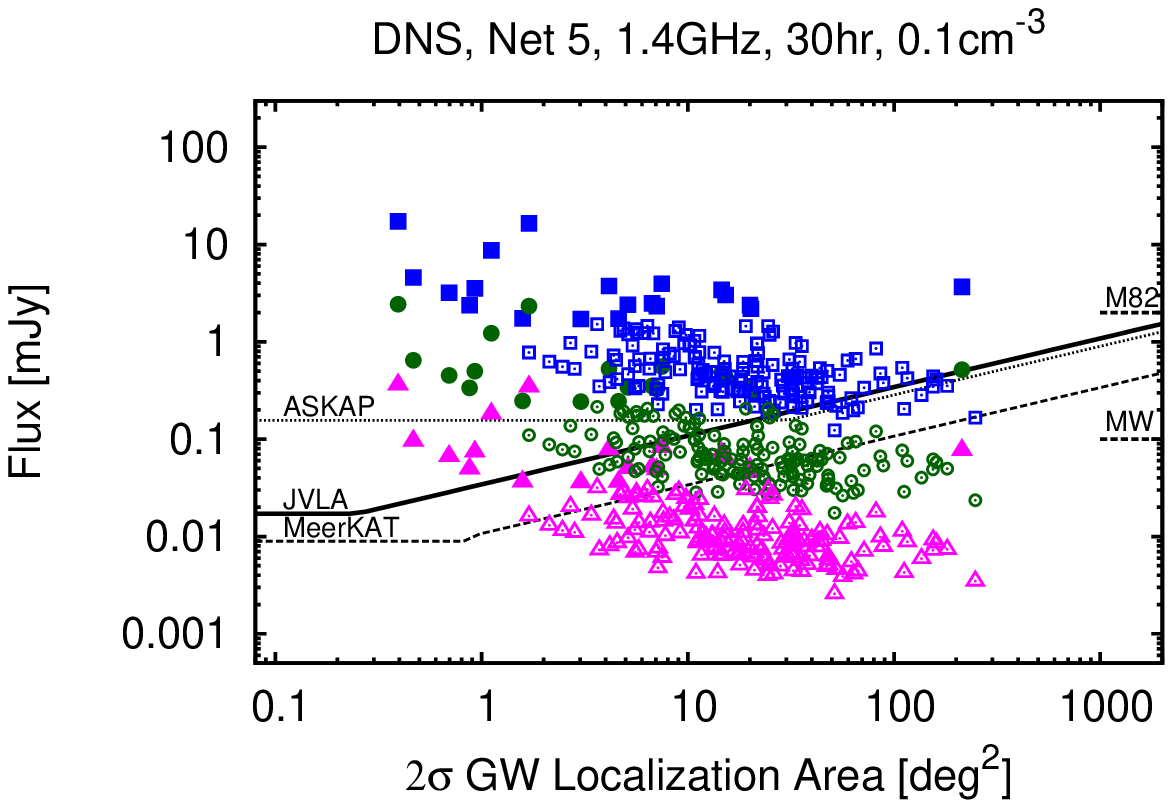}
\includegraphics[bb=50 50 410 302,width=80mm]{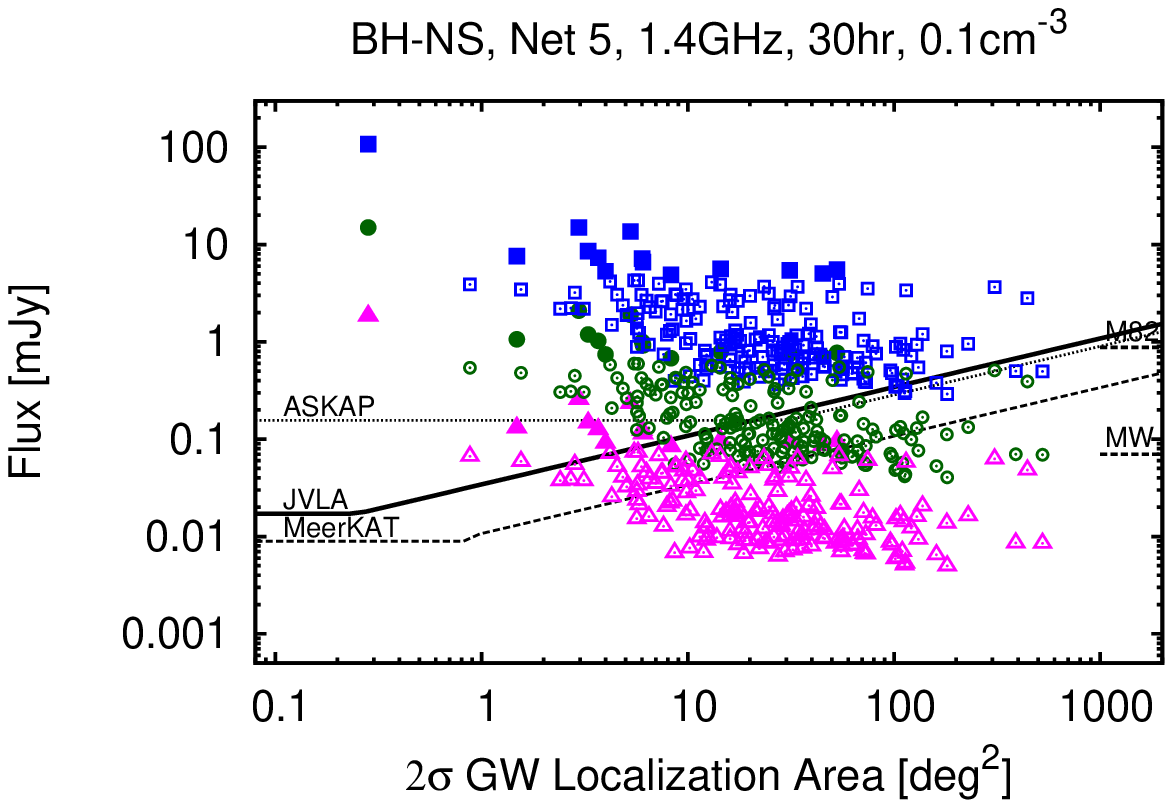}\\
\includegraphics[bb=50 50 410 302,width=80mm]{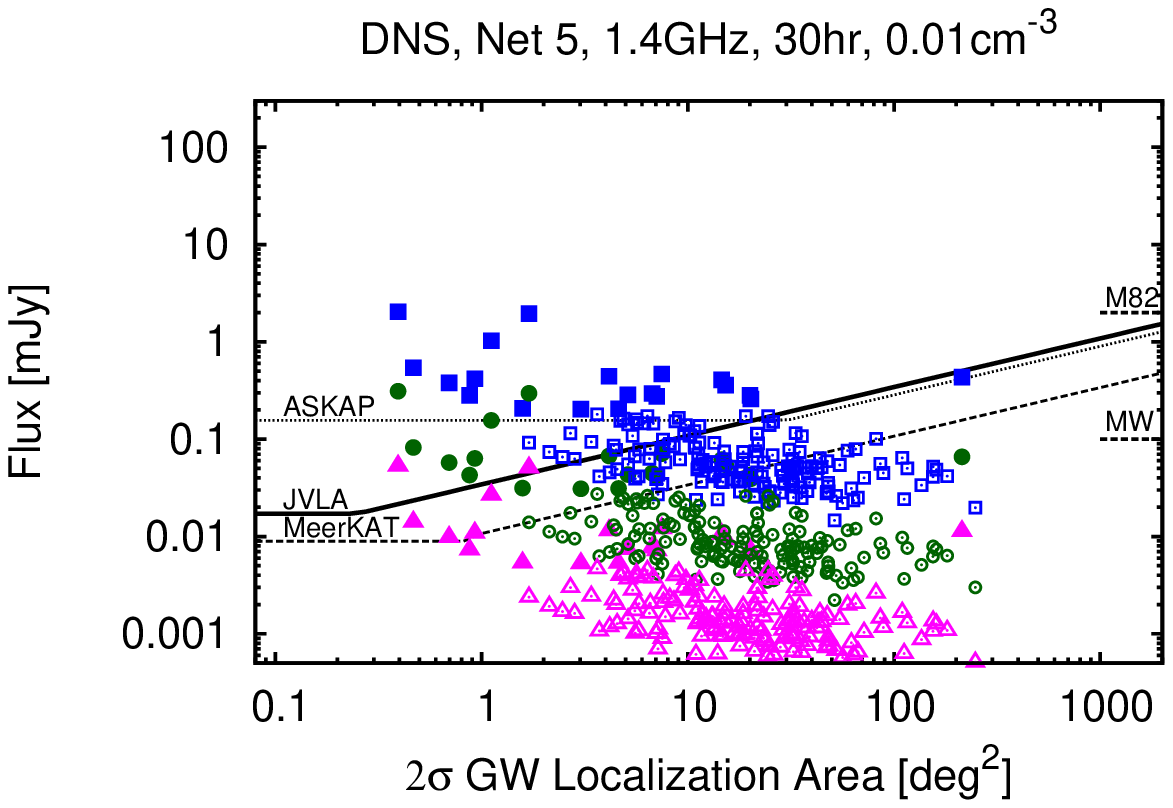}
\includegraphics[bb=50 50 410 302,width=80mm]{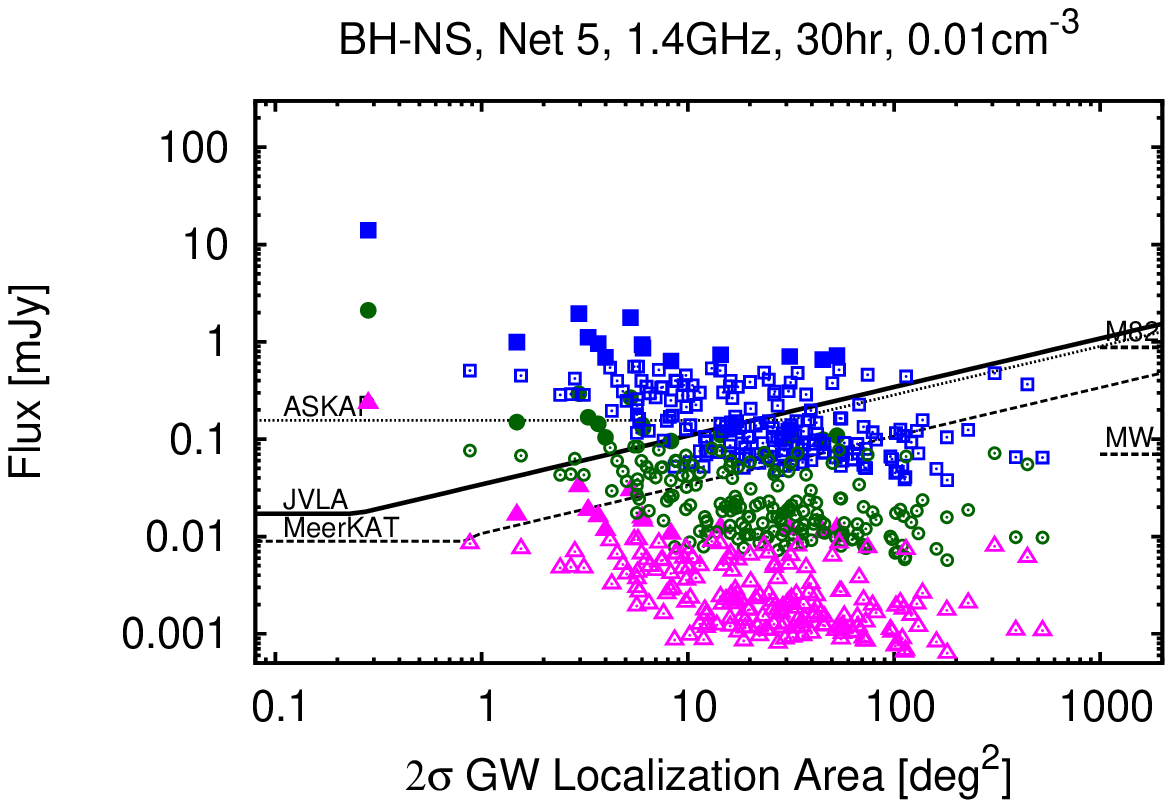}
\end{center}
\caption{The same as Fig.~\ref{fig:3det14} but for the $5$-detector network.
}
\label{fig:5det14}
\end{figure*}

For a given detection likelihood, localization area
of a GW event, and radio luminosity, one can set an optimized 
survey allocation time $T$ based on Figs.~\ref{fig:3det14}--\ref{fig:5det14}. 
For instance, in order to achieve a detection likelihood of $\sim 50\%$ for GW Net 3 (Net 5), 
the survey allocation time can be estimated as
\begin{eqnarray}
T \sim \begin{cases}
4~(60)~{\rm hr}~\left(\frac{\Delta \Omega_{\rm GW}}{\rm 10\, deg^2}\right)^3
\left(\frac{\dot{\Omega}}{\rm 14\,deg^2/hr} \right)^{-1} \\
~~~~\times 
\left(\frac{L_{\nu}}{\rm 10^{28}\,erg/s/Hz}\right)^{-2}~~~~~(\Delta \Omega_{\rm GW}\leq \Omega_c),\\
35~(60)~{\rm hr}~\left(\frac{\Delta \Omega_{\rm GW}}{\rm 10\, deg^2}\right) 
\left(\frac{\dot{\Omega}}{\rm 14\,deg^2/hr} \right)^{-1} \\
~~~~\times 
\left(\frac{L_{\nu}}{\rm 10^{28}\,erg/s/Hz}\right)^{-2}~~~~~(\Delta \Omega_{\rm GW} > \Omega_c),
\end{cases}
\end{eqnarray}
where $\dot{\Omega}$ is the survey speed at $1$-$\sigma$ rms noise of $0.1$\,mJy and 
$\Omega_c\sim 30~(10)~{\rm deg^2}$. This estimation is valid only in the case of 
$\Delta \Omega_{\rm GW}>\Omega_{\rm FoV}$. Note that $T$ is rather sensitive
to the radio luminosity of the source. 

Figure~\ref{fig:3detjet} shows the results of the detectability of orphan radio afterglows.
The scatter in the maximum flux densities is much larger than the
corresponding case for the long-lasting radio remnants
due to the viewing angle effects. Note that the dependence of the detection likelihood 
on the circum-merger density is somewhat weaker than that of the long-lasting remnants
in this density range. 

Figure~\ref{fig:5det14} shows the same cases as before though this
time in the instance of GW Net 5.  The detection likelihood does not change significantly 
because the gain in the radio sensitivity due to better localizations 
compensates with the loss in the radio flux brightness due to the increase 
of the GW detectable distances. 

The detectability at $150$~MHz is limited by
the confusion noise~($\sim 0.6$~mJy) for the configuration employed in this work. 
For detecting the radio counterparts by the LOFAR with a similar detection likelihood
as those at GHz band, the confusion noise should be reduced by at least an order of magnitude.  
Note that, however, the actual confusion limit may be lower than that
we use in this work~\citep{Heald}. Furthermore, it will be
reduced by increasing the angular resolution in the near future~(see e.g.,
\citealt{Shimwell}).

The detection likelihoods that we obtain here change for 
different choice of the microphysics parameters. The dependence of the 
peak flux on those parameters is discussed in Sec.~\ref{subsec:synchrotron}.
For instance, in the case of $\epsilon_b = 0.01$, 
the peak flux densities at $1.4$~GHz is lower by an 
order of magnitude than those with $\epsilon_b=0.1$, e.g.,
the detection likelihood of DNS$_{m}$ with $n=1~{\rm cm^{-3}}$
and $\epsilon_b = 0.01$ by JVLA is $20\%$. 
For the orphan afterglows, the detectability depends also on the initial 
jet half-opening angle.
One can expect 
a higher detection likelihood for a wider jet-half opening angle
because the probability that an observer is located within
the initial opening angle is higher.

{\it Detectability with two GW detectors and/or use of GW volumes}: The advanced LIGO detectors have been operating in their first
science run (O1) from September 2015 to January
2016 \citep{Aasi:2013}. For their second run, the two detectors will undergo further
upgrades and will operate jointly in the second half of 2016 with the first upgraded version of advanced Virgo \citep{Aasi:2013}. 
We consider radio detectability of GW mergers using only
two advanced LIGO detectors \citep{Kasliwal:2014,Singer:2014}. In this case, the GW localization
areas will be as large as several hundreds to a thousand of ${\rm deg^{2}}$ 
for events at smaller distances of $\lesssim 80$~Mpc and $\lesssim 120$~Mpc for DNS and
BH-NS mergers respectively. In spite of the poor GW localizations, the radio detection likelihood 
is generally higher than the $3$ and $5$-detector networks because of
the smaller GW detectable distances,
though of course we also expect far fewer merger events in the smaller
GW detectable volumes. In particular, for these large GW sky
localizations, recent works have shown how the
use of GW volumes, together with local Universe Galaxy catalogs
(either tracing H-I or H-$\alpha$ -- see e.g., \citealt{KasliwalPhD:2011}),
can provide optimal targeted ranked galaxy follow-up
strategies or can substantially reduce the number of astrophysical
false-positives using the spatial coincidence within or nearby local
galaxies (e.g., NKG13, \citealt{gehrels2016ApJ},
\citealt{Singer:2016}). For radio follow-up of
long-lasting flares and afterglows, we emphasise that the combined use of GW
volumes and galaxy catalogs are critical, in particular, for two
reasons. Firstly, targeted galaxy follow-up will be optimum when the
GW areas cover hundreds of deg.$^2$ because of
the small FoV relative to the mapping speed of some radio telescopes
(Table~\ref{tab:Lspeed}). Secondly, the
GW strain will provide accurate Bayesian-derived distance measures on
the days timescale comparable to the timescale for our suggested first
observational epoch in the radio; we discuss this in detail in the following Sec.~\ref{sec:identification}.

\section{Radio counterpart identification}
\label{sec:identification}


When estimating the radio detectability of GW mergers in Sec.~\ref{sec:detectability},
we have not taken into account: i) the radio emission of the host galaxy 
that may significantly contaminate the merger, and ii) any astrophysical false-positive transients 
and variables that may mimic a 
neutron star binary merger in the huge swaths of the searched
sky. For instance, these transients and variables span a variety of sources from tidal disruption events,
different flavours of supernovae, long GRBs to active galactic nuclei~(AGN). In this section
we now discuss the challenges posed first by the host galaxy contamination and second by the
astrophysical false positive transients. We then provide strategies to overcome them.

\begin{figure*}[t]
\begin{center}
\includegraphics[bb=50 50 554 770,width=60mm,angle=270]{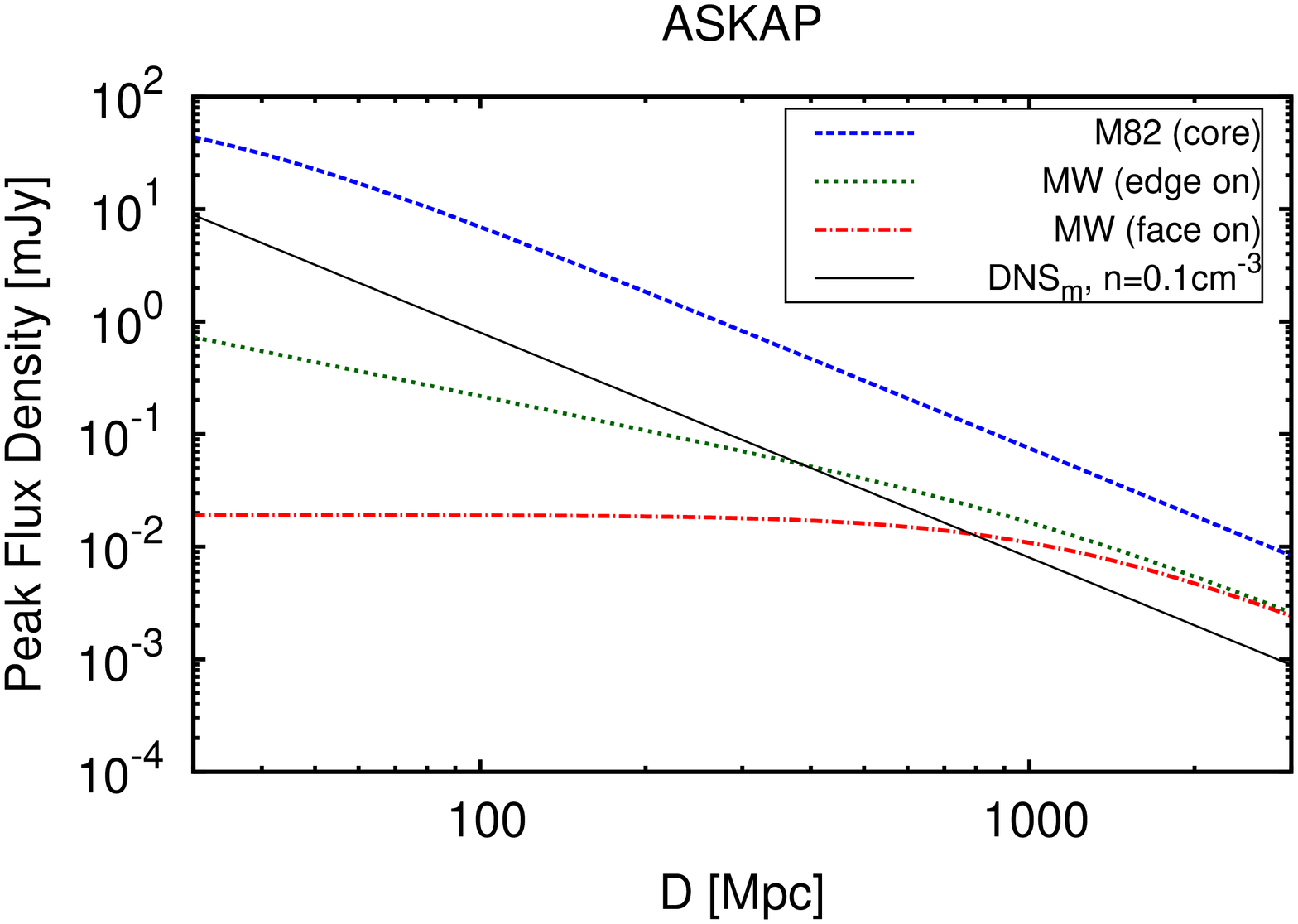}
\includegraphics[bb=50 50 554 770,width=60mm,angle=270]{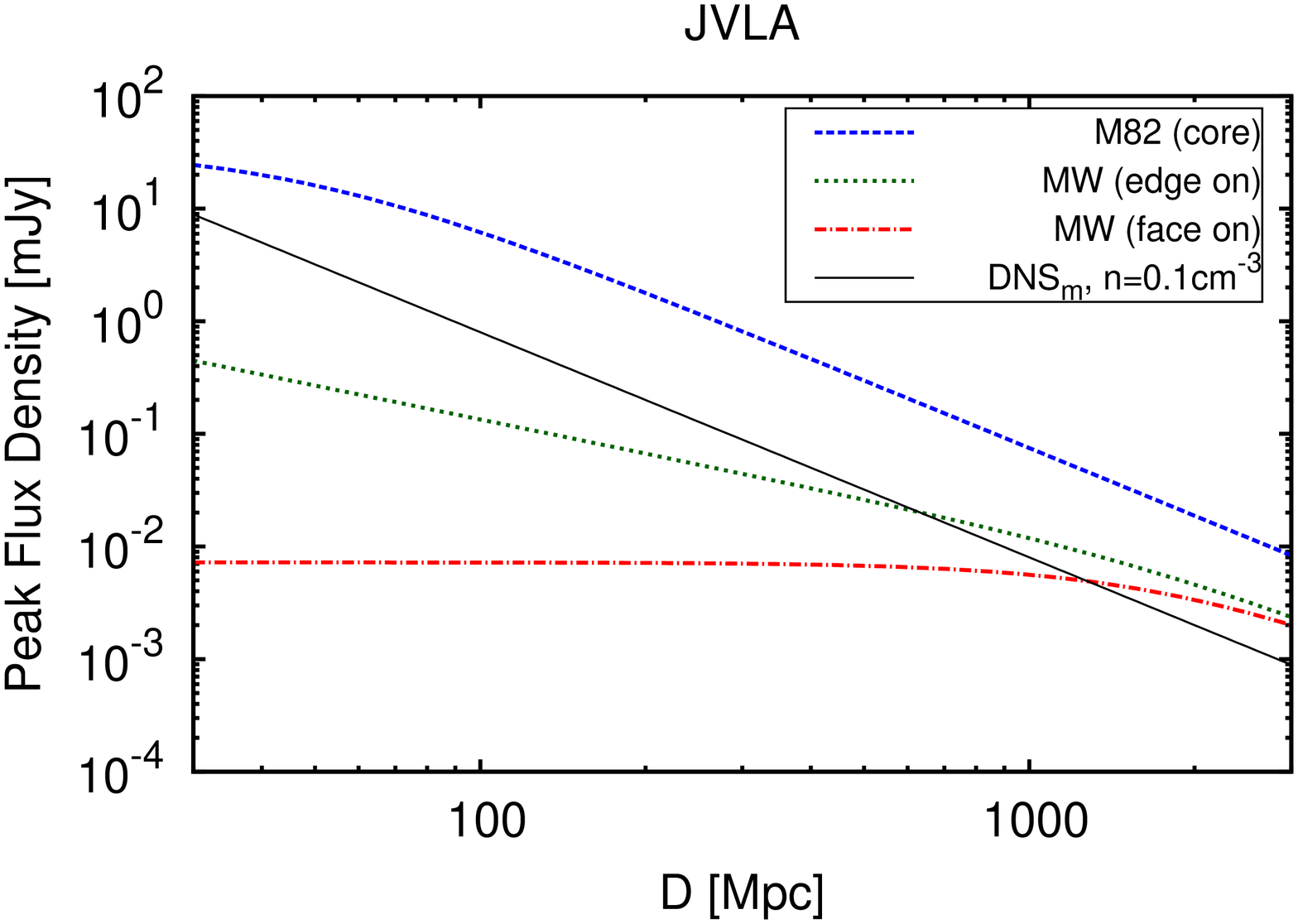}
\end{center}
\caption{The peak flux densities at $1.4$~GHz of star-forming galaxies and 
a long-lasting radio remnant~(DNS$_{m}$ with $n=0.1~{\rm cm^{-3}}$)  in an image
as a function of the distance for the AKSAP~(left panel) and JVLA B-configuration~(right panel). 
As examples, we choose a Milky Way-like galaxy with a luminosity of 
$3\cdot 10^{28}~{\rm erg/s/Hz}$ and diameter of $40$~kpc and
an M82-like galaxy with $10^{29}~{\rm erg/s/Hz}$ 
emitted by a small compact region with a diameter of $1$~kpc. Note that 
the radio counterparts can be typically separated from the  
radio bright cores since mergers typically take place at a few kpc 
away from their host centers. In such cases, the counterparts are detectable.     
For a Milky Way-like galaxy, the face-on~(dot-dashed) and edge-on~(dotted) 
cases are shown.
}
\label{fig:host}
\end{figure*}

\subsection{Host galaxy contamination}
\label{sec:host}

The host galaxies of DNS and BH-NS mergers exhibit radio emission, 
which may contaminate the emission from the radio counterparts of GW events.
For example, the $1.4$~GHz radio luminosities of M33, the Milky Way,
and M82 are $10^{27.5}$, $10^{28.5}$, and $10^{29}~{\rm erg/s/Hz}$
\citep{beuermann1985A&A,condon1990ApJS}
 and these values are comparable or even brighter than the expected luminosities of 
 the radio counterparts (see Table~\ref{tab:models}). 
Here we discuss what is the probability that host
galaxy contamination may prevent identifying GW-radio counterparts.

Galaxies bright in the radio band are either star-forming galaxies or those associated with AGN.
Since the former have radio emission extending spatially much more than the radio
counterparts, i.e., small surface brightness, the contamination of a star-forming galaxy
can be reduced significantly if the angular resolution of a radio facility
is high enough to spatially resolve a galaxy. On the contrary, AGNs and star burst galaxies like M82
have compact radio emitting regions at the centers of galaxies, i.e.,
large surface brightnesses. 
It is also important for mergers in these hosts to spatially 
resolve the galaxy scales to distinguish the radio counterparts
from the compact core of the hosts. In the following, we discuss these
different galaxy types of contamination
separately.

For spatially-extended sources like normal star-forming galaxies, 
the peak flux density $S_p$
in an image is the total flux density $S$ divided by the area of a galaxy~(see e.g. \citealt{Condon15}).
The peak flux density with angular resolution of $\theta$ is given
by: 
\begin{eqnarray}
\frac{S_{p}}{S} = \begin{cases}
 \left(\frac{\theta^2}{\theta^2 + \phi^2} \right) &(\rm{a~face~on~galaxy}),\\
 \left(\frac{\theta^2}{\theta^2 + \phi^2} \right)^{1/2} &(\rm{an~edge~on~galaxy}),
  \end{cases}
\end{eqnarray}
where $\phi$ is the angular diameter of a galaxy. A GW-radio counterpart is detectable if
the ratio of its flux density  to $S_p$ of the host is
larger than a threshold which is determined by a false-positive probability based on the 
statistics of the variabilities of radio sources. Identifying radio
counterparts will be possible when the flux densities of the
counterparts are larger than the peak flux densities of the hosts. 

In Fig.~\ref{fig:host}, we show the peak flux density at $1.4$~GHz of
a Milky Way-like galaxy with a luminosity of $3\cdot 10^{28}~{\rm erg/s/Hz}$ 
and diameter of $40$~kpc \citep{beuermann1985A&A}
and an M82-like galaxy with a bright compact region
with $10^{29}~{\rm erg/s/Hz}$ and $1$~kpc \citep{condon1990ApJS}. Also shown is the peak flux density
of a long-lasting radio remnant, DNS$_{m}$ with $n=0.1~{\rm cm^{-3}}$, which is independent
of the angular resolution of radio telescopes.
For ASKAP, which has an angular resolution of $7^{\prime \prime}$, the peak flux density of
DNS$_{m}$ with $n=0.1~{\rm cm^{-3}}$ is brighter than those of Milky Way-like galaxies
out to $800$~Mpc~($300$~Mpc) for the face-on~(edge-on) case.
For the JVLA B configuration~($\theta=4.3^{\prime \prime}$), the flux density of merger remnants are 
brighter than the peak flux densities of Milky Way-like galaxies out to a distance of $500$~Mpc
even in the edge-on case. Thus, host contamination will not be a serious problem for
Milky Way-like galaxies.

For star-burst galaxies and AGNs, they have bright radio emitting
compact cores with a scale of $\sim 1$~kpc. 
The radio counterparts in such hosts are identifiable if either they are spatially separated
from bright compact cores or the counterparts themselves are brighter than these cores. 
It is important to note that 
more than $90\%$ of sGRBs have  projected physical offsets of $>1$~kpc from their host centers~\citep{berger2014}.  
This suggests that mergers typically take place outside the core regions and
will be detectable if a telescope has angular resolution high enough to separate
a radio counterpart from the core of the host, e.g., $1$~kpc at $200$~Mpc
corresponds to $\sim 1^{\prime \prime}$.
For such a galaxy at $200$~Mpc, the fractions that radio counterparts are 
contaminated by the bright cores are estimated as $0.1$, $0.3$, and $0.7$
for the JVLA A configuration~($\theta=1.3^{\prime \prime}$), B configuration~($4.3^{\prime \prime}$), 
and ASKAP~($7^{\prime \prime}$) respectively.
Here we use the distribution of projected physical offsets of sGRBs~\citep{berger2014}.

Now we turn to estimate the population of  galaxies
that have peak flux densities brighter than a range of GW-radio counterparts. 
Based on the local radio luminosity function of star forming galaxies~\citep{condon2002ApJ},
the number densities of galaxies brighter than 
$L_{1.4}~({\rm erg/s/Hz})\simeq (10^{27},~10^{28},~10^{29})$
are estimated as $n_{>L}~({\rm Mpc^{-3}})\simeq (7\cdot 10^{-3},~3\cdot 10^{-3},~3\cdot 10^{-4})$,
respectively.
Using the number density of galaxies of $n_{\rm gal}\simeq 0.01~{\rm Mpc^{-3}}$,
the estimated fractions of star forming galaxies brighter than  
$L_{1.4}=(10^{27},~10^{28},10^{29})$
are $f_{>L}\simeq (0.7,~0.3,~0.03)$, respectively. 
The same estimates can be done for AGNs using
the AGNs' radio luminosity function \citep{Mauch07}.  
The number densities of radio bright AGNs are 
$n_{>L}\simeq  (10^{-3},~3\cdot 10^{-4},~10^{-4})$
and the fractions are $f_{>L}\simeq (0.1,~0.03,~0.01)$.
Therefore the majority of merger 
events likely take place in star-forming galaxies fainter 
than the Milky Way, for which host contamination will
not be a serious problem. 
$5$--$10\%$ of events may
occur in bright star bursts and AGNs. Even for such cases,
telescopes with high angular resolution can identify the radio
counterparts by separating them from the radio bright regions of hosts.

We can also estimate the population of radio bright star forming galaxies 
hosting merger events based on the star formation rates~(SFRs) of sGRB hosts.
To do so, we use the phenomenological relation between the SFRs and the radio luminosities of star forming galaxies:  
\citep{carilli1999ApJ,condon2002ApJ}:
\begin{eqnarray}
L_{\nu} \approx 1.2 \times 10^{28}~{\rm erg/s/Hz}\left(\frac{SFR}{1M_{\odot}/{\rm yr}}\right)\nonumber \\
~~~~~~\times \left(\frac{\nu}{1.4~{\rm GHz}}\right)^{-0.8},
\end{eqnarray}
where a Kroupa initial mass function with $M \leq 100M_{\odot}$ is assumed.
Applying this relation to sGRB hosts
of which their SFRs are estimated through the 
luminosities of the hosts in the rest-frame $B$-band~\citep{berger2009ApJ,berger2014}. 
The estimated fraction of galaxies hosting sGRBs brighter than 
$10^{28}~{\rm erg/s/Hz}$ is $\sim 0.5$. None of them are brighter than
$10^{29}~{\rm erg/s/Hz}$ and fainter than $10^{27}$~erg/s/Hz. 
These estimates are consistent with those estimated  from 
the radio luminosity function of local galaxies.

We classify the radio counterparts as {\bf bright}, {\bf marginal},
and {\bf faint} events as shown
in the last column of Table~\ref{tab:detectability}.
Here we define the bright events as those with a luminosity of $L>10^{29}~{\rm erg/s/Hz}$,
which are brighter than M82.
The faint events are defined as those with a luminosity of $L<5\cdot 10^{27}~{\rm erg/s/Hz}$, and
the marginal events as those that have luminosities in between these values. 
The radio facilities with low angular
resolution $\theta \sim 7^{\prime \prime}$ will be able to detect most of
the medium events taking place in Milky Way-like galaxies.

\subsection{False positives: Extragalactic radio transients and variables}

{\it Radio transients.} 
There are various kinds of extragalactic astrophysical phenomena associated with
relativistic or mildly-relativistic explosions. They produce
synchrotron radio emission on timescales of a week to years; see \cite{metzger2015ApJ} for
a comprehensive study. Such events may incorrectly be identified as the radio counterparts of GW mergers.  
Radio transient surveys have already been conducted at flux densities
up to $\sim 0.2~{\rm mJy}$~(e.g., \citealt{Bannister2011,ofek2011ApJ,thyagarajan2011ApJ,frail2012ApJ,mooley2013ApJ,Mooley16}).
While many of them have not detected any radio transients,
\cite{Bannister2011} have found $15$ 
in $3000~{\rm deg^{2}}$ at $10$~mJy.  
\cite{Mooley16} also have found a few radio transients in $50~{\rm deg^{2}}$ at $0.2$~mJy.
Two of them are arising from Galactic flaring stars and the others likely from 
variable AGNs.
The derived upper limit on the areal densities of extragalactic radio transients 
is $<0.4~{\rm deg^{-2}}$ for flux densities $\gtrsim 0.2~{\rm mJy}$ at $1.4~{\rm GHz}$
on timescales between a week and three months~\citep{mooley2013ApJ} and 
$<0.04~{\rm deg^{-2}}$ for $\gtrsim 0.5~{\rm mJy}$ at $3~{\rm GHz}$~\citep{Mooley16}.
Therefore, we expect there to be less than one
radio transient per square degree for flux densities of $\gtrsim 0.1~{\rm mJy}$.
Thus, the number of radio false positive transients 
is much smaller than that of optical--infrared counterparts. For comparison, an areal density of 
extragalactic optical--infrared false positives at a depth of 24th apparent magnitude
is $\sim 60~{\rm deg^{-2}}$~(e.g. NKG13).

\begin{figure}[t]
\begin{center}
\includegraphics[bb=50 50 482 302,width=85mm]{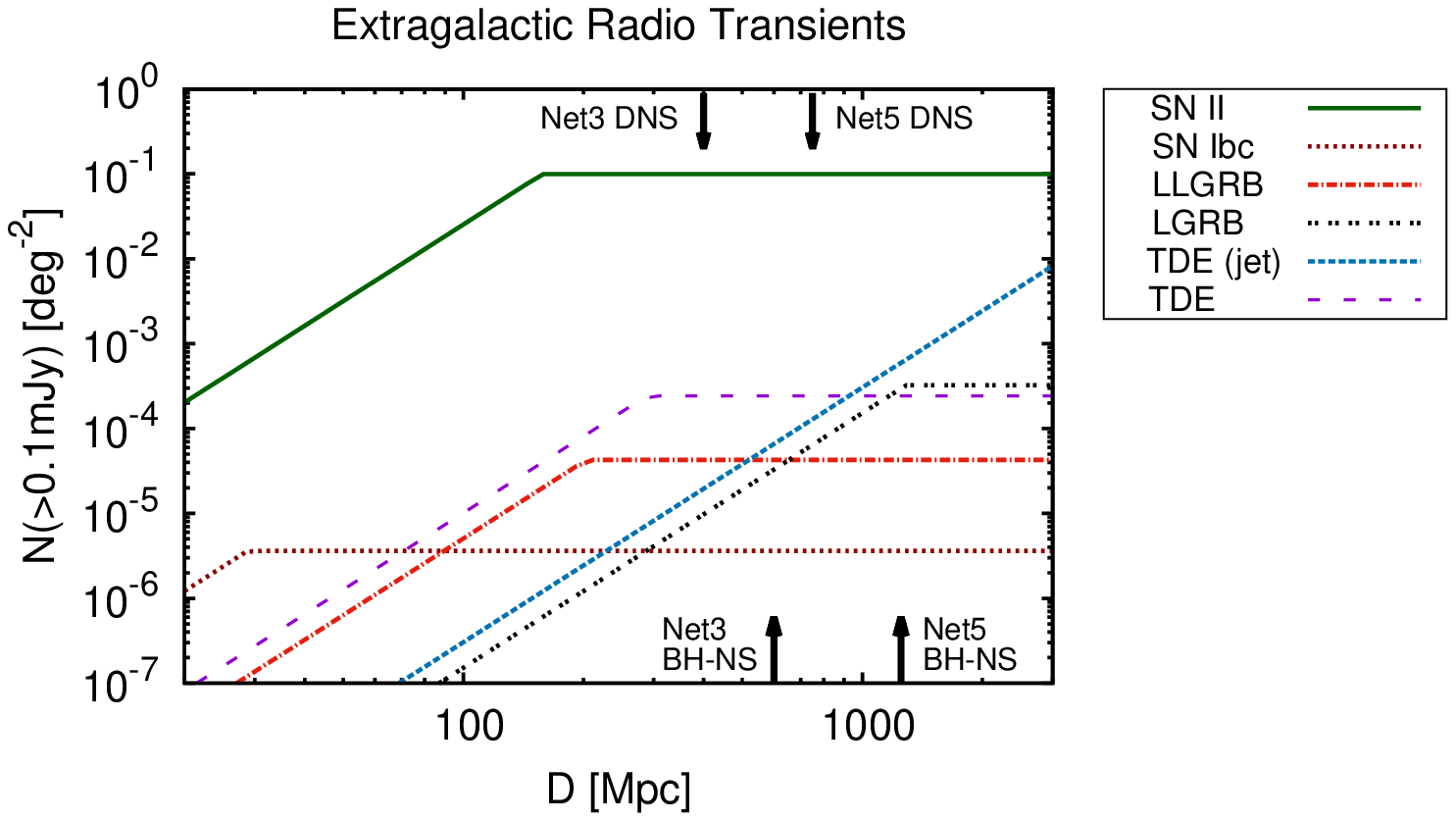}
\end{center}
\caption{The areal densities of radio transients with flux densities
brighter than $0.1~{\rm mJy}$ as a function of the source distance
including: 
type II SNe~(solid; \citealt{levinson2002ApJ,galyam2006ApJ,chevalier1998ApJ,weiler2002ARA&A}),
type Ibc SNe~(dotted; \citealt{berger2003ApJ,soderberg2006ApJ}),
low luminosity GRBs~(dot-dashed; \citealt{soderberg2006Nature,rodolfo2015MNRAS}),
off-axis long GRBs~(double dotted; \citealt{vaneerten2010ApJ,ghirlanda2014PASA}), strong jet TDEs~(dashed; \citealt{zauderer2011Nature,
burrows2011Nature,berger2012ApJ}; 
TDEs~(long-dashed; \citealt{vanvelzen2016Sci,holoien2015,alexiander2015}). 
Also shown are the maximum detectable distances of the GW networks for DNS and BH-NS mergers.
}
\label{fig_tr}
\end{figure}

We estimate the areal densities of radio transients based on the known radio bright astrophysical 
phenomena~(see also \citealt{metzger2015ApJ,Mooley16}). Table~\ref{tab:transients} summrizes 
relevant values of the radio transients used for the estimates of false positives. 
Figure~\ref{fig_tr} shows the expected areal densities
of extragalactic radio transients brighter than $0.1~{\rm mJy}$ within given distances. 
The areal densities of type Ibc supernovae~(SNe Ibc; \citealt{berger2003ApJ,soderberg2006ApJ}), 
low luminosity GRBs~(LLGRB; \citealt{soderberg2006Nature,rodolfo2015MNRAS}),
and tidal disruption events~(TDEs) without strong jets~\citep{vanvelzen2016Sci,holoien2015,alexiander2015} 
are so small that it will be quite rare to detect them as false positive transients.
Although off-axis long GRBs~(LGRBs; \citealt{vaneerten2010ApJ,ghirlanda2014PASA}) and 
tidal disruption events with strong jets
(TDE~(jet); \citealt{zauderer2011Nature,burrows2011Nature,berger2012ApJ}) can be 
false positive transients, they will be identified earlier through
their optical counterparts or can be filtered by identifying their host galaxies since the typical distance of these events is far 
beyond the detectable distance of the GW networks.

A certain fraction of type II 
supernovae have bright radio luminosities of $10^{26}$ -- $10^{28}~{\rm erg/s/Hz}$
on timescales of $100$ -- $1000$~days~\citep{chevalier1998ApJ,weiler2002ARA&A}.
According to the identification of a radio supernova in a radio survey without any other counterparts
\citep{levinson2002ApJ, galyam2006ApJ}, the areal density of radio supernovae is
roughly estimated as $~{0.1~\rm deg^{-2}}$ at $0.1~{\rm mJy}$. 
Thus there will be a few to tens of type II radio supernovae in a GW localization field.
There are several ways to identify radio supernovae.
First, they can be clearly identifiable as supernovae
if the associated supernovae are observable in the optical bands. 
The ongoing and upcoming optical transient surveys are
powerful methods to ensure prior optical identification of such supernovae. A fraction of supernovae, however,
will be missed in optical surveys due to strong dust extinction. Indeed, a supernova SN 2008iz in M82 is
discovered only in the radio bands \citep{brunthaler2009A&A,brunthaler2010A&A}.
Even if associated supernovae are not identifiable, 
they can be distinguished from merger events using timescale arguments,
which are significantly longer than those of GW-radio counterparts, and 
radio spectral properties.
Because radio bright supernovae take place in high circumstellar densities,
their radio spectra are affected strongly by synchrotron
self-absorption and free-free absorption. 
Such strong absorption features should be absent in the radio 
signals arising from compact binary mergers at frequencies 
above $1~{\rm GHz}$.

\begin{table}[t]
\caption{Astrophysical False Positive Transients.}
\begin{center}
\scalebox{0.8}{\begin{tabular}{lccccccc} \hline \hline
Transients & $R~[{\rm Gpc^{-3}yr^{-1}}]$ & $L_{1.4\rm{GHz}}$~[${\rm erg~s^{-1}Hz^{-1}}$] & $T$~[yr]&  Ref. \\ \hline
Type II radio SN & $3\cdot 10^{4}$ & $10^{27.5}$ & $10$ &  [1]\\
Type Ib/c SN & $5000$ & $10^{27}$& $0.3$ &  [2]\\
LLGRB &$500$& $5\cdot 10^{27}$ & $0.1$ & [3]\\
Orphan LGRB &$15$ & $2\cdot 10^{29}$& $3$ & [4]\\
TDE~(strong jets) & $1$ & $10^{31}$& $3$& [5]\\
TDE & $200$ & $10^{28}$& $0.5$& [6]\\
\hline \hline\\
\label{tab:transients}
\end{tabular}}
{\scriptsize \\
References;\\
$[1]$ \cite{levinson2002ApJ,galyam2006ApJ,chevalier1998ApJ,weiler2002ARA&A}, 
$[2]$ \cite{berger2003ApJ,soderberg2006ApJ},
$[3]$ \cite{soderberg2006Nature,rodolfo2015MNRAS},
$[4]$ \cite{vaneerten2010ApJ,ghirlanda2014PASA}, 
$[5]$ \cite{zauderer2011Nature,burrows2011Nature,berger2012ApJ},
$[6]$ \cite{vanvelzen2016Sci,holoien2015,alexiander2015}.
}
\end{center}
\end{table}

{\it Variable radio sources:}
The observed flux densities of persistent extragalactic radio sources vary with time 
due to either intrinsic variabilities or interstellar scintillation.
If the variability of these sources are large enough~($\gtrsim 30\%$)
on timescales between days and a few years, such sources
will be detected as false positives.
According to radio variable studies~(e.g., \citealt{ofek2011ApJb,thyagarajan2011ApJ,mooley2013ApJ,Mooley16}), 
the population of radio variables on these timescales 
with flux densities at $1.4$~GHz between $0.3~{\rm mJy}$ and $100~{\rm mJy}$
is about $1~\%$ or less of the total persistent radio sources and these 
variables are mainly AGNs. 
The areal density of persistent radio point sources with flux densities larger than
$0.1$~mJy is $\sim 1000~{\rm deg^{-2}}$ \citep{Huynh05}
and roughly half of them are AGNs, therefore 
hundreds to thousands of radio variables are expected to be 
in GW localization areas. Most variable AGNs can be rejected by using
their redshift information that will be beyond their GW localization volumes.
However, some fraction of them will remain as false positives and
they are divided into two groups: (i) AGNs inside the GW localization volumes and
(ii) AGNs outside the GW localization volumes but behind the host galaxy
candidates. In what follows, we 
discuss these two cases of AGN false positives separately.

\begin{figure*}[t]
\begin{center}
\includegraphics[bb=50 50 410 302,width=80mm]{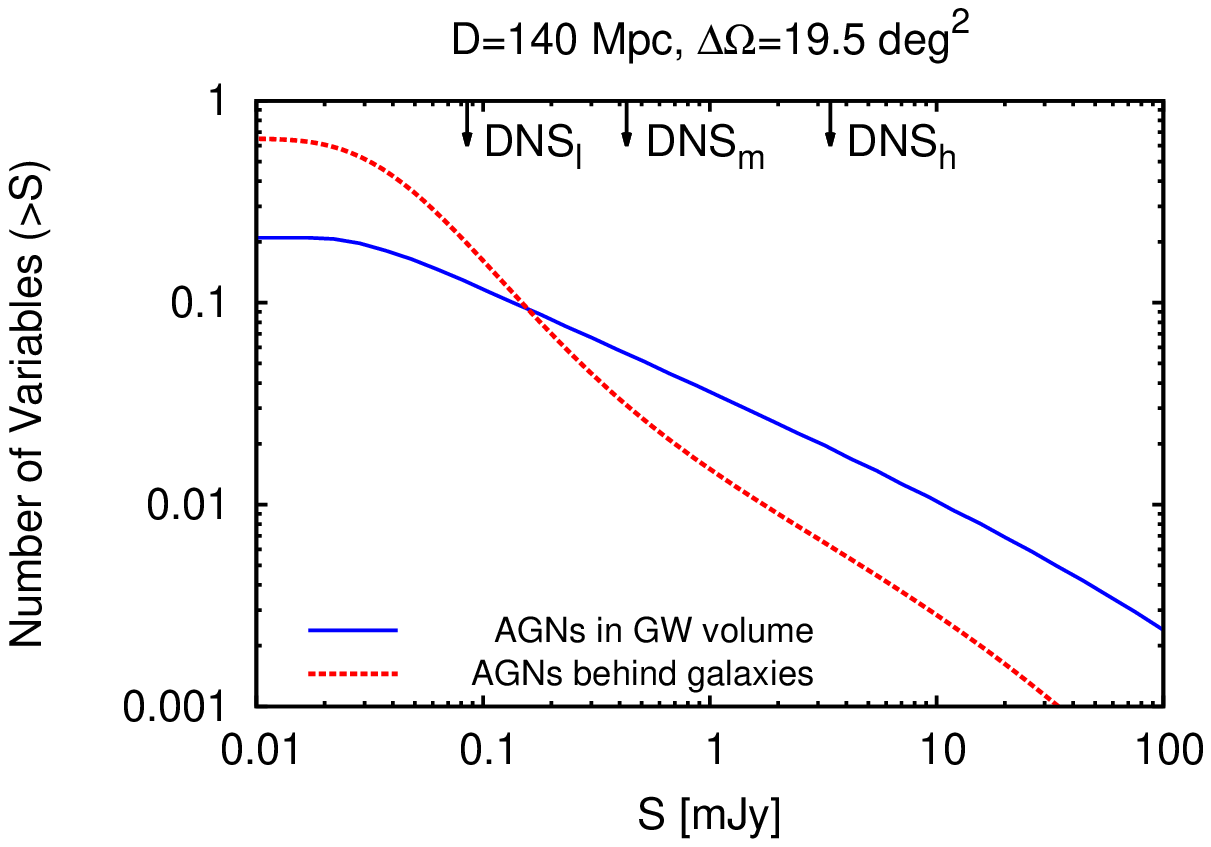}
\includegraphics[bb=50 50 410 302,width=80mm]{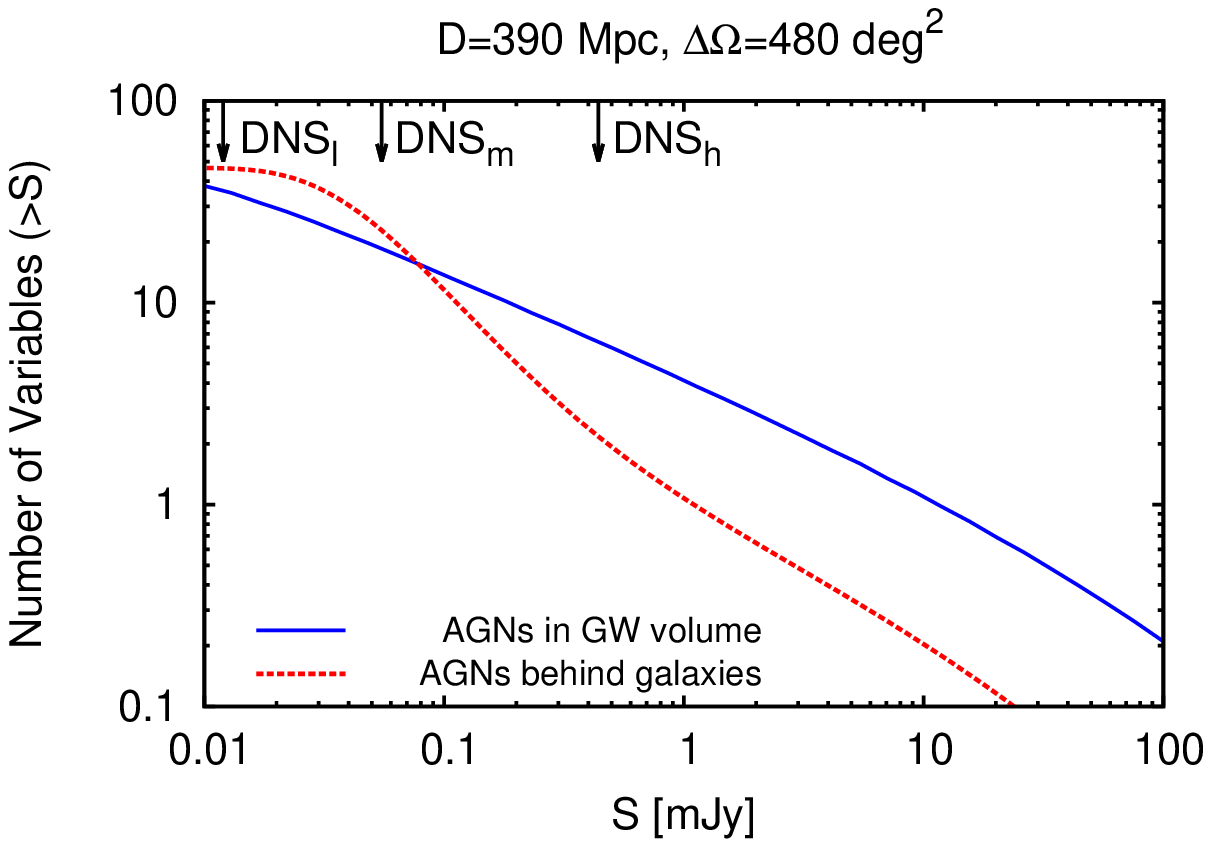}
\end{center}
\caption{
The number of radio variables behind the galaxies within a GW localized volume at $1.4$~GHz
as a function of the flux densities. 
Two specific cases for DNS mergers using GW Net 3 are shown: a merger 
at a distance of $140$~Mpc with
a localization area of $19.5~{\rm deg^{2}}$~(an optimistic case; left panel)
and that at $390$~Mpc with $480~{\rm deg^{2}}$~(a pessimistic case; right panel). 
Also shown are the flux densities of
DNS$_{h}$, DNS$_{m}$, and DNS$_{l}$ with a circum-binary density of $0.1~{\rm cm^{-3}}$.
Note that this analysis does not incorporate the degree of variability.
In reality, less AGNs contribute to false positives for brighter radio counterparts (see the text for details).  
}
\label{fig:var}
\end{figure*}

As discussed in Sec.~\ref{sec:host}, the number of AGNs inside the GW localization volumes
can be estimated based on the local radio luminosity function of AGNs~\citep{Mauch07},
which gives $\sim 3~{\rm deg^{-2}}~(D/450{\rm~Mpc})^3$ at $0.1$~mJy. Given a fraction
of variable sources $\lesssim 1\%$, the number of false positives due to radio variables
inside the GW localization volumes is $\lesssim 3~(D/450{\rm~Mpc})^3(\Delta \Omega_{\rm GW}/100~{\rm{deg^{2}}})$
at $0.1$~mJy. Figure~\ref{fig:var} shows the number of these false positives 
as a function of the flux densities~(blue-solid lines)
for two examples using GW Net 3:  a well-localized merger event~(an optimistic case) at $D=140$~Mpc
with $\Delta \Omega_{\rm GW} = 19.5~{\rm deg^{2}}$~(left panel) and 
a poorly-localized one~(a pessimistic case) at $D=390$~Mpc and with 
$\Delta \Omega_{\rm GW} = 480~{\rm deg^{2}}$~(right panel).
Here we assume $1\%$ of AGNs are variable. Also shown are the flux densities of 
DNS$_{h}$, DNS$_{m}$, and DNS$_{l}$ with a circum-binary density of $0.1~{\rm cm^{-3}}$.
At the flux densities of these models, the expected number of the false positives is 
$6$~--~$40$ for the poorly localizable GW events. On the contrary, this
number is significantly reduced as $0.02$~--~$0.1$ for
the well-localizable GW event
because of a relatively small GW localization volume:
$\Delta \Omega_{\rm GW} = 19.5~{\rm deg^{2}}$ and $D=139^{+79}_{-21}$~Mpc.

AGNs outside the GW localization volumes but 
behind the host galaxy candidates of the GW merger events
will prove more problematic. Assuming each galaxy has
a disk shape with a diameter of $50$~kpc, $\sim 1\%$ of the sky is covered by
galaxies inside a distance of $450$~Mpc so that we expect the number of the false positives due to AGNs behind 
those galaxies as $\sim 5~(\Delta \Omega_{\rm GW}/100~{\rm deg^{2}})$ at $0.1$~mJy. 
The expected number of radio variables behind galaxies
inside of the GW localization volumes as a function of the flux densities
is shown in Fig.~\ref{fig:var}~(red lines).
Here the population of the background radio sources derived by \cite{Huynh05} is used.
The sky areas covered by the host galaxy candidates within the
GW localization volumes are estimated using the number density of galaxies 
$0.01~{\rm Mpc^{-3}}$. The number of the false positives linearly declines with the flux densities
around $0.1$~mJy. For the poorly-localizable case, the expected numbers of these variables
are $1$, $20$, and $50$ at the flux densities of DNS$_{h}$, DNS$_{m}$, and DNS$_{l}$
with $n=0.1~{\rm cm^{-3}}$ respectively.
For the well-localizable case, these values are 
$6\cdot10^{-3}$, $0.03$, and $0.2$.

Some of false positives due to AGNs are removable by using multi-epoch
observations if they do not fade away. In addition,
there are several ways to identify AGNs using: 
(i) radio source catalogs which
will be available thanks to existing and upcoming radio all-sky
surveys, (ii) the locations of the radio counterparts in the host galaxies
compared to the AGN central cores, 
and (iii) AGNs have flat radio spectra around $1$~GHz, 
which is different from those of the radio counterparts.
Here the method (ii) is valid only for AGNs inside
the GW localization volumes. Note that the analysis here does not
incorporate the degree of variability. The population of
variable AGNs decreases with the modulation index $\Delta S/S=2|S_1-S_2|/(S_1+S_2)$,
where $S_1$ and $S_2$ are the flux densities in two epochs.
For instance, \cite{Mooley16} find that only one 
out of $3700$ radio sources is highly variable as $\Delta S/S\approx 1$, 
thereby the number of false positives due to radio variables
is significantly reduced for brighter radio counterparts.
Note that, however,
the population of radio variable sources depends on
the sensitivity, observed frequency, time scale, and direction of the sky
and such analysis still remains unqualified at the flux densities of
GW-radio counterparts.  
Therefore a critical understanding of the properties of radio variable
sources is necessary to identify GW radio counterparts, especially in
the era of GW astronomy where we may have tens of GW detections per year.

The above discussions are based on the assumption that
we have a galaxy catalog covering the GW localization volumes.
This significantly reduces the number of false positives. 
However, the spectroscopic galaxy catalogs currently available are 
incomplete, in particular, beyond $200$~Mpc. This incompleteness
that is not so critical for DNS merges, will be crucial for identifying the radio counterparts to BH-NS mergers, 
of which the detectable distances in GWs are as high as $1$~Gpc. 
Therefore, deeper optical observations will be necessary to 
complete galaxy catalogs out to the edge of the GW localization volumes
when identifying the radio-GW counterparts for such cases.

\section{Comparison with previous works}
\label{sec:Discussion}

To compare our radio counterpart detectability results with
previous works  we translate the detection likelihood
to a detection rate for a given merger rate density $R$. 
For DNS$_{m}$ with $n=0.1~{\rm cm^{-3}}$, the expected radio detection rates
are $\sim 7$ and $20~{\rm yr^{-1}}~(R/500~{\rm Gpc^{-3}~yr^{-1}})$ for the JVLA
and MeerKAT in GW Net 3~(see Eqn.~(7) in NKG13).
\cite{metzger2015ApJ} studied the detectability of extragalactic radio transients
and found the expected detection rate 
of long-lasting radio remnants arising from the DNS merger ejecta is
$\lesssim 0.03~{\rm yr^{-1}}~(R/500~{\rm Gpc^{-3}~yr^{-1}})$ 
for a three-year survey with ASKAP and JVLA.
This rate is much lower than the one we find for a number of reasons. 
Our work focuses on the follow-up surveys of GW merger events so that 
the observations are optimized and can reach the sensitivity as deep as
$\sim 0.1~{\rm mJy}$. On the contrary, \cite{metzger2015ApJ} considered 
{\it blind surveys} that
can detect radio transients with much higher  flux densities of $1$~--~$5$~mJy. 
Moreover, many events are missed in \cite{metzger2015ApJ} due to the 
variability criterion for detections because the peak timescale of 
the signals is too long compared to the duration of the surveys, i.e,
one cannot recognize the radio signals as transients.
This reduces the detection rate by an order of magnitude.
Note also that the ejecta model of \cite{metzger2015ApJ} is an outflow with
a single velocity component with $v=0.2c$, 
which gives a longer peak timescale and a fainter peak flux density
than an outflow with multi-velocity components as given by Eqn.~(\ref{eq:pheno}).
If one takes the ejectas' multi-velocity components into account,
the detectability of neutron star binary mergers in blind surveys may increase.

Long-lived magnetars have been proposed to explain the 
prompt GRB emission or late-time X-ray activities of sGRBs
based either on the spin down luminosity \citep{fan2006MNRAS, troja2007ApJ,
metzger2008MNRAS, rowlinson2013MNRAS, lu2015ApJ, siegel2016ApJa,siegel2016ApJb, gao2015},
or on outflows powered by differential rotation~\citep{shibata2011ApJ, kiuchi2012PRD, siegel2014ApJ}.
In fact, numerical simulations have recently shown strong amplification of magnetic fields at 
merger~\citep{price2006Sci,giacomazzo2013ApJ, giacomazzo2015ApJ,kiuchi2015PRD}. 
The spin-down magnetar model predicts that the ejecta can derive a large amount of kinetic energy 
$\sim 10^{52}$~erg from the magnetar itself. For typical merger ejecta masses,
such ejecta expand with relativistic velocities and result in bright radio emissions.

\cite{metzger2014MNRAS} constrained such magnetar models 
using late-time radio observations of sGRBs and ruled out 
a magnetar remnant in GRB 050724 and 060505.
More recently, \cite{horesh2016ApJ} provided strong constraints on 
magnetar activity  of the macronova candidates: GRB 060614 and 130603B.
In addition, radio transient surveys can put strong constraints on the
the formation rate of the magnetars~\citep{metzger2015ApJ}. 
The current limit on the rate is about $5~{\rm Gpc^{-3}\,yr^{-1}}$, 
which is already much lower than the expected neutron star 
merger rate by one to two orders of magnitude.
At the typical distance of GW merger events,  
the expected radio flux densities of the magnetar models 
are $\sim 100$~--~$1000$~mJy so that the radio follow-up observations
will easily  detect such signals. Moreover, these signals are sufficiently brighter than
the typical radio luminosity of the galaxies and false positive
transients and variables. Therefore, identification of post-merger
magnetar emission from GW events will be relatively straightforward.

\section{Recommendations for the radio surveys}
Here we briefly summarize our recommendations for the radio surveys.
We propose that  radio follow-up observations of GW mergers at $1.4$~GHz 
take place  in five epochs separated by logarithmic time intervals: within a day after the detection, at
$\lesssim 10$~days, at $\sim 30$~days, at $\sim 100$~days, at $\sim300$~days and at $\gtrsim 1000$~days.  
At $3$~GHz, a similar strategy can be employed. The radio surveys at $3$~GHz 
have an advantage of high angular resolution, which reduces the radio contamination 
of the host galaxies, while the peak fluxes
are lower than those at $1.4$~GHz by a factor of $\sim 2$.   
At $150$~MHz, we recommend conducting intensive surveys at late times~(a year or later) because 
the radio light curves at the low frequencies arise at later times.
Importantly, the peak flux densities are higher than those at $1.4$~GHz. 
Therefore, the late-time observations at $150$~MHz will be quite important for both the detections and
confirmations of the GW radio counterparts. 
Finally, the comparison with the optical data will be necessary to identify
the redshift of the radio counterpart candidates otherwise astrophysical false positives mainly radio variable AGNs
contaminate significantly.

\section{Summary and Discussion}
\label{sec:Conclusion}

We have explored optimized strategies for detecting long-lasting radio signals arising from compact binary mergers
following detection of GW events. 
To do so, we first simulated GW merger events and constructed  mock catalogs of
detectable GW events. We then computed the expected radio flux densities
assuming different ejecta models and circum-merger densities for each detectable GW event. 
We focused on synchrotron radiation 
arising from  (i) sub-relativistic merger ejecta~(long-lasting radio remnants) 
and (ii) ultra-relativistic jets~(orphan GRB afterglows).
The radio flux densities depend on the unknown ejectas' kinetic energy and velocity distributions and
the circum-merger densities.
Here we take into account uncertainties in the ejecta's kinetic energy and velocities according to
numerical simulations of compact binary mergers and three different circum-merger densities of
$0.01~{\rm cm^{-3}}$, $0.1~{\rm cm^{-3}}$, and $1~{\rm cm^{-3}}$ based on the Galactic DNS population.

Based on the derived light curves, we propose that  radio follow-up
observations of GW mergers at $1.4$~GHz take place  in five epochs separated
by logarithmic time intervals: within a day after the detection, at
$\lesssim 10$~days, at $\sim 30$~days, at $\sim 100$~days, at $\sim
300$~days and at $\gtrsim 1000$~days. We compare the expected radio
flux density of each GW merger event with the sensitivities of a slew of radio facilities 
assuming that each radio telescope searches the radio counterparts in the GW localization areas. 
Assuming  a total observation time of $30$~hr for each epoch,
we derive the detection likelihood of each ejecta model. 
Note that we have not taken into account
Northern and Southern hemisphere considerations of the GW sky
localizations, and hence the following relative detectability fractions
should be reduced by approximately a factor of two.
For the sub-relativistic merger ejecta, the JVLA will detect $5$--$90\%$
of the GW events with a GW network of three detectors for circum-merger densities of $0.1~{\rm cm^{-3}}$.
For the orphan GRB afterglows with a canonical~(large) kinetic energy,
$\sim 10~(40)\%$ of the GW events will be detected by the JVLA.
We find that the detection likelihood does not change significantly for a GW network with 
five detectors because the gain in the radio sensitivity 
due to better localizations somehow compensates the loss in radio flux densities
due to the increase in the GW detectable distance.

The detection likelihood increases if we conduct follow-up observations
only for the well-localizable GW events.
For instance, the JVLA can detect more than $60\%$ ($15\%$) of the GW events
of which GW localization areas are better than $20~{\rm deg^{2}}$
for the ejecta model with medium kinetic
energies, velocities, and a density of $0.1~{\rm cm^{-3}}$ ($0.01~{\rm cm^{-3}}$).
Note that the probability of localizing a GW DNS merger to within a sky area of 
$20~{\rm deg^{2}}$ is about $25\%$ for
a GW network of three detectors.

The detectable radio signals at $150$~MHz appear at later times~($\sim 3$ years or later)
and the sensitivity of wide-field searches may be limited by the confusion noise.
However, it is quite important to search for the radio counterparts 
at low frequency since the spectral properties at low frequency of 
radio counterparts are quite different from those of other radio transients
and variables, e.g., radio supernovae and AGNs. 
Such observations will be significant to discriminate 
between radio counterparts from other astrophysical false positives. Therefore, we suggest 
that low frequency arrays, such as LOFAR, search with high angular resolution 
for the radio counterparts of GW events
for which the radio counterpart candidates have been detected earlier by other
radio facilities at higher frequencies.

Looking into the next decade, the Square Kilometer Array~(SKA) will achieve much faster survey speeds than those
used  in this work. It will hence detect radio counterparts
of GW mergers far more efficiently.
Here we discuss prospects for SKA-mid and focus on the detectability prospects
based on DNS$_{m}$ at $1.4~{\rm GHz}$ with the GW Net 5.
In order to detect most of the radio counterparts of GW merger events at
$n=0.1~{\rm cm^{-3}}$, 
the required survey speed is $100$ times faster than that of the JVLA.
The survey speeds of SKA-mid will reach this value. 
Therefore we expect a significant progress in studies of radio
counterparts once SKA and GW Net 5 coincidently both become operational.

We discuss the possible contamination of the host galaxies,
which can be divided into normal star-forming galaxies that spatially 
extend on a radial scale of $\sim 10$~kpc and AGNs and star-bursts that have 
central radio bright compact regions on a radial scale of $\sim 1$~kpc. For both 
cases, resolving the hosts will greatly reduce the contamination. For instance,
the radio counterparts are identifiable for DNS$_{m}$ with a density of $0.1~{\rm cm^{-3}}$
in Milky Way-like galaxies out to $400$~Mpc. 
While the probability is low, mergers may 
take place in radio bright AGNs and star bursts. In order to be
identifiable, the radio counterparts must be separable spatially
from the galaxies' bright compact regions.
For such a galaxy at $200$~Mpc, we estimate that
the fractions that the radio counterpart are masked due to galaxy contamination
are $0.1$, $0.3$, and $0.7$
for the JVLA A configuration, B configuration, and ASKAP respectively. 

Astrophysical false positives, include extragalactic radio transients
and variables, may mimic the radio counterparts.
We estimate that a few to tens of radio transients 
will be detectable at $0.1$~mJy as false positives and most of them maybe type II supernovae. 
These false positives can be rejected by using the optical identification of
supernovae, their longer timescales, and strong-absorption feature in the radio spectra.

The false positives due to radio variables will be more problematic. A few hundreds to thousands of 
variables~(mainly AGNs) will be detectable 
at flux densities comparable to the detectable  GW - radio counterparts in
a GW localization area of $\sim 100~{\rm deg^{2}}$.
Most of these false-positives will be located at distances far beyond
the GW detectable distance and they can be rejected as merger 
candidates by identifying their distances or redshifts.
However, there will be a few to tens of radio variables in
the GW-localization volumes. Moreover some fraction of radio 
variables beyond the GW-horizon distance will be located behind 
the host galaxy candidates. They can be rejected by using multi-epoch variability test, the location of the host,
and the flatter spectra than that of the radio counterparts.
It is worth emphasizing that the number of false positives is significantly reduced
for well-localized events.
We expect there to be less than one false positive for such an event.
Note the number of false positives due to variables
per unit sky area depends on the sensitivity, degree of variability,
observed frequency, and the sky direction,
so that a better understanding of the statistical properties of radio variables 
will be important for identifying radio counterparts.

In summary, while there are uncertainties in the ejectas' kinetic energy, velocity, and
circum-merger density, a certain fraction of GW merger events will be detectable with
current and upcoming radio facilities. In addition, identifying radio counterparts will not be so difficult thanks to the
relatively quite radio transient sky. 
We therefore advocate radio counterpart surveys for GW merger events of which 
EM counterparts in other wavelengths~(e.g., X-ray and optical) 
have not been detected using not only the GW sky error but the GW
distance and the use of galaxy catalogs thanks to the relatively long
timescales of the radio emission.

\acknowledgements
This work was supported by the I-CORE Program of the Planning and Budgeting Committee and The Israel Science Foundation (grant No 1829/12) and by an ISF-CNSF grant.
S.M.N. acknowledges generous support from the Radboud University
Excellence Fellowship. Part of this research was carried out at the Jet Propulsion Laboratory, 
California Institute of Technology, under a contract with the National Aeronautics and Space Administration.

We are very grateful to Dale Frail for careful reading of the manuscript and for
initially encouraging us to embark on this project. We thank Keith Bannister, Paz Beniamin, Robert Braun, 
Jess Broderick, Aaron Chippendale, Assaf Horesh, Rob Fender, Jason Hessels, David Kaplan, 
Mansi Kasliwal, Shri Kulkarni, Duncan Lorimer, Kunal Mooley, Tara Murphy,
Steve Myers, and Antonia Rowlinson for useful discussions. We thank the anonymous referee
for suggestions that improve our work.

\begin{figure}[t]
\begin{center}
\includegraphics[bb=50 50 482 302,width=85mm]{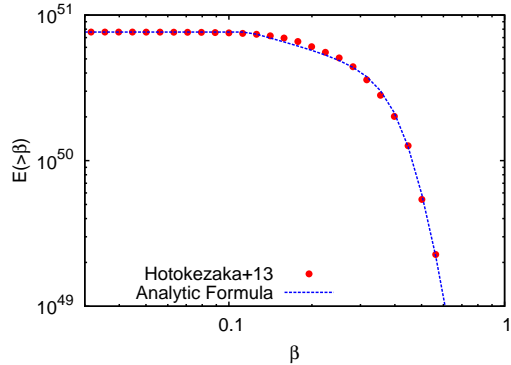}
\end{center}
\caption{Ejecta kinetic energy above a given velocity. Red points
show a result of a numerical relativity simulation by \cite{hotokezaka2013PRDa}.
Also shown as a blue line is an analytic formula given by Eqn.~(\ref{eq:pheno}).
}
\label{fig_E}
\end{figure}

\begin{appendix}
We show the ejecta kinetic energy profile in Fig.~\ref{fig_E}
taken from a numerical-relativity simulation by \cite{hotokezaka2013PRDa}.
Here a result of an equal-mass DNS merger simulation 
with a total mass of $2.8M_{\odot}$ and a neutron star equation of state APR4
is shown as an example. We also depict an analytic formula, a power law with
an exponential cut off~(see Eqn.~\ref{eq:pheno}). It can be seen that this 
formula describes the result of the simulation well. Note that a significant
fraction of material, more than $25\%$ in terms of the kinetic energy, 
has velocities above $0.4 c$. These fast components contribute to the 
radio light curves at early times. As a result, the flux densities 
at early times are brighter than those expected from an ejecta model
with a single velocity component.

\end{appendix}
\vspace{1cm}



\end{document}